\documentclass[%
  preprint,
  superscriptaddress,
  nofootinbib,
  nobibnotes,
  amsmath,amssymb,
]{revtex4-2}
\usepackage{hyperref}
\usepackage{multirow}
\usepackage{listings}
\usepackage{datetime2}
\usepackage{tikz}
\usepackage[compat=1.1.0]{tikz-feynhand}
\usetikzlibrary{external}
\usepackage[compat=1.1.0]{tikz-feynman}
\usepackage{ulem}
\usepackage{setspace} 
\usepackage{here}
\usepackage{fancyvrb}

\usepackage[title,titletoc]{appendix} 

\graphicspath{ {./figures/} {./tikz/} }

\definecolor{codegreen}{rgb}{0,0.6,0}
\definecolor{codegray}{rgb}{0.5,0.5,0.5}
\definecolor{codepurple}{rgb}{0.58,0,0.82}
\definecolor{backcolour}{rgb}{0.95,0.95,0.92}

\newcommand{\HELAS}{\mbox{\sc Helas}}
\newcommand{\MadGraph}{\mbox{\sc MadGraph}}
\newcommand{\MG}{\mbox{\sc MG5}}

\newcommand{\MadEvent}{\mbox{\sc MadEvent}}
\newcommand{\ALOHA}{\mbox{\sc Aloha}}

\newcommand{\shat}{\mbox{$\hat{s}$}}

\newcounter{romannum}
\newcommand{\Rnum}[1]{\setcounter{romannum}{#1} \Roman{romannum}}

\begin{document}

\vspace*{-2cm}
\begin{flushright}
  KEK-TH-2809, IRMP-CP3-26-05
\end{flushright}
%
\title{\Large Multi-channel phase space \\with Feynman-diagram-gauge amplitudes}
%
\author{Kaoru Hagiwara}
\affiliation{\footnotesize KEK Theory Center and Sokendai, Tsukuba, Ibaraki 305-0801, Japan}
\author{Junichi Kanzaki}
\email{junichi.kanzaki@gmail.com}
\affiliation{\footnotesize Research Support and Industry-Academic Collaboration Center, Iwate University, Morioka, Iwate 020-8550, Japan}
%
\author{Fabio~Maltoni}
\affiliation{\footnotesize Centre for Cosmology, Particle Physics and Phenomenology (CP3),
Universit\'{e} Catholique de Louvain, B-1348 Louvain-la-Neuve, Belgium}
\affiliation{\footnotesize  Dipartimento di Fisica e Astronomia, Universit\`{a} di Bologna, Via Irnerio 46, 40126 Bologna, Italy}
\affiliation{\footnotesize  INFN, Sezione di Bologna, Via Irnerio 46, 40126 Bologna, Italy}
\author{Kentarou Mawatari }
\affiliation{\footnotesize  Faculty of Education,  Iwate University, Morioka, Iwate 020-8550, Japan}
\affiliation{\footnotesize  Graduate School of Arts and Sciences, Graduate School of Science and Engineering, Iwate University, Morioka, Iwate 020-8550, Japan}
\author{Ya-Juan Zheng}
\affiliation{\footnotesize  International College, The University of Osaka, Toyonaka, Osaka 560-0043, Japan}
%
\begin{abstract}
Multi-channel phase space with a single Feynman diagram enhancement is a powerful tool for high-energy physics event generation if a diagram with a singular propagator dominates the total scattering amplitude at the corresponding singular kinematical region, and when the interference among amplitudes is not larger than the square of each amplitude.
These conditions are satisfied in the Feynman-diagram-gauge amplitudes for both unbroken (QED and QCD) and broken (EW) gauge theories. 
We illustrate the usefulness of this method in lepton collider processes that are challenging to accurately simulate at very high energies, i.e., 
$ l\bar{l} \to \nu_l \bar{\nu}_l t\bar{t} H $,
$ l \bar{\nu}_l t\bar{b} H $, and 
$ l \bar{l} t\bar{t} H $,
in the SMEFT  with a complex top-Yukawa coupling.
The total cross sections of the latter two processes contain lepton-mass singularities arising from $t$-channel photon-exchange diagrams. To address this issue, we develop a phase-space parametrization that accurately generates the distributions of forward-emitted charged leptons. We modify the \HELAS\ library to evaluate helicity amplitudes in the singular region so that vertices at very small invariant momentum square of order $m_e^2$ can be accurately evaluated even at multi-TeV energies.
\end{abstract}
\maketitle
\setcounter{tocdepth}{2}
\tableofcontents

\newpage
\section{Introduction}
\label{sec:intro}

Matrix-element event generators play a central role in precision collider phenomenology, providing first-principles predictions for total rates, fully differential distributions and parton-level events for fully exclusive simulations across a wide range of processes. They are indispensable tools for the interpretation of current collider data and for the design and physics case of future facilities, where increasingly complex final states and higher energies demand reliable and efficient numerical integration of scattering amplitudes. Over the past decades, substantial progress has been achieved in the automation of matrix-element calculations including next-to-leading order corrections in QCD and electroweak interaction and event generation, leading to widely used general-purpose frameworks that combine exact fixed-order matrix elements with sophisticated Monte Carlo integration and sampling techniques~\cite{Campbell:2022qmc}.

Looking ahead, future collider concepts pose new and qualitatively different challenges for event generation. 
At future lepton colliders operating at multi-TeV energies, and in particular at muon colliders or plasma-accelerator--based machines reaching several tens of TeV in the center-of-mass frame, scattering processes will be characterized by extreme kinematical configurations. 
These include strongly forward or collinear particle production, enhanced sensitivity to small invariant masses and momentum transfers, and large hierarchies among physical scales.  In such regimes, large logarithmic enhancements associated with collinear lepton and photon radiation can develop. 
For sufficiently inclusive observables, these logarithms can be systematically resummed into effective lepton and photon parton distribution functions, as demonstrated in Refs.~\cite{Frixione:2025wsv,Frixione:2025guf}.  However, for fully exclusive observables such resummations are not applicable. 
Standard phase-space parametrizations may become inefficient or numerically unstable, and the faithful generation of rare but phenomenologically relevant regions of phase space, see e.g.\ Ref.~\cite{Ruhdorfer:2024dgz}, becomes a critical issue.

The construction of efficient phase-space mappings is a long-standing and still highly active topic in event generation and cross-section integration. Early foundational work established the general structure of multi-particle phase space and its recursive decomposition~\cite{BycklingKajantie1973}, while subsequent developments introduced adaptive and multi-channel techniques tailored to the singularity structure of perturbative amplitudes~\cite{Kleiss:1986gy,Kleiss:1994qy}. These ideas underpin many automatic phase-space generators and matrix-element tools, including dedicated algorithms and libraries developed to handle complex multi-particle final states~\cite{Papadopoulos:2000tt,vanHameren:2002tc,Maltoni:2002qb,Mangano:2002ea,Gleisberg:2003xi,Kersevan:2004yh,vanHameren:2010gg,Mattelaer:2021xdr}. More recently, renewed interest in phase-space construction has led to novel approaches, ranging from analytically inspired parametrizations~\cite{Larkoski:2020thc} to machine-learning--based importance sampling and multi-channel strategies~\cite{Klimek:2018mza,Bothmann:2020ywa,Gao:2020vdv,Gao:2020zvv,Butter:2022rso,Heimel:2022wyj,Bothmann:2023siu,Heimel:2023ngj,Heimel:2026hgp}. In this broader context, it remains highly valuable to develop physically motivated phase-space parametrizations that directly exploit the structure of Feynman amplitudes and their dominant singularities, especially in the challenging kinematic regimes anticipated at future high-energy colliders.

In this paper, we build upon the idea of single-diagram-enhanced (SDE) multi-channel phase-space (MCPS) integration, originally proposed in Ref.~\cite{Maltoni:2002qb} to achieve efficient event generation for high-energy particle collisions. This strategy was instrumental in transforming the matrix-element calculator \MadGraph~\cite{Stelzer:1994ta} into the event generator \MadEvent~\cite{Maltoni:2002qb}, later embedded in the  \textsc{MadGraph5\_aMC@NLO} framework~\cite{Alwall:2014hca}. The basic idea can be expressed as follows. 

The differential cross section of the 2 to ($n-2$) particle
scattering process
\begin{align}
  a(p_1) + b(p_2) \to c_1(p_3) + c_2(p_4) + \cdots + c_{n-2}(p_{n})
  \label{eq:FD}
\end{align}
at leading order can be expressed as
\begin{align}
  d\sigma = \frac{1}{F}\; d\Phi_{n-2} \sum_{h_i,c_i} |{\cal M}(a+b\to c_1+\cdots +c_{n-2})|^2 ,
\end{align}
where ${\cal M}$ is the helicity amplitudes of the process, $h_i/c_i$ is the helicity/color quantum number of the external particle labeled by $i$ ($i=1,2,\cdots,n$), and $d\Phi_{n-2}$ is the ($n-2$)-body phase space.
The invariant flux factor 
\begin{align}
  F = 4 ((p_1\cdot p_2)^2-(m_1)^2(m_2)^2)) \times n_1 \times n_2
\end{align}
accounts for Fermi's plane wave normalization factor times the helicity/color multiplicity ($n_1$ and $n_2$) of the initial particles $a$ and $b$.
In the following, we do not show explicitly the flux factor $1/F$ and the summation over helicity/color of the external particles $\sum_{h_i,c_i}$,
which are common to all the processes studied in this paper.

When the helicity amplitudes are calculated from Feynman diagrams, they are expressed as a summation over all the contributions of individual Feynman amplitudes:
\begin{align}
 {\cal M}  = \sum_{k=1}^{N_{d}} {\cal M}_k ,
\label{eq:amp_tot}
\end{align}
where $N_{d}$ is the number of diagrams.
We can symbolically express the differential cross-section as follows: 
\begin{align}
  d\sigma = d\Phi_{n-2}\, \bigg| \sum_{k=1}^{N_{d}} {\cal M}_k \bigg|^2 .
\end{align}
The core idea is to decompose the phase-space integral into channels associated with individual Feynman diagrams, each equipped with a phase-space parametrization adapted to its dominant singular structure, namely
\begin{subequations}
	  \label{eq:dsigma_mcps}
  \begin{align}
    \label{eq:dsigma_mcps_a}
    d\sigma & = \sum_{k=1}^{N_{d}} d\sigma_k , \\
    \label{eq:dsigma_mcps_b}
    d\sigma_k & = d\Phi_{n-2}\, |{\cal M}_k|^2 R , \\
    \label{eq:dsigma_mcps_c}
    R & = \frac{|\sum_j {\cal M}_j|^2}{\sum_l |{\cal M}_l|^2} .
  \end{align}
\end{subequations}
The correct differential cross-section by the sum in Eq.~\eqref{eq:dsigma_mcps_a} is guaranteed by the identity
\begin{align}
  1 = \sum_{k=1}^{N_{d}} \frac{|{\cal M}_k|^2}{\sum_l |{\cal M}_l|^2} .
\end{align}
The $(n-2)$-body phase space integral of the squared matrix elements is split into $N_{d}$ channels when the number of Feynman diagrams is $N_{d}$.
The integral of each channel, Eq.~\eqref{eq:dsigma_mcps_b}, can be done efficiently if the singular behavior of $|{\cal M}_k|^2$ can be accounted for by appropriately choosing the integration variables, and the ratio $R$ is of the order of unity in the region of the phase space where $|{\cal M}_k|^2$ is singular.

This approach has been proven effective for a broad class of applications, in particular for multi-particle production processes at the LHC.
At higher energies, however, its performance is limited by gauge cancellations among interfering amplitudes, a generic feature of formulations based on covariant gauge propagators for gauge bosons, including the photon, the gluon, and the electroweak bosons $W$ and $Z$.

In \MadGraph~\cite{Stelzer:1994ta}, \MadEvent~\cite{Maltoni:2002qb}, and in their successive releases~\cite{Alwall:2007st,Alwall:2011uj,Alwall:2014hca}, the scattering amplitudes are calculated by \HELAS~\cite{Hagiwara:1990dw,Murayama:1992gi} libraries (which are currently automatically generated by \ALOHA~\cite{deAquino:2011ub}), which adopt the Feynman gauge for the photon and the gluon, while the unitary gauge is used for the weak bosons.
They are both covariant gauges, and the amplitudes suffer from subtle gauge theory cancellation.

Recently, a non-covariant gauge called the Feynman-diagram (FD) gauge has been introduced for the photon and the gluon~\cite{Hagiwara:2020tbx} and for the weak bosons~\cite{Chen:2022gxv,Chen:2022xlg}. 
A common feature of the FD gauge amplitudes is that they are free from so-called subtle `gauge cancellation' among interfering amplitudes and the examples presented in Refs.~\cite{Hagiwara:2020tbx,Chen:2022gxv,Chen:2022xlg,Furusato:2024ghr} give the following observations:
(i) A single diagram, or a set of diagrams with a common gauge-boson propagator dominates the total amplitude in the collinear final-state configurations where the propagator is large.
(ii) The ratio $R$ of Eq.~\eqref{eq:dsigma_mcps_c} is found to be of the order of unity at all kinematical configurations.
Therefore, we can expect efficient event generation based on SDE MCPS integration by using the scattering amplitudes calculated in the FD gauge.

In this work we present  a dedicated scheme to perform an SDE MCPS integration by using modular phase space generation units with a phase space parametrization tailored for each Feynman diagram. The structure of the paper is as follows. 

The modular units are presented in Sec.~\ref{sec:units}. In Sec.~\ref{sec:integration}, we explain the integration method by using the process
for lepton colliders ($l=e$ or $\mu$)
\begin{align}
  \label{eq:ll_vvtth}
  l\bar{l} \to\nu_l\bar{\nu}_l t\bar{t} H ,
\end{align}
which was studied extensively in Refs.~\cite{Chen:2022yiu,Barger:2023wbg,Cassidy:2023lwd,Hagiwara:2024xdh}, and present the numerical results in Sec.~\ref{sec:vvtth}.

In Sec.~\ref{sec:lvtbh}, we explain how amplitudes with the light charged lepton  mass singularities due to nearly on-shell photons can be evaluated without numerical loss of effective digits.
This is achieved by introducing a special phase space parametrization and 
by modifying the original \HELAS\ codes~\cite{Murayama:1992gi} that calculate off-shell wave functions.
We apply the singular phase space parametrization and the new \HELAS\ codes
to study the processes
\begin{align}
  \label{eq:ll_lvtbh}
  l\overline{l} &\rightarrow l\overline{\nu}_l t\overline{b} H ,\\
  \label{eq:ll_lltth}
  l\overline{l} &\rightarrow l\overline{l} t\overline{t} H ,
\end{align}
for $l=e$ and $\mu$ in Sec.~\ref{sec:lvtbh} and Sec.~\ref{sec:lltth}, respectively.

Section~\ref{sec:summary} summarizes our findings.
Appendix~\ref{app:modular_units} presents Fortran programs for the modular phase-space units.
Appendix~\ref{app:helas} 
presents necessary modifications to \HELAS\ subroutines for singular vertices.

\section{Modular units of $n$-body phase space volume}
\label{sec:units}

\subsection{Overview of general $n$-body phase space}
With the SDE MCPS integration (see Sec.~2 of Ref.~\cite{Maltoni:2002qb}), the total cross section of the process under consideration, $\sigma_\mathrm{tot}$, is expressed as the sum~\eqref{eq:dsigma_mcps_a} of the single-channel contribution, $\sigma_k$, labeled by each Feynman diagram. Here $\sigma_k$ of Eq.~\eqref{eq:dsigma_mcps_b} can be expressed as
\begin{align}
    \sigma_k & = \int d\Phi_{n-2} \,|{\cal M}_k|^2\, R ,
  \label{eq:sigma_mcps}
\end{align}
where the ($n-2$) body phase space dependence of the matrix elements of the $k$-th diagram ${\cal M}_k$ and that of the ratio $R$ of Eq.~\eqref{eq:dsigma_mcps_c} are explicitly shown.
When the amplitude of each diagram is calculated in the FD gauge~\cite{Hagiwara:2020tbx,Chen:2022gxv,Chen:2022xlg}, the ratio $R$ approaches a constant of order unity in the phase space (region) where the squared amplitude $|{\cal M}_{k}|^2$ is large due to its propagator factors.
This feature allows us to increase the accuracy of the $\sigma_k$ calculations by generating appropriately weighted phase space for each Feynman diagram.  
Thus, the total cross section, the sum of all the $\sigma_k$, can be calculated accurately, and we can expect all kinematical distributions to be accurately reproduced because all propagator singularities are taken care of.
This section describes our method for generating an appropriate phase-space parametrization for each Feynman diagram channel.

The invariant phase space is defined as 
\begin{align}
  \label{eq:general-nbody-ps2}
d\Phi_{n-2}=
(2\pi)^4\delta^4\bigg(p_1+p_2-\sum_{k=3}^n  p_k\bigg)
\prod\limits_{k=3}^{n}
\frac{d^3 p_k}{(2\pi)^3 2E_k}
\end{align}
with $E_k=\sqrt{m_k^2+|{\rm p}_k|^2}$.
By using the recursive relation
\begin{align}
  \label{eq:general-nbody-ps3}
  d\Phi_{n-2}\bigg(p_1+p_2 = \sum_{k=3}^n p_k\bigg) 
  &= 
  d\Phi_2\bigg(p_1+p_2 = p_3+
 \sum_{k=4}^n p_k\bigg) \nonumber\\
 &
 \quad\times\frac{d(\sum_{k=4}^n p_k)^2}{2\pi}
 d\Phi_{n-3}\bigg(\sum_{k=4}^n p_k=p_4+\sum_{k=5}^n p_k\bigg)
\end{align}
successively, we obtain the following expression
\begin{align}
  \label{eq:general-nbody-ps4}
  d\Phi_{n-2}\bigg(p_1+p_2 = \sum_{k=3}^n p_k\bigg) 
  &=
  d\Phi_2\bigg(p_1+p_2 = p_3+
 \sum_{k=4}^{n}p_k\bigg) \nonumber\\
 &\quad
 \times\prod\limits_{m=4}^{n-1}
 \frac{d{M}^2_{(m,\cdots,n)}}{2\pi}
 d\Phi_2\bigg(\sum_{k=m}^n p_k=p_m+\sum_{k=m+1}^n p_k\bigg).
\end{align}
Here the invariant mass of the sub-system of particles $m$ to $n$:
\begin{align}
  \label{eq:general-nbody-ps5}
    M_{(m,\cdots,n)} = \sqrt{\bigg( \sum_{k=m}^{n} p_k \bigg)^2 }   
\end{align}
are ordered as
\begin{align}
	\label{eq:boundary}
    \sum_{k=m}^n m_k < M_{(m,\cdots, n)} < M_{(m-1, m,\cdots, n)} -m_{m-1} 	
\end{align}
for $m=4$ to $m=n-1$. The two-body phase space can be expressed as
\begin{align}
  \label{eq:two-body}
  d\Phi_2(P=p_j+p_k) =
  \dfrac{1}{8\pi} \overline{\beta}\bigg(\dfrac{p_j^2}{P^2},\dfrac{p_k^2}{P^2}\bigg)
  \dfrac{d\cos\theta_j}{2} \dfrac{d\phi_j}{2\pi}
\end{align}
with
  \begin{align}
    \label{eq:beta}
	\overline{\beta}(x,y) 
     = \sqrt{(1-x-y)^2 - 4xy}
  \end{align}
in the rest frame of $P=p_j+p_k$.
We define the total c.m. energy of the colliding particles as
\begin{align}
\sqrt{s} = \sqrt{ (p_1+p_2 )^2 } = \sqrt{ \bigg(\sum_{k=3}^{n} p_k \bigg)^2 }
\end{align}
with the constraint
\begin{align}
\sqrt{s}>\sum_{k=3}^n m_k   .    
\end{align}
The angular integrals in the rest frame of each sub-system
are all defined in the region:
\begin{align}	
	\label{eq:boundary2}
        -1 \le \cos\theta_m \le 1 ,\quad  
    -\pi \le \phi_m \le \pi ,   
\end{align}
for $m=3$ to $m=n-1$.
The total phase space is then a positive definite function only of the total energy $\sqrt{s}$ and the final state masses 
\begin{align}
    \Phi_{n-2}\left(\sqrt{s};m_k\right),
\end{align}
which is used to test all the phase space parameterizations given below.

For each channel characterized by the square $ |{\cal M}_k|^2 $ of the $k^{\rm th}$ Feynman diagram, we modify the integration variables $ M^2_{(m,\cdots, n)} $ and $ \cos\theta_m $ such that the corresponding Jacobian cancels the singular propagator factor in $ |{\cal M}_k|^2 $.
When the amplitudes are evaluated in the FD gauge, we can expect that the phase space integral in the $k^{\rm th}$ channel, Eq.\,\eqref{eq:sigma_mcps}, is smooth because the singular behavior of $ |{\cal M}_k|^2 $ is moderated by the Jacobian factor, and the ratio $R$ is expected to be a constant of order unity throughout the singular region.

\subsection{Building blocks of the phase space generator}
\label{sec:PSgen}

In the expression~\eqref{eq:general-nbody-ps4} for the $(n-2)$-body phase space, we have $(n-4)$ integration over ${M}^2_{(m,\cdots,n)}$ and $(n-3)$ two-body phase space. We therefore propose a subroutine to generate ${M}^2_{(m,\cdots,n)}$ in the boundary~\eqref{eq:boundary}, labeled as {\texttt{dshat}}, and two subroutines for generating the two-body phase space in the boundaries~\eqref{eq:boundary2}, \texttt{ph2s} and \texttt{ph2t}.
We find that the above three modules can handle all patterns of Feynman diagram propagator structures.
Let us explain the main features of the three subroutines, while the explicit Fortran codes are presented in Appendix~\ref{app:modular_units}. 

\paragraph{Subroutine \texttt{dshat} generates \shat\ values of a subsystem, where $\hat{s}= M^2_{(m,\cdots,n)}$.\\}

\begin{figure}
	\includegraphics[width=0.55\textwidth]{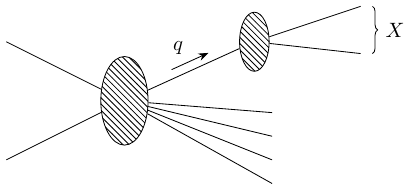}
    \caption{$s$-channel propagator ($q^2=m_X^2>0$).}
    \label{fig:s-channel-1}
\end{figure}

The subroutine gives $\hat{s}/2\pi$ in ${\rm GeV}^2$ units in the region
\begin{align}
  \hat{s}_{\rm min}<\hat{s}_{}<\hat{s}_{\rm max},
\end{align}
where the boundaries are found from Eq.~\eqref{eq:boundary}. If a propagator appears in the $s$-channel as in Fig.~\ref{fig:s-channel-1} with $q^2=\hat{s}$, this subroutine takes care of the propagator factor by giving the Jacobian
\begin{align}
	\label{eq:jacob}
J=\frac{(\hat{s}-m^2)^2+(m\Gamma)^2}{\hat{s}}.
\end{align}
In the standard model (SM), the Breit--Wigner distributions of the weak bosons and the top quark are taken care of.
Note that for massless gauge bosons and massless fermions, the Jacobian~\eqref{eq:jacob} reduces to $\hat s$.   

\paragraph{Subroutine \texttt{ph2s} generates two-body phase space with an $s$-channel splitting.\\}

\begin{figure}
    \centering
    \includegraphics[width=0.25\textwidth]{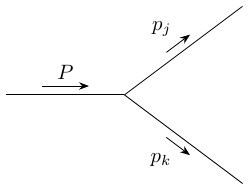}
    \caption{
$s$-channel 2-body splitting,  $\Big(P=p_j+p_k$ with $P^2,\,p_j^2,\,p_k^2>0$, 
	$\sqrt{P^2}>\sqrt{p_j^2}+\sqrt{p_k^2}\Big)$.
	}
    \label{fig:s-channel-2}
\end{figure}

This subroutine takes care of the two-body phase space Eq.~\eqref{eq:two-body} in the rest frame of the particle when the pair is created by an $s$-channel propagator or by a contact interaction as in Fig.~\ref{fig:s-channel-2}.

The subroutine does not give an extra Jacobian, because the two-body splitting in the parent rest frame are described by Wigner's $d$-functions which are powers of our integration variables, $\cos\theta$, and the azimuthal angle $\phi$ enters only as a phase factor.

The four momentum $p_j$ and $p_k$ are parametrized in the parent rest frame, $P^\mu=(\sqrt{P^2},0,0,0)$, as
\begin{subequations}
  \begin{align}   p_j^\mu&=\frac{\sqrt{P^2}}{2}\Big(1+\frac{p_j^2-p_k^2}{P^2},\bar{\beta}\sin\theta\cos\phi,\bar{\beta}\sin\theta\sin\phi,\bar{\beta}\cos\theta\Big) ,\\ p_k^\mu&=\frac{\sqrt{P^2}}{2}\Big(1-\frac{p_j^2-p_k^2}{P^2},-\bar{\beta}\sin\theta\cos\phi,-\bar{\beta}\sin\theta\sin\phi,-\bar{\beta}\cos\theta\Big) ,
  \end{align}
\end{subequations}
where $\bar\beta$ is the normalized c.m.\ momentum of Eqs.~\eqref{eq:two-body} and \eqref{eq:beta}, and the angular variables are in the region of Eq.~\eqref{eq:boundary}.
The four momenta are generated in the parent rest frame, and then transformed to the Lorentz frame where the helicity amplitude are calculated.

\paragraph{Subroutine \texttt{ph2t} generates two-body phase space with a $t$-channel propagator.\\}

\begin{figure}
 \raisebox{15mm}{(a)}\quad
 \includegraphics[width=0.32\textwidth]{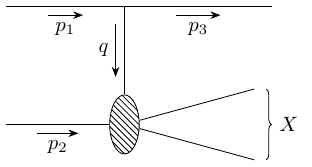}\qquad
 \raisebox{15mm}{(b)}\quad
 \includegraphics[width=0.32\textwidth]{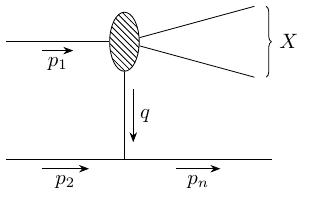}
  \caption{2-body splitting in the $t$-channel ($q^2<0$).}
  \label{fig:t-channel}
\end{figure}

When a $t$-channel propagator appears in the $2\to2$ ($a+b\to c+X$) subamplitudes, the four momenta of the final particles, $c$ and $X$ in Fig.~\ref{fig:t-channel} are parametrized in the rest frame of the colliding particles ($a$ and $b$) by choosing the polar axis along the colliding particle momentum direction, $\vec{p}_1$.
In this frame, the azimuthal angle $\phi$ is again a phase factor, and we modify the $\cos\theta$ variable to obtain a Jacobian, that cancels one power of the $t$-channel propagator.

In case of a subdiagram Fig.~\ref{fig:t-channel}(a), where the final four momentum $p_3$ can be one, or a set of final particles, we parametrize the four momenta as
\begin{subequations}
  \begin{align}  
    p_1^\mu&=\frac{\sqrt{P^2}}{2}\Big(1+\frac{p_1^2-p_2^2}{P^2},0,0,\beta_{i}\Big),\\
    p_2^\mu&=\frac{\sqrt{P^2}}{2}\Big(1-\frac{p_1^2-p_2^2}{P^2},0,0,-\beta_{i}\Big),\\   
	p_3^\mu&=\frac{\sqrt{P^2}}{2}\Big(1+\frac{p_3^2-p_X^2}{P^2},\beta_f\sin\theta\cos\phi,\beta_f\sin\theta\sin\phi,\beta_f\cos\theta\Big),\\
	p_X^\mu&=\frac{\sqrt{P^2}}{2}\Big(1-\frac{p_3^2-p_X^2}{P^2},-\beta_f\sin\theta\cos\phi,-\beta_f\sin\theta\sin\phi,-\beta_f\cos\theta\Big),
  \end{align}
 \label{eq:4mom}
\end{subequations}
with
\begin{align}
    \beta_{i}=\bar{\beta}\Big(
    \frac{{ p}_1^2}{P^2},\,\frac{{p}_2^2}{P^2}
    \Big)
    \quad
    {\rm and}\quad
    \beta_f
    =\bar{\beta}
    \Big(
    \frac{{p}_3^2}{P^2},\,
    \frac{p_X^2}{P^2}
    \Big)
    \label{eq:betabar}
\end{align}
in the rest frame of $P=p_1+p_2$.
The momentum transfer is given by
\begin{align}
  t=q^2=(p_1-p_3)^2.
\end{align}
We first note that $p_1^2$ and $p_2^2$ can be zero (for one-shell photons or gluons), $m_l^2$ for initial leptons, or negative values of order $m_l^2$ when they are virtual photons emitted from charged leptons. Likewise, for the final state particles, $p_3^2$ and $p_X^2$ can be zero if they are real photons or gluons, $m_f^2$ if fermions, or positive values of order $m_f^2$ when $X$ in a fermion pair. 
All the above cases appear in the three processes we study in this paper, \eqref{eq:ll_vvtth}-\eqref{eq:ll_lltth}, and we obtain a parametrization which takes care of all singular regions.

The inverse of the propagator factor $t'$ for the diagram Fig.~\ref{fig:t-channel}(a) can be expressed as 
\begin{align}
   t'= m^2-t=m^2-(p_1-p_3)^2=t'_{\rm min}+2|{ \vec{ p}}_1||{ \vec{ p}}_3|(1-\cos\theta)
    \label{eq:m2-q2}
\end{align}
with the minimum of the
absolute value of the inverse propagator factor
\begin{align}
 t'_{\rm min}
 =m^2-{ p}_1^2-{ p}_3^2+2(p_1^0E_3^{}-|{\vec{p}}_1||{\vec p}_3|).
 \label{eq:tmin}
\end{align}
When the exchanged particle is a photon ($m=0$) or a light lepton $(m=m_l)$, $t'_{\rm min}$ should be evaluated carefully to avoid loss of effective digits.
We evaluate the last term in Eq.~\eqref{eq:tmin} as
\begin{align}
    p_1^0E_3^{}-|\vec{p}_1||\vec{p}_3|
    =\frac{p_1^2 E_3^2+(p_1^0)^2p_3^2}{p_1^0E_3+|\vec{p}_1||\vec{p}_3|}.
	 \label{eq:tmin2}
\end{align}
By using the above expression, $t'_{\rm min}$ of Eq.~\eqref{eq:tmin} becomes a sum of terms of order $p_1^2$ and $p_3^2$ when the exchanged particles is a photon ($m^2=0$) or a lepton ($m^2=m_l^2$), and we can avoid significant loss of effective digits which appears in the original expression~\eqref{eq:tmin}.

When neglecting terms of order $m^2/s$, $t'_{\rm min}$  takes the following simple form 
as a function of the virtual particle (photon or lepton) energy fraction $x$,
when one of the external particles ($\gamma$ or $l$) is on-shell:
\begin{align}\label{eq:tmin3}	
    t'_{\rm min}=
	\left\{ 
	  \begin{aligned}   
	     & t_{\rm min}=m^2\frac{x^2}{1-x}  && {\rm for}\ l\to l+\gamma^*, \\   
	     & m^2(1-x)^2  && {\rm for}\ l\to \gamma+l^*, \\  
		 & m^2\frac{1}{1-x}  && {\rm for}\ \gamma\to l+\bar l^*.
	  \end{aligned} 
	  \right.  
\end{align}
The exact numerical expression~\eqref{eq:tmin} with~\eqref{eq:tmin2}
gives accurate $t’_{\rm min}$ value even when external particles are slightly off-shell. 

Once $t'_{\rm min}$ is calculated accurately, the inverse of the propagator factor takes the value in the range
\begin{align}
    t'_{\rm min}<t'<t'_{\rm max}=t'_{\rm min} + 4|{\vec{ p}}_1||\vec{p}_3|,
\end{align}
when $\cos\theta$ takes value in the range~\eqref{eq:boundary2}. We choose the logarithm of the inverse propagator as the integration variable
\begin{align}
    \ln t'_{\rm min} < \ln t' <\ln t'_{\rm max},
	\label{eq:lnt'}
\end{align}
so that the Jacobian
\begin{align}
J=\frac{t'}{2|{\vec{p}}_1||{\vec{ p}}_3|}=\frac{m^2-q^2}{2|{\vec{p}}_1||{\vec{ p}}_3|}
\end{align}
cancels one power of the $t$-channel propagator factor in $|{\cal M}_k|^2$. Only one power of the propagator should be canceled because the gauge-theory amplitudes give only logarithmic singularities in the collinear ($\cos\theta\sim1$) region. 

In case the $t$-channel propagator is emitted from the $b$-particle, see Fig.~\ref{fig:t-channel}(b), the inverse propagator factor takes the form of Eq.~\eqref{eq:m2-q2} in which $(1-\cos\theta)$ factor is replaced by $(1+\cos\theta)$, and $p_3$ replaced by $p_X$. 
When multiple propagators of photons and light charged leptons appear, we should use the \texttt{ph2t} subroutines where one or both of the incoming particle momenta ($p_1$ and $p_2$) can be space-like. 
We note here that one of $p_1^0$ and $p_2^0$ can be negative in the $p_1+p_2=P$ rest frame, when the incoming momenta are significantly space-like.
Because of this, we cannot replace the factor of $p_1^0$ by $E_1$ in Eqs.~\eqref{eq:tmin} and \eqref{eq:tmin2}.
It is also worth noting that it is only in the frame where the incoming particle momenta are chosen along the $z$-axis, as in Eq.~\eqref{eq:4mom}, the azimuthal angle $\phi$ is free from the singularities. 

\begin{figure}
\begin{tabular}{cccc}	 	
   \raisebox{9mm}{ \includegraphics[width=0.25\textwidth]{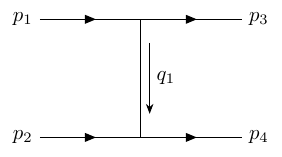}} &
   \raisebox{4mm}{ \includegraphics[width=0.25\textwidth]{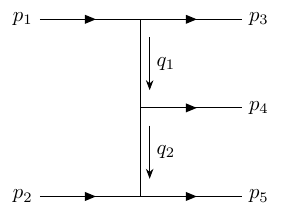} }&
   \raisebox{20mm}{$\quad\cdots\quad$} &
    \includegraphics[width=0.25\textwidth]{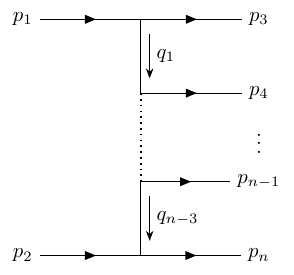} \\	
	(a) & (b) && (c)
\end{tabular}  
  \caption{Diagrams with multiple $t$-channel splittings.}
  \label{fig:t-channel-elements}
\end{figure}

When multiple $t$-channel propagators appear in a diagram, as shown in Fig.~\ref{fig:t-channel-elements}, the subroutine \texttt{ph2t} should be called multiple times.
It is essential that all $2\to2$ four momenta are parametrized in the c.m. frame of the $2\to2$ process where the incoming momentum is chosen along the $z$-axis.
Although the ordering of implementing $t$-channel propagators is not as critical as the choice of the Lorentz frame, we call \texttt{ph2t} from outside to inside.
In the example shown in Fig.~\ref{fig:t-channel-elements}(c), we first call \texttt{ph2t} to obtain the four momentum $q_1=p_1-p_3$, for a given invariant mass of the $p_2+\cdots+p_n$ system, and then obtain $q_{n-3}=p_n-p_2$ by using \texttt{ph2t} for the $2\to2$ scattering for $q_1+p_2=p_n+(p_4+\cdots+p_{n-1})$ in the $q_1+p_2$ rest frame.
This ordering allows us to study vector-boson-fusion amplitudes when $q_1$ and $q_{n-3}$ in Fig.~\ref{fig:t-channel-elements}(c) are photons, gluons, or weak bosons.

\paragraph{Subroutine \texttt{ph\#c} generates \#-body phase space with a contact interaction.\\}

\begin{figure}
    \includegraphics[width=0.3\textwidth]{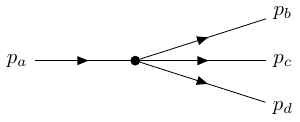}\qquad
    \includegraphics[width=0.3\textwidth]{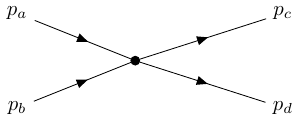}
  \caption{Diagrams with a 4-point contact vertex.}
  \label{fig:s-channel-elements}
\end{figure}

When a diagram contains an $n$-point contact vertex, there appear 1 to ($n-1$) splitting and 2 to ($n-2$) scattering, as shown in Fig.~\ref{fig:s-channel-elements} for $n=4$. 
Phase space units for those diagrams are named \texttt{ph\#c}, where \# is the number of final state particles, $\#=n-1$ for the 1 to ($n-1$) splitting, and $\#=n-2$ for the 2 to ($n-2$) scattering.
We use the standard \#-body phase space without the Jacobian factor for all \texttt{ph\#c} subroutines. 

We note here that in the SM effective field theory (SMEFT) with higher dimensional operators, as studied e.g. in Ref.~\cite{Hagiwara:2024xdh} for the process $l\bar l\to\nu_l\bar\nu_l t\bar tH$ with CP-violating (CPV) top-Yukawa coupling, amplitudes with dimension-five and dimension-six vertices dominate the cross section at high energies.
Although it is possible to introduce a Jacobian factor like $s^{4-m}$, with $m$ being the mass dimension of the vertex operator, to account for the overall energy dependence of the squared amplitude, we have not implemented it because the standard adaptive Monte Carlo (MC) integration programs such as {\sc Vegas}~\cite{Lepage:1977sw} and {\sc Bases}~\cite{Kawabata:1995th} can easily account for the effect.

\subsection{Brief notes on the phase-space library convention}
\label{sec:PSnote}

Our phase-space library is written in Fortran~90.
In the library, all real variables are declared as ``\texttt{double precision (real64)}"  and all integer variables are ``\texttt{integer (int32)}" in the library.

A unique feature of the library is using arrays of size five, \texttt{p(0:4)}, to represent four vectors.
The first four elements, \texttt{p(0)} to \texttt{p(3)}, give the four momentum components, $p^0$ to $p^3$, respectively. 
The fifth element, \texttt{p(4)}, gives the invariant square value of the four-vector
\begin{align}
  {\texttt p(4)}=p^2=(p^0)^2-(p^1)^2-(p^2)^2-(p^3)^2.
\end{align}
This fifth element must be given in advance for all four vectors of external particles so that we can evade loss of effective digits due to cancellation when the $p^2$ value is tiny as compared to the square of energy, $(p^0)^2$.
Without this $5$-component prescription, double precision calculation of about 13 digits accuracy is lost when we study $l\bar l$ collisions at a few TeV or higher.%
\footnote{The loss of numerical accuracy in the original \HELAS\ code was first observed in Ref.~\cite{Hagiwara:1990gk},
where single weak-boson productions at a TeV $e^+e^-$ collider were studied. 
The prescription given in Refs.~\cite{Hagiwara:1990gk,Murayama:1992gi} based on the BRST invariance~\cite{Becchi:1975nq,Tyutin:1975qk} for evaluating the amplitudes led to the first FD gauge paper~\cite{Hagiwara:2020tbx} for QED and QCD, whereas the special phase-space parametrization introduced in Refs.~\cite{Hagiwara:1990gk,Murayama:1992gi} for the initial collinear splitting is upgraded in this paper for processes with multiple splittings.
}
This problem can easily be avoided if we give $p^2$ ($m^2$'s for external on-shell particles) as the fifth element of the four-momentum, \texttt{p(4)}, by using the expression~\eqref{eq:tmin2}.

Another notable feature of the library is that the modules that generate two-body phase space always perform calculations at an appropriate center-of-momentum system.
In particular, the module \texttt{ph2t} that handles two-body phase space with a $t$-channel propagator always appears with a set of coordinate transformations between the center-of-momentum frame and the laboratory frame (where the helicity amplitudes are calculated) since the module requires two incoming particles to have three momenta along the $z$-axis, as in Eq.~\eqref{eq:4mom}.
By including these coordinate transformations, each module's independence can be maintained, and the full phase space can be obtained simply by calling the modular phase-space units, \texttt{dshat}, \texttt{ph2s}, and \texttt{ph2t}, successively.

\section{Single-diagram-enhanced multi-channel integration}
\label{sec:integration}

In this section, we introduce our parametrization of the SDE MCPS and event generation, using as an example the process $l\overline{l} \rightarrow \nu_l\overline{\nu}_l t\overline{t}H$, Eq.~\eqref{eq:ll_vvtth}, whose helicity amplitudes have recently been obtained in the FD gauge with an SMEFT operator~\cite{Hagiwara:2024xdh} that allows for CP-violating top–Higgs couplings~\cite{Barger:2023wbg,Cassidy:2023lwd,Hagiwara:2024xdh}.
Each phase-space parametrization yields an appropriate Jacobian factor, which typically cancels one power of the propagator factor appearing in the squared amplitude of the corresponding Feynman diagram.  
The number of Feynman diagrams in the FD (unitary) gauge are tabulated in Table~\ref{tab:ngraphs} for the three processes, Eqs.~\eqref{eq:ll_vvtth} to \eqref{eq:ll_lltth}, in both the SM and the SMEFT.
The SMEFT model, which will be described in detail in Sec.~\ref{sec:vvtth}, 
includes higher-dimensional vertices.

\begin{table}[b]
  \caption{Number of Feynman diagrams in the FD (unitary) gauge.}
  \label{tab:ngraphs}
  \begin{tabular}{c|rrcrr}
	\hline  
    \parbox[c][3ex]{7em}{Processes}  & \multicolumn{2}{c}{SM}  &\hspace*{5mm}& \multicolumn{2}{c}{SMEFT} \\ \hline
    $l\bar{l} \rightarrow \nu_l\bar{\nu}_l t\bar{t} H$ 
                                     & 89 & (87) && 118 & (89) \\
    $l\bar{l} \rightarrow l\bar{\nu}_l t\bar{b} H$
                                     & 64 & (60) && 86 & (60) \\
    $l\bar{l} \rightarrow l\bar{l} t\bar{t} H$
                                     & 172 & (172) && 216 & (174) \\ \hline 
  \end{tabular}
\end{table}  

Although each phase-space parametrization is coded manually in this work, we envisage that this step can eventually be automated within the process of Feynman diagram generation by amplitude generators.  
We therefore explain our method in detail below, so that it can be applied in future developments.  
  
Our method can be summarized as follows:\\[-4mm]

\noindent
\textbf{(1) Generate a FD-gauge amplitude program, ``\texttt{matrix.f},'' to calculate helicity amplitudes from the four-momenta and helicities of external particles using {\bfseries\scshape MadGraph5}\_aMC@NLO} {\bfseries\scshape (MG5)}~\cite{Alwall:2014hca}.%
  \footnote{The FD gauge is implemented in the MG5 version of 3.6.0 or higher~\cite{Hagiwara:2024xdh}.}

For the process~\eqref{eq:ll_vvtth} in the SM (SMEFT), \MG\ generates 89 (118) Feynman diagrams in the FD gauge%
\footnote{ 
The number of diagrams depends on the gauge choice; see, e.g., Table~\ref{tab:ngraphs} for the unitary gauge.} 
and the corresponding code to evaluate the amplitudes.\\[-4mm]

\noindent
\textbf{(2) Classify Feynman diagrams by topology.}
  
For each Feynman diagram, we identify its topology.  
We find that the 89 SM diagrams of the process~\eqref{eq:ll_vvtth} fall into 12 topologies, shown in Table~\ref{fig:phase_space_topologies_sm}, while an additional 11 topologies appear in SMEFT, listed in Table~\ref{fig:phase_space_topologies_cpv}.  
They are categorized according to the number of $t$-channel propagators.
The arrows along the seven external lines indicate the positive energy flow.
The momenta $p_1$ and $p_2$ are incoming, while $l$ and $\bar{l}$ correspond to the process~\eqref{eq:ll_vvtth}.
The remaining five momenta (3 to 7) are outgoing, i.e., in the final state.
Sideways connecting lines represent $s$-channel propagators, whereas vertical lines represent $t$-channel propagators.
It is important to count the number of $t$-channel propagators, $N_t$, for the topologies shown in Table~\ref{fig:phase_space_topologies_sm}.
The additional topologies arising in the 118 diagrams of the SMEFT are listed in Table~\ref{fig:phase_space_topologies_cpv}, where one of the vertices in each topology receives a contribution from higher-dimensional interactions.\\[-4mm]
  
\begin{table}
  \caption{Topologies of the 89 Feynman diagrams for the process
  $l\bar{l} \rightarrow \nu_l\bar{\nu}_l t\bar{t}H$ in the SM.
  The topologies are ordered according to the number of $t$-channel propagators $N_t$.}
  \label{fig:phase_space_topologies_sm}
  \vspace*{2mm}	
\begin{tabular}{c|cccc}	 
 \hline $N_t$ &  \multicolumn{4}{c}{Topologies of Feynman diagrams} \\ \hline	
 \raisebox{10mm}{0} &
  \includegraphics[width=0.23\textwidth]{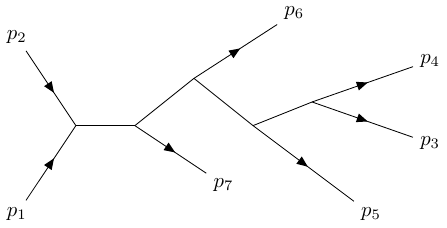} &
  \includegraphics[width=0.23\textwidth]{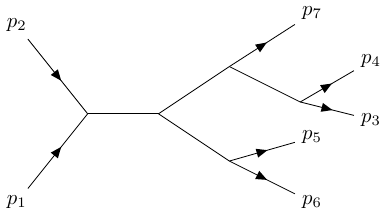} &
  \includegraphics[width=0.23\textwidth]{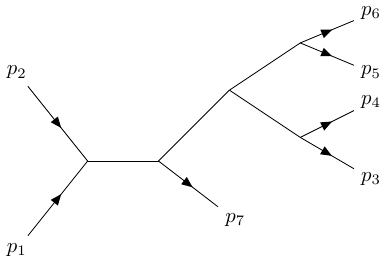} &
  \includegraphics[width=0.23\textwidth]{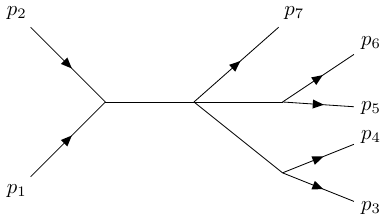} \\ \hline 
 \raisebox{10mm}{1} &
  \includegraphics[width=0.23\textwidth]{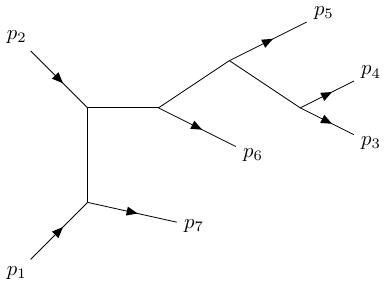} &
  \includegraphics[width=0.23\textwidth]{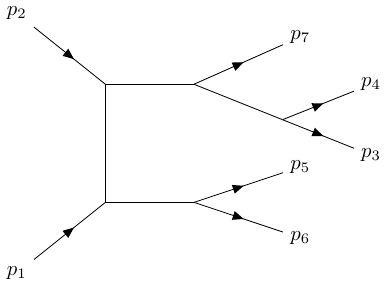} \\ \hline
 \raisebox{10mm}{2} &
  \includegraphics[width=0.23\textwidth]{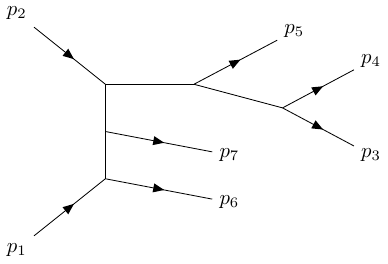} &
  \includegraphics[width=0.23\textwidth]{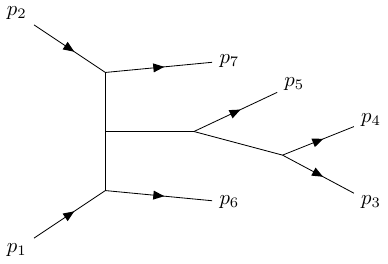} &
  \includegraphics[width=0.23\textwidth]{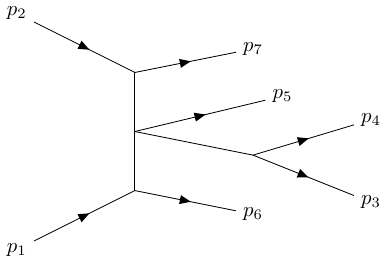} \\ \hline 
 \raisebox{10mm}{3} &
  \includegraphics[width=0.23\textwidth]{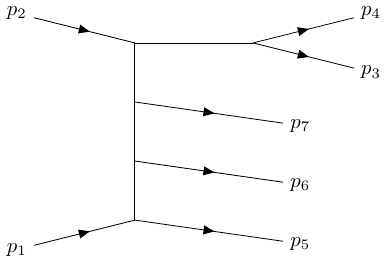} &
  \includegraphics[width=0.23\textwidth]{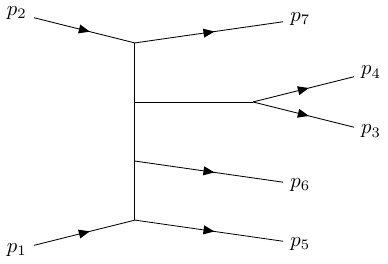} \\ \hline 
 \raisebox{10mm}{4} &
  \includegraphics[width=0.23\textwidth]{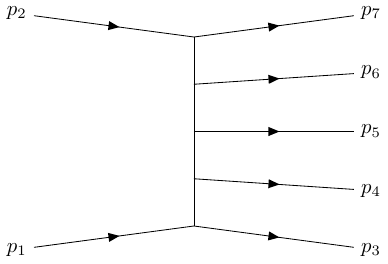} \\ \hline       
\end{tabular}
\end{table}  

\begin{table}
  \caption{Additional topologies for the $l\bar{l} \rightarrow \nu_l\bar{\nu}_l t\bar{t}H$  amplitudes in the SMEFT with a dimension-6 operator in Eq.~\eqref{eq:SMEFTLag}.
  One of the vertices in each topology has mass dimension of 5 or 6.}
  \label{fig:phase_space_topologies_cpv}	
  \vspace*{2mm}	
\begin{tabular}{c|cccc}	 
 \hline $N_t$ &  \multicolumn{4}{c}{Topologies of Feynman diagrams} \\ \hline	
 \raisebox{0mm}{0} &
  \includegraphics[width=0.23\textwidth]{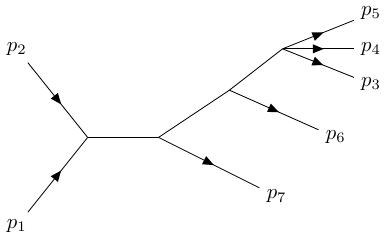} &
  \includegraphics[width=0.23\textwidth]{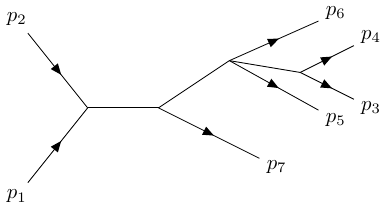} &
  \includegraphics[width=0.23\textwidth]{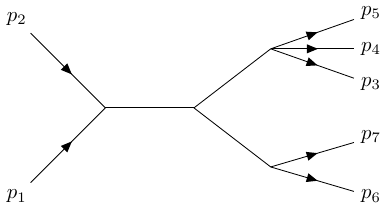} &
  \includegraphics[width=0.23\textwidth]{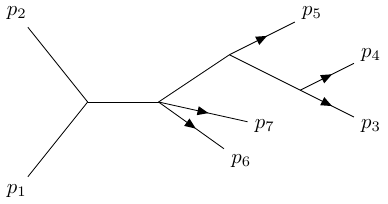} \\
 &\includegraphics[width=0.23\textwidth]{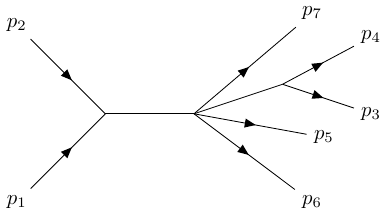} \\ \hline 
 \raisebox{10mm}{1} &
  \includegraphics[width=0.23\textwidth]{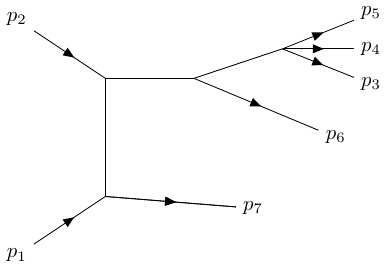} &
  \includegraphics[width=0.23\textwidth]{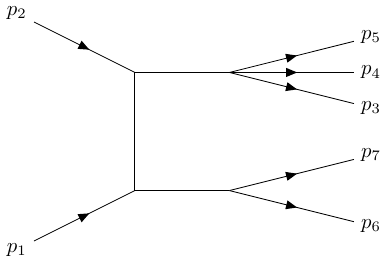} \\ \hline
 \raisebox{10mm}{2} &
  \includegraphics[width=0.23\textwidth]{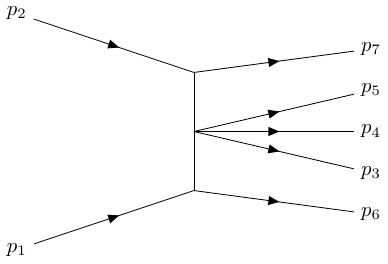} &
  \includegraphics[width=0.23\textwidth]{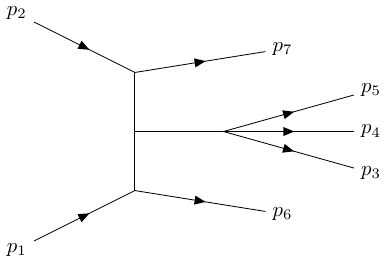} &
  \includegraphics[width=0.23\textwidth]{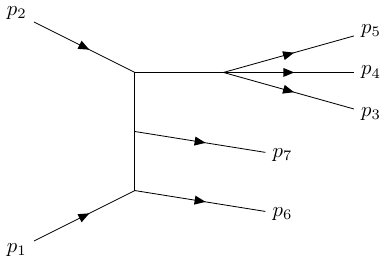} \\ \hline 
 \raisebox{10mm}{3} &
  \includegraphics[width=0.23\textwidth]{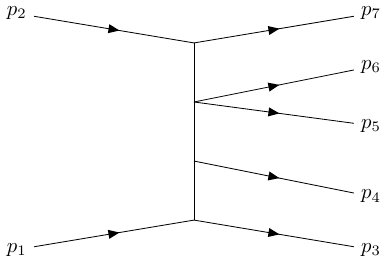} \\ \hline
\end{tabular}
\end{table}

\noindent
\textbf{(3) Obtain the phase-space parametrization for each Feynman diagram based on its topology.}

We now describe how the phase-space code is constructed for the topology of a given Feynman diagram.
\begin{enumerate}
  \item[(3-1)] Successively merge all $s$-channel splittings in the diagram according to the mass ordering of Eq.~\eqref{eq:boundary},  and compute $\shat$ using \texttt{dshat} for the corresponding subsystem.  

  \item[(3-2)] If $t$-channel splittings remain after step (3-1), generate the corresponding $t$-channel phase space using \texttt{ph2t}.  

  \item[(3-3)] Once all $t$-channel phase spaces are generated, construct the $s$-channel phase space for all $s$-channel splittings using \texttt{ph2s}.  

  \end{enumerate}
  
\begin{figure}
\begin{tabular}{cc}	 	
 \includegraphics[width=0.4\textwidth]{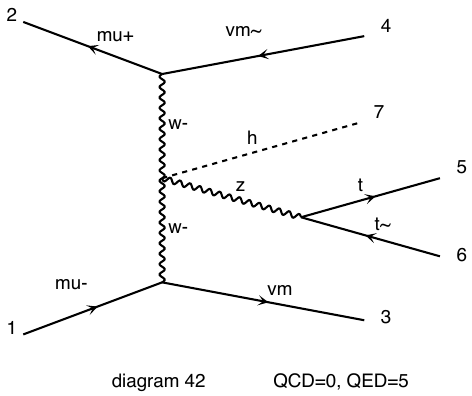} &
 \raisebox{10mm}{\includegraphics[width=0.4\textwidth]{ex1_sm_t2_03}} 
\end{tabular}
  \caption{One of the 89 Feynman diagram (diagram 42 generated by MG5) for $l\bar{l} \rightarrow \nu_l\bar{\nu}_l t\bar{t}H$ in the SM, and its corresponding topology (the third one with $N_t=2$ in Table~\ref{fig:phase_space_topologies_sm}).}
  \label{fig:ex1_topex}
\end{figure}

As an illustration, we consider diagram 42 of the process generated by \MG, shown in Fig.~\ref{fig:ex1_topex}(left).  
This diagram contains two $t$-channel $W$ propagators, and its phase-space topology corresponds to the third topology with $N_t=2$ in Table~\ref{fig:phase_space_topologies_sm}.

As preparation, we match the particle momentum labels in the diagram topology to the external momenta defined by \MG, as summarized in Table~\ref{tab:ass_external_ex1_topex}.  
Once this identification is made, all external and internal particles along the topology lines are uniquely fixed for each diagram.  
Although the overall phase-space structure is common to diagrams of the same topology, the boundaries and Jacobians of the internal propagators are determined by this identification.

We now derive the five-body phase space for the chosen diagram step by step.
In the example below, we use the momentum labels of the topology diagram (right of Fig.~\ref{fig:ex1_topex}) that are used in the subroutines.

\begin{figure}
\begin{tabular}{ccl}
 \raisebox{15mm}{(S1)} &
 \includegraphics[height=0.135\textheight]{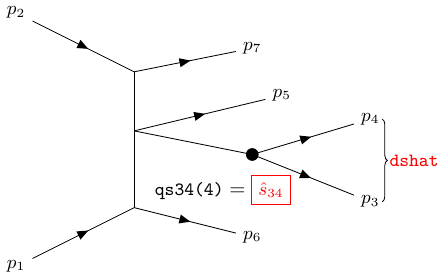} &
   \raisebox{15mm}{\begin{tabular}{l}
    \texttt{call dshat(z1,am(3)+am(4),}\\
    \hspace*{11mm}\texttt{w-am(5)-am(6)-am(7),mv1,wv1,} \\
    \hspace*{11mm}\texttt{qs34,dshat\_1, ierr)}	
	 \end{tabular}}
\\
 \raisebox{15mm}{(S2)} &
 \includegraphics[height=0.135\textheight]{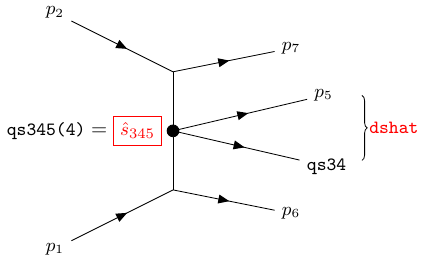} &
   \raisebox{15mm}{\begin{tabular}{l}
    \texttt{call dshat(z2,sqrt(qs34(4))+am(5),}\\
    \hspace*{11mm}\texttt{w-am(6)-am(7),zero,zero,} \\
    \hspace*{11mm}\texttt{qs345,dshat\_2, ierr)}	
	 \end{tabular}}
\\
 \raisebox{15mm}{(S3)} &
 \includegraphics[height=0.135\textheight]{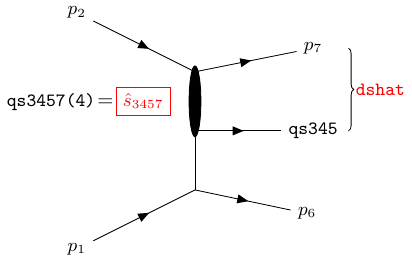} &
   \raisebox{15mm}{\begin{tabular}{l}
    \texttt{call dshat(z3,sqrt(qs345(4))+am(7),}\\
    \hspace*{11mm}\texttt{w-am(6),zero,zero,} \\
    \hspace*{11mm}\texttt{qs3457,dshat\_3, ierr)}	
	 \end{tabular}}
\\
 \raisebox{15mm}{(S4)} &
 \includegraphics[height=0.135\textheight]{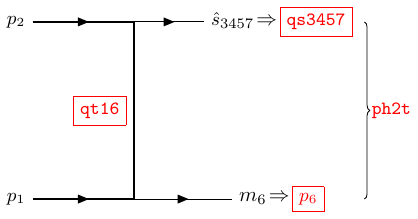} &
   \raisebox{15mm}{\begin{tabular}{l}
    \texttt{call ph2t(z4,z5,p(0,1),p(0,2),}\\
    \hspace*{11mm}\texttt{p(0,6),qs3457,qt16,mv2,} \\
    \hspace*{11mm}\texttt{ph2t\_1, ierr)}	
	 \end{tabular}}
\\
 \raisebox{15mm}{(S5)} &
 \includegraphics[height=0.135\textheight]{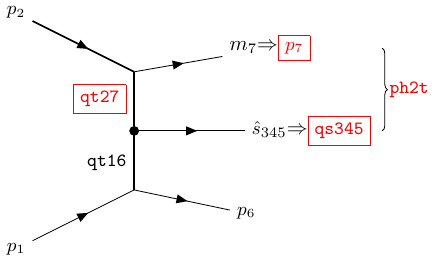} &
   \raisebox{15mm}{\begin{tabular}{l}
    \texttt{call ph2t(z6,z7,p(0,2),qt16,}\\
    \hspace*{11mm}\texttt{p(0,7),qs345,qt27,mv3,} \\
    \hspace*{11mm}\texttt{ph2t\_2, ierr)}	
	 \end{tabular}}
\\
 \raisebox{15mm}{(S6)} &
 \includegraphics[height=0.135\textheight]{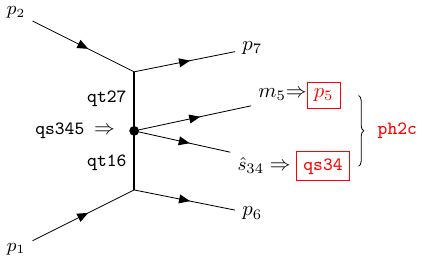} &
   \raisebox{15mm}{\begin{tabular}{l}
    \texttt{call ph2c(z8,z9,qt16,qt27,}\\
    \hspace*{11mm}\texttt{p(0,5),qs34,ph2c\_1)}	
	 \end{tabular}}
\\
 \raisebox{15mm}{(S7)} &
 \includegraphics[height=0.135\textheight]{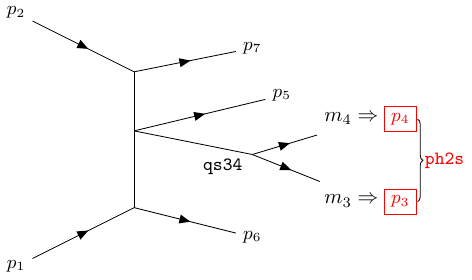} &
   \raisebox{15mm}{\begin{tabular}{l}
    \texttt{call ph2s(z10,z11,qs34,}\\
    \hspace*{11mm}\texttt{p(0,3),p(0,4),ph2s\_1)}	
	 \end{tabular}}
\end{tabular}
  \caption{Steps to obtain the phase-space parameterization for the Feynman diagram in Fig.~\ref{fig:ex1_topex}.}
  \label{fig:step}
\end{figure}

As shown in Fig.~\ref{fig:step}, the procedure involves seven steps:
\begin{itemize}
\setlength{\itemsep}{-2.25mm}
 \item[(S1)] Apply \texttt{dshat} to $3+4$ to generate $\shat_{34}$.
 \item[(S2)] Apply \texttt{dshat} to \texttt{qs34} and $5$ to generate $\shat_{345}$.	
 \item[(S3)] Apply \texttt{dshat} to \texttt{qs345} and $7$ to generate $\shat_{3457}$.
 \item[(S4)] Apply \texttt{ph2t} to ``$1+2\rightarrow 6+\texttt{qs3457}$'' to obtain the four-vectors of $6$ and \texttt{qs3457}.
 \item[(S5)] Apply \texttt{ph2t} to ``$\texttt{qt16}+2\rightarrow\texttt{qs345}+7$'' to obtain the four-vectors of \texttt{qs345} and $7$.
 \item[(S6)] Apply \texttt{ph2c} to \texttt{qs345} to generate the four-vectors of \texttt{qs34} and $5$.
 \item[(S7)] Apply \texttt{ph2s} to \texttt{qs34} to generate the four-vectors of $3$ and $4$.    
\end{itemize}	

\begin{table}
  \caption{Momentum assignments of external particles for diagram 42.}\label{tab:ass_external_ex1_topex}
  \begin{tabular}{c|ccccccc}
  \hline 	  
  \parbox[c][3ex]{10em}{External particles} & \parbox{1.5em}{$l$} & \parbox{1.5em}{$\bar l$} & \parbox{1.5em}{$\nu_l$} & \parbox{1.5em}{$\bar{\nu}_l$} & \parbox{1.5em}{$t$} & \parbox{1.5em}{$\bar{t}$} & \parbox{1.5em}{$H$} \\ \hline
  Momenta defined by MG5 & $p_1$ & $p_2$ & $p_3$ & $p_4$ & $p_5$ & $p_6$ & $p_7$ \\
  Momenta matched to the topology diagram & $p_1$ & $p_2$ & $p_6$ & $p_7$ & $p_4$ & $p_3$ & $p_5$ \\ \hline  
  \end{tabular}
\end{table}

After (S1) and (S2), all $s$-channel splittings are merged, leaving two $t$-channel propagators.  
To apply the two-body phase space \texttt{ph2t} in step (S4), we first generate $\shat_{3457}=(p_3+p_4+p_5+p_7)^2$ using \texttt{dshat} in step (S3).  
In step (S4), we call \texttt{ph2t} to obtain the four-vectors of $p_6$ and $p_{3457}=p_3+p_4+p_5+p_7$, denoted as \texttt{qs3457}.  
In step (S5), \texttt{ph2t} generates the four-vectors of $p_7$ and $p_{345}=p_3+p_4+p_5$.  
Next, in step (S6), \texttt{ph2c} is applied to \texttt{qs345}, producing $p_5$ and $p_{34}=p_3+p_4$.  
Here, \texttt{ph2c} is equivalent to \texttt{ph2s} but without the Jacobian factor of Eq.~\eqref{eq:jacob}.  
Finally, in step (S7), \texttt{ph2s} generates the four-vectors of $p_3$ and $p_4$.

The overall phase-space factor is the product of all subroutine outputs:
\begin{align}
 \Phi = \texttt{dshat\_1}\cdot \texttt{dshat\_2} \cdot \texttt{dshat\_3} \cdot \texttt{ph2t\_1} \cdot \texttt{ph2t\_2} \cdot \texttt{ph2c\_1} \cdot \texttt{ph2s\_1}.
\end{align}

\begin{lstlisting}[label={list:ex1_topex},caption={ex1\_topex.f90},language=Fortran,float]
  use, intrinsic :: iso_fortran_env
  real(real64), parameter :: zero = 0.d0
 
  real(real64), intent(in) :: z1, z2, z3, z4, z5, z6, z7, z8, z9, z10, z11
  real(real64), intent(inout) :: p(0:4,7)
  real(real64), intent(out) :: phfact

  real(real64) :: am(7)
  
  real(real64) :: dshat_1, dshat_2, dshat_3
  real(real64) :: ph2t_1, ph2t_2, ph3c_1, ph2s_1

  real(real64) :: qs34(0:4), qt16(0:4), qt27(0:4)
  real(real64) :: qs345(0:4), qs3457(0:4)

  real(real64) :: mv1, wv1
  real(real64) :: mv2, mv3

  ! dshat_1
  call dshat(z1,am(3)+am(4), w-am(5)-am(6)-am(7),mv1,wv1, qs34,dshat_1, ierr)

  ! dshat_2
  call dshat(z2, sqrt(qs34(4))+am(5), w-am(6)-am(7), zero,zero, qs345,dshat_2, ierr)

  ! dshat_3
  call dshat(z3, sqrt(qs345(4))+am(7), w-am(6), zero,zero, qs3457,dshat_3, ierr)
       
  ! ph2t_1
  call ph2t(z4,z5, p(0,1),p(0,2), p(0,6),qs3457, qt16, mv2, ph2t_1, ierr)

  ! ph2t_2
  call ph2t(z6,z7, p(0,2),qt16, p(0,7),qs345, qt27, mv3, ph2t_2, ierr)

  ! ph2c_1
  call ph2c(z8,z9, qt16,qt27, p(0,5),qs34, ph2c_1)

  ! ph2s_1
  call ph2s(z10,z11, qs34, p(0,3),p(0,4), ph2s_1)

  phfact = dshat_1*dshat_2*dshat_3*ph2t_1*ph2t_2*ph2c_1*ph2s_1

  
\end{lstlisting}

The corresponding Fortran implementation is given in List~1, which shows the code for phase-space generation following the steps above.  
The arrays of ``five-vectors,'' declared as \texttt{qt16(0:4)}, \texttt{qt27(0:4)}, and \texttt{qs34(0:4)}, store information on the internal propagators of this topology.  
They are assigned to internal lines in the order indicated in the figure.  

Let us briefly trace the Fortran code.  
In step (S1), \texttt{dshat} generates $\shat_{34}=(p_3+p_4)^2$ from the final-state particle masses (Fig.~\ref{fig:step} (S1)).  
For collision energy $W$, the parameters \texttt{what\_min} and \texttt{what\_max} are set to $m_3+m_4$ and $W-m_5-m_6-m_7$, respectively (line\#20 of List~\ref{list:ex1_topex}).  
The merged five-vector is stored in \texttt{qs34(0:4)}, with the fifth component containing the generated $\shat_{34}$.  
Afterward, merging $p_5$ and \texttt{qs34} yields $\shat_{345}$, generated by \texttt{dshat} within the range $m_5+\sqrt{\shat_{34}}$ to $W-m_6-m_7$ (line\#23).  
This value is stored in the fifth component of \texttt{qs345(0:4)}.  
At this stage, the topology contains two $t$-channel elements (Fig.~\ref{fig:step} (S2)).

The two $t$-channel elements are then processed sequentially.  
First, we compute $\shat_{3457}$ for the subsystem of $p_7$ and \texttt{qs345} with \texttt{dshat} (Fig.~\ref{fig:step} (S3)), storing the result in \texttt{qs3457(4)} (line\#26).  
Its allowed range is from $m_7+\sqrt{\shat_{345}}$ to $W-m_6$.  
The corresponding system,
\begin{align}
  p_1 (m_1^2)+ p_2(m_2^2) \rightarrow p_6 (m_6^2) + \texttt{qs3457} (\shat_{3457}),
\end{align}
is then evaluated with \texttt{ph2t} (line\#29), which generates the four-vectors of $p_6$ and \texttt{qs3457} (Fig.~\ref{fig:step} (S4)).  
\texttt{ph2t} also computes the five-vector of the propagator \texttt{qt16}, used as input to the next \texttt{ph2t} call.  
The second $t$-channel system,
\begin{align}
  \texttt{qt16} + p_2(m_2^2) \rightarrow  \texttt{qs345} (\shat_{345}) + p_7 (m_7^2),
\end{align}
is then processed, and the four-vectors of \texttt{qs345} and $p_7$ are generated (line\#32).

\noindent
\textbf{(4) Organize diagrams with the same internal propagators.}

In the FD gauge, Feynman amplitudes have the property that the sum of all diagrams sharing the same internal propagator dominates the total amplitude in the kinematical region where the propagator becomes singular.  
It is therefore desirable to organize Feynman diagrams such that those with identical propagators are grouped together.  
Since diagrams generated by \MG\ do not provide such sub-amplitudes automatically, we first identify all possible internal propagators using the external particle momentum labels.

\begin{table}
  \caption{Categorization of internal propagators.}\label{tab:categorize_internals_ex1_topex}
  \begin{tabular}{ccc|ccc|ccc}
	  \hline
    \parbox[c]{2em}{\#}  & Particle & 
    \begin{tabular}{c}
    Momentum \\ combination
    \end{tabular}
    & \parbox[c]{2em}{\#} & Particle &
    \begin{tabular}{c}
    Momentum \\ combination
    \end{tabular}
    & \parbox[c]{2em}{\#} & Particle & 
    \begin{tabular}{c}
    Momentum \\ combination
    \end{tabular}
    \\ \hline 
     1 & $\gamma$ & (1,2) &
     12 & $\nu_{l}$ & (1,2,4) &
     23 & $l$ & (2,5,6)  \\ 
     2 & $Z$ & (1,2) &
     13 & $t$ & (1,2,5) &
     24 & $t$ & (3,4,6) \\ 
     3 & $W$ & (1,3) &
     14 & $t$ & (1,2,6) &
     25 & $t$ & (3,4,6) \\ 
     4 & $W$ & (2,4) & 
     15 & $Z$ & (1,2,7) &
     26 & $Z$ & (3,4,7) \\ 
     5 & $Z$ & (3,4) &
     16 & $l$ & (1,3,4) &
     27 & $\nu_l$ & (3,5,6) \\ 
     6 & $\gamma$ & (5,6) &
     17 & $b$ & (1,3,6) &
     28 & $\nu_l$ & (4,5,6) \\ 
     7 & $Z$ & (5,6) &
     18 & $W$ & (1,3,7) &
     29 & $\gamma$ & (5,6,7) \\ 
     8 & $H$ & (5,6) &
     19 & $l$ & (1,5,6) &
     30 & $Z$ & (5,6,7) \\ 
     9 & $t$ & (5,7) &
     20 & $l$ & (2,3,4) &
     31 & $H$ & (5,6,7) \\ 
     10 & $t$ & (6,7) &
     21 & $b$ & (2,4,5) \\ 
     11 & $\nu_{l}$ & (1,2,3) &
     22 & $W$ & (2,4,7) \\ \hline
  \end{tabular}
\end{table}

Table~\ref{tab:categorize_internals_ex1_topex} lists all possible propagators for the process~\eqref{eq:ll_vvtth},  $l\bar{l} \rightarrow \nu_l\bar{\nu}_l t\bar{t} H$,  based on the particle momentum labels given by \MG\ (shown in Fig.~\ref{fig:ex1_topex}(left) and Table~\ref{tab:ass_external_ex1_topex}).  
There are 31 possible propagators in the SM and the SMEFT.

We introduce a compact notation for the propagator four-momenta
\begin{align}
	q_{ij\cdots k}=q_i+q_j+\cdots+q_k,
\end{align} 
where $q_k$ denotes the incoming four-momentum for $k=1,2$, and the negative of the outgoing momentum for $k \geq 3$:
\begin{align}
	q_{k}&=p_k  && {\rm for}\ k=1,2, \\
	q_{k}&=-p_k && {\rm for}\ k=3,\cdots,n.
\end{align} 

With this notation, overall four-momentum conservation reads
\begin{align}
  q_{123\cdots n}=\sum_{k=1}^n q_k = \sum_{k=1}^2 p_k -\sum_{k=3}^n p_k =0.
\end{align}
Thus, propagator momenta can be expressed using combinations of external four-momenta with a reduced number of indices.  
For the process~\eqref{eq:ll_vvtth} with $n=7$, only three-particle momentum combinations appear in Table~\ref{tab:categorize_internals_ex1_topex}.  
Propagators involving either $1$ or $2$ (but not both) are space-like ($t$-channel), while all others are time-like ($s$-channel).

In Table~\ref{tab:categorize_internals_ex1_topex}, propagators 3, 4, 18, and 22 correspond to $t$-channel $W$ exchange, 16, 19, 20, and 23 to $t$-channel $\mu$, 17 and 21 to $t$-channel $b$, and the rest to $s$-channel propagators.  
In the following, we denote diagrams containing both propagators 3 and 4 as vector-boson-fusion (VBF) amplitudes, those containing 3 but not 4 as $W^-l^+$ collision, and those containing 4 but not 3 as $l^-W^+$ collision.

\begin{figure}
  \includegraphics[height=0.25\textwidth]{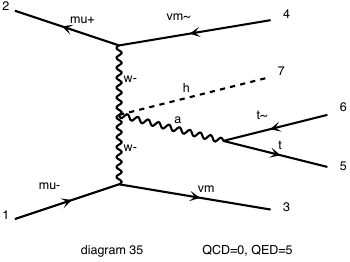} 
  \includegraphics[height=0.25\textwidth]{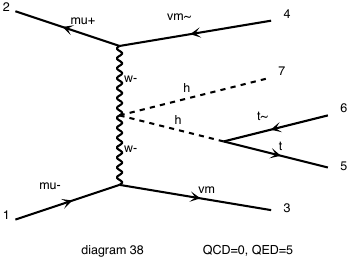} 
  \includegraphics[height=0.25\textwidth]{ex1_mg5_d042.pdf} 
  \caption{The diagrams with the same phase-space topology.}
  \label{fig:top_Nt2}
\end{figure}

\begin{table}
  \caption{Assignments of internal propagators.}\label{tab:top_Nt2}
\begin{tabular}{c|ccc}
	  \hline
    \parbox[c]{11em}{propagator momentum}  & \parbox[c]{6em}{diagram 35} & \parbox[c]{6em}{diagram 38} & \parbox[c]{6em}{diagram 42}\\ \hline
	(1,3) & $W$ ($q_{13}$)  & $W$ ($q_{13}$) & $W$ ($q_{13}$)  \\ 
	(2,4) & $W$ ($q_{24}$)  & $W$ ($q_{24}$)  & $W$ ($q_{24}$) \\ 
	(5,6) & $\gamma$ ($q_{56}$)  & $H$ ($q_{56}$) & $Z$ ($q_{56}$) \\ \hline
\end{tabular}
\end{table}

Once all internal propagators are labeled (as in Table~\ref{tab:categorize_internals_ex1_topex}), the Jacobian factors can be fixed for each diagram within a given phase-space topology.  
For the topology of Fig.~\ref{fig:ex1_topex}(right), three diagrams are identified, as shown in Fig.~\ref{fig:top_Nt2}, and distinguished by their propagator particle in Table~\ref{tab:top_Nt2}.

The only difference among these three diagrams arises from the Jacobian factor of the $q_{56}$ propagator in step (S7), inside the \texttt{ph2s} subroutine.  
In this particular case, since $q_{56}^2>(2m_t)^2$ is always much larger than $m_H^2$ or $m_Z^2$, introducing the Jacobian~\eqref{eq:jacob} yields negligible merit.  
Hence, all three diagrams can be described by the same phase space.  
We do not attempt further refinements, such as reducing the number of phase-space channels, in this report.

\section{Numerical results for $l\bar{l}\to\nu_l\bar{\nu}_l t\bar{t}H$}
\label{sec:vvtth}

In the previous section, we explained the integration method using the process $l\bar l\rightarrow\nu_l\bar{\nu}_{l}t\bar{t}H$,
where higher-dimensional vertices are also included within the framework of the SMEFT.
In this section, we demonstrate the integration procedure using {\sc Bases}~\cite{Kawabata:1995th} and present the corresponding numerical results for this process.
We note that this process was studied extensively at the total cross-section level in Ref.~\cite{Hagiwara:2024xdh}, where a comparison between the FD and unitary gauges was performed.
It should also be noted that, in the unitary gauge, accurate calculation of the cross section becomes very difficult, especially for $\sqrt{s}>10$~TeV, because of unphysical and delicate gauge cancellations, leading to poor convergence of the Monte Carlo integration.

As a demonstration, we introduce a dimension-six operator in the SMEFT~\cite{Peccei:1990uv,Zhang:1994fb,Whisnant:1994fh,Chen:2022yiu,Liu:2023yrb,Barger:2023wbg,Cassidy:2023lwd}
\begin{align}
 {\cal L} ={\cal L}_{\rm SM}+ 
\left\{ \frac{C}{\Lambda^2}~Q_3^\dagger t_R \tilde{\phi}
~\left(\tilde{\phi}^\dagger\tilde{\phi}-\frac{v^2}{2}\right) + {\rm h.c.} \right\},
\label{eq:SMEFTLag}
\end{align}
where $Q_3 = (t_L, b_L)^T$ and
$\tilde{\phi} =( (v+H-i\pi^0)/\sqrt{2},\, -i\pi^-)^T$.
When we take the coefficient as~\cite{Peccei:1990uv,Barger:2023wbg}
\begin{align}
\frac{C}{\Lambda^2}
=\frac{\sqrt{2}(g_{\rm SM}-ge^{i\xi})}{v^2},
\label{eq:coef}
\end{align}
the phenomenological Lagrangian
\begin{align}
  {\cal L}_{ttH} 
  = -g(e^{i\xi}t_L^\dagger t_R+e^{-i\xi}t_R^\dagger t_L)H
  = -g\, \bar{t} (\cos\xi + i\sin\xi \gamma_5) tH
  \label{eq:tthLag}
\end{align}
for CPV top-quark Yukawa coupling is obtained.
In the following, we take
\begin{align}
 g = g_{\rm SM}^{} = \frac{m_t}{v},
\end{align}
so that the only non-SM parameter is the CP phase $\xi$, where $\xi=0$ recovers the SM.

\begin{figure}
  \includegraphics[width=0.14\textwidth,clip]{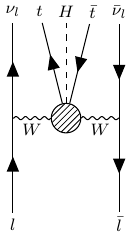} 
  \quad
  \includegraphics[width=0.14\textwidth,clip]{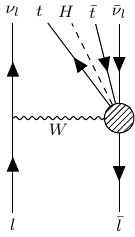} 
  \includegraphics[width=0.14\textwidth,clip]{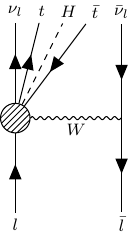} 
  \quad
  \includegraphics[width=0.14\textwidth,clip]{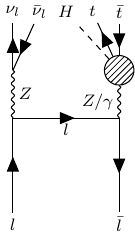} 
  \includegraphics[width=0.14\textwidth,clip]{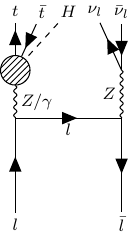} 
  \quad
  \includegraphics[width=0.14\textwidth,clip]{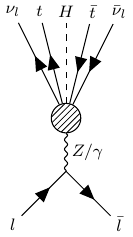} 
\hspace*{-0.1cm}(A)\hspace*{3.5cm}(B)\hspace*{4.6cm}(C)\hspace*{3.5cm}(D)
\caption{The Feynman diagrams for the process 
  $l\bar{l}\to\nu_l\bar{\nu}_l t \bar{t} H$ are classified into four categories. (A) $W^-W^+$F, (B) $W^-l^+/l^-W^+$, (C) ann-$l$, and (D) ann-$V$ ($V=Z/\gamma$).
In (C), there are additional diagrams where the Higgs boson is emitted from the $Z$ boson that gives $\nu_l\bar{\nu}_l$.
 }
\label{fig:proc1}
\end{figure}

In order to study the interference patterns among the Feynman diagrams, we classify the diagrams of the process $l\bar{l}\rightarrow\nu_l\bar{\nu}_{l}t\bar{t}H$ into four categories shown in Table~\ref{tab:ex_categories}, which are illustrated in Fig.~\ref{fig:proc1}.

\begin{table}
  \caption{Four categories of Feynman diagrams for $l\bar l\to\nu_l\bar\nu_l t\bar tH$, $l\bar\nu_l t\bar bH$, $l\bar l t\bar tH$.}
  \label{tab:ex_categories}
  \begin{tabular}{l|l} \hline
    \multicolumn{1}{c|}{Category name} & \multicolumn{1}{c}{Diagram type} \\ \hline
    (A) $V_1V_2$F & \hspace{2mm}Vector boson fusion (VBF) \\
    (B) $Vl^+/l^-V$ & \hspace{2mm}$Vl^+$ and $l^-V$ collision \\
    (C) ann-$f$ & \hspace{2mm}$l\bar{l}$ annihilation with $t$-channel fermion exchange \\
    (D) ann-$V$ & \hspace{2mm}$l\bar{l}$ annihilation with $s$-channel vector-boson ($Z/\gamma$) exchange \\ \hline
  \end{tabular}
\end{table}

We summarize the numbers of Feynman diagrams in the FD gauge for each group in Table~\ref{tab:proc1}.
The numbers of Feynman diagrams in the SM and those with higher-dimensional vertices (dimension-five, dim5, or dimension-six, dim6) are shown separately.%
\footnote{
We do not distinguish between the SM vertices and the non-SM vertices of dimension-four (dim4), such as the $ttH$ vertices from the Lagrangian~\eqref{eq:tthLag}, 
because they give the same energy scaling of the scattering amplitudes in the FD gauge~\cite{Hagiwara:2024xdh}.
}  
Because all the dim5 and dim6 vertices are proportional to the coefficient of the SMEFT operator $C/\Lambda^2$~\eqref{eq:coef}, 
those amplitudes vanish in the SM limits of $C/\Lambda^2=0$, or $(g,\xi)=(g_{\rm SM},0)$.
Therefore, only the amplitudes with dim4 vertices contribute in the SM. 
The numbers in parentheses indicate the diagram IDs generated by \MG.

\begin{figure}
    \includegraphics[width=0.495\textwidth,clip]{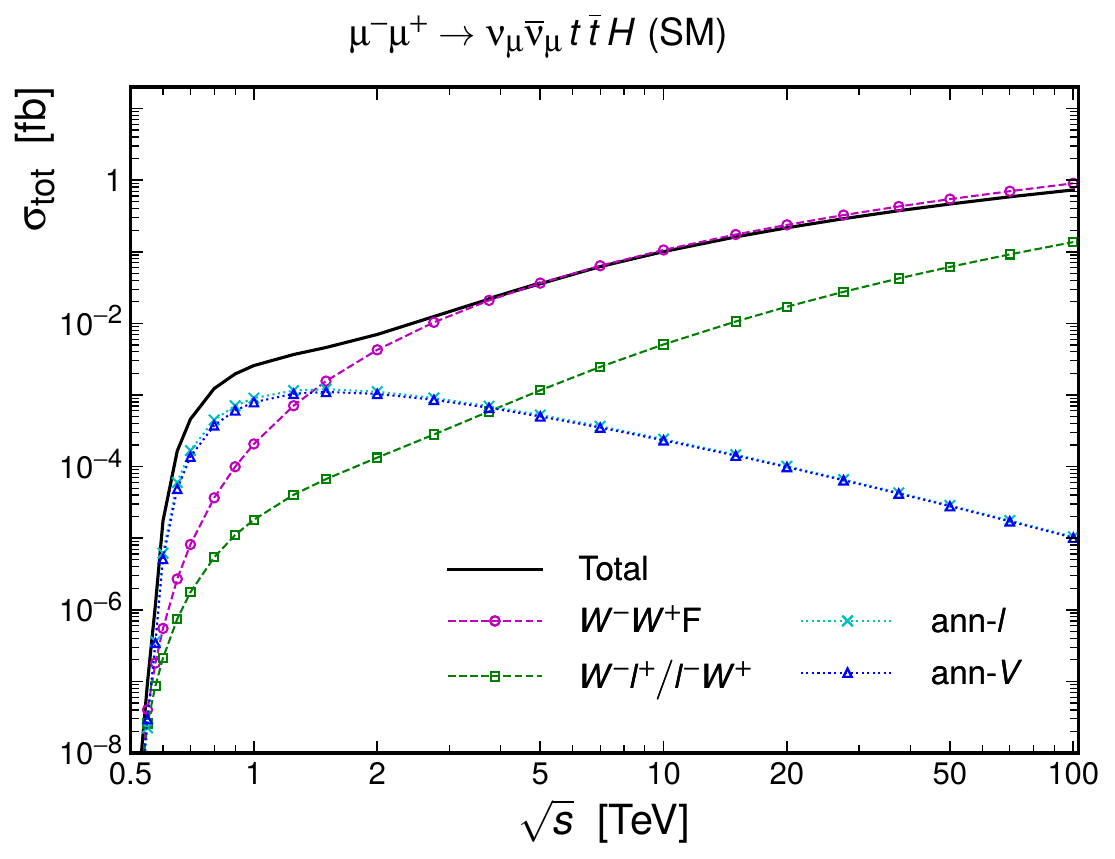}
    \includegraphics[width=0.495\textwidth,clip]{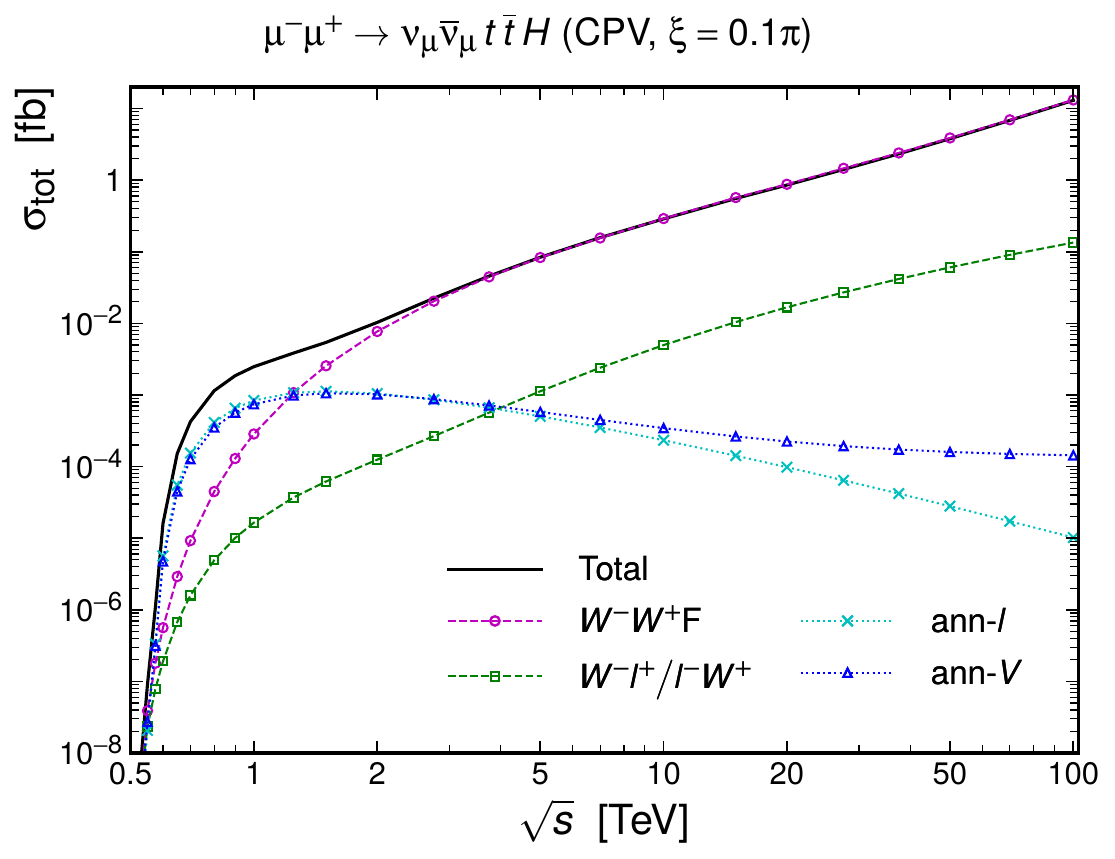}
\caption{Collision-energy dependence of the total cross section for $\mu^-\mu^+\rightarrow\nu_\mu\bar{\nu}_{\mu}t\bar{t}H$ in the SM (left) and in the CPV SMEFT with $(g,\xi)=(g_{\rm SM},0.1\pi)$ (right).
The solid line represents the observable total cross section, while the dashed and dotted lines represent the contributions from each diagram category, as defined in Table~\ref{tab:ex_categories}, calculated in the FD gauge.}
  \label{fig:xsec_proc1}
\end{figure}

In Fig.~\ref{fig:xsec_proc1}, we show the total cross section of $\mu^-\mu^+\to\nu_\mu\bar{\nu}_{\mu}t\bar{t}H$ as a function of the collision energy in the SM (left) and in the CPV SMEFT with $(g,\xi)=(g_{\rm SM},0.1\pi)$ (right). 
The observable total cross section is shown by the black solid line, where all Feynman amplitudes in the FD gauge, i.e., 89 (118) diagrams in the SM (SMEFT), are summed over in the amplitudes, as in Eq.~\eqref{eq:amp_tot}.

\begin{table}
\caption{Number of Feynman diagrams in each category classified in Fig.~\ref{fig:proc1}, where the SMEFT model in the FD gauge is considered. 
The numbers in parentheses indicate the diagram IDs generated by \MG.}
	\label{tab:proc1}
\begin{tabular}{c|lllclrllrl}
\hline
 & \multicolumn{10}{c}{{\bf No. of Feynman diagrams} (diagram ID)} \\
 \multicolumn{1}{c|}{category} &\hspace*{0.1cm}& \multicolumn{1}{c}{total} &\hspace*{0.5cm} & dim4 &\hspace*{0.3cm}& \multicolumn{2}{c}{dim5} &\hspace*{0.3cm}& \multicolumn{2}{c}{dim6} \\ \hline
 (A) && {\bf 30} (47-76) && {\bf 21} && {\bf 8} & (49, 51-53, 65, 70, 75, 76) && {\bf 1} & (47) \\
 (B) && {\bf 26} (77--102) && {\bf 22} &&  {\bf 4} & (88, 89, 101, 102) && {\bf 0} \\
 (C) && {\bf 16} (103--118) && {\bf 14} &&  {\bf 2} & (117, 118) && {\bf 0} & \\ 
 (D)    && {\bf 46} (1--46) && {\bf 32} && {\bf 13} & (3, 6, 10, 13, 16--19, 26, 32, 36, 45, 46) && {\bf 1} & (7) \\ \hline
\end{tabular}
\end{table}

Thanks to the absence of unphysical gauge cancellation among amplitudes in the FD gauge,
the total cross section can be calculated efficiently as a summation over contributions from the 89 (118) phase-space channels, as in Eq.~\eqref{eq:dsigma_mcps}.

Also shown in Fig.~\ref{fig:xsec_proc1} are the partial contributions from subsets of diagrams classified in Fig.~\ref{fig:proc1}, where only the diagrams within each group are considered, and the corresponding amplitudes are summed before taking the absolute square. 
For example, the $W^-W^+$F contribution, shown by a magenta-dashed line with circles, is obtained
as a summation over contributions from the 21 (30) phase-space channels for the $W^-W^+$F amplitudes  
in the SM (SMEFT):
\begin{align}
  \sigma(W^-W^+{\rm F})= \sum_k^{W^-W^+{\rm F}} \int d\Phi_5\, |{\cal M}_k|^2
  \frac{\bigg|\displaystyle\sum_j^{W^-W^+{\rm F}} {\cal M}_j\bigg|^2}{\displaystyle\sum_l^{W^-W^+{\rm F}} |{\cal M}_l|^2}.
\end{align}
Because 
\begin{align}
  \sum_k^{N_d}{\cal M}_k = \sum_j^{W^-W^+{\rm F}}{\cal M}_j
  +\sum_j^{W^-l^+/l^-W^+}{\cal M}_j+\sum_j^{{\rm ann-}l}{\cal M}_j
  +\sum_j^{{\rm ann-}V}{\cal M}_j,
\end{align}
the sum over the four curves gives the total cross section (black-solid curve)
if there is no significant interference among diagrams between different categories.
In other words, if we observe difference between the total cross section and the sum of the contributions from each category diagrams,
it tells significant interference among diagrams belonging to different categories.
Because one of the power of our MCPS integration with the FD gauge amplitudes over the effective real particle distribution method
(which treats contribution from each diagram category as a distinct process)
is its capability of studying interference among amplitudes belonging to different diagram groups, 
we study the mechanisms of such interference effects carefully.

In Fig.~\ref{fig:xsec_proc1}(left) for the SM, one can see that the $t$-channel double and single $W$-exchange contributions, denoted as $W^-W^+$F and $W^-l^+/l^-W^+$, respectively, increase with the collision energy, and the $W^-W^+$F contribution dominates the total cross section for $\sqrt{s}\gtrsim 3$~TeV. 
On the other hand, the annihilation contributions, denoted as ann-$l$ and ann-$V$, are significant only near the threshold region and decrease as $1/s$, as expected. 
At very high energies, the $W^-W^+$F contribution becomes slightly larger than the total cross section. 
This can be explained by the destructive interference between the $W^-W^+$F and $W^-l^+/l^-W^+$ amplitudes, as discussed in more detail later. 

We note that, in the unitary gauge, the $W^-W^+$F and $W^-l^+/l^-W^+$ contributions are much larger than the total cross section in the high-energy region, requiring delicate cancellations to obtain the physical result; see Ref.~\cite{Hagiwara:2024xdh} for details. 
This leads to a serious problem for numerical evaluation. 
In contrast, each FD amplitude behaves well, and such delicate cancellations are absent even at high energies. 
Thanks to our integration method, which takes advantage of this property of the FD amplitudes, we are able to obtain stable and reliable numerical results even at $\sqrt{s}=100$~TeV.

For the CPV case for $(g,\xi)=(g_{\rm SM},0.1\pi)$, shown in Fig.~\ref{fig:xsec_proc1}(right), the overall behavior is similar to that in the SM, 
and one can hardly observe difference at $\sqrt{s}\lesssim 1$~TeV. 
On the other hand, for $\sqrt{s}\gtrsim 3$~TeV, the enhancement of the $W^-W^+$F and ann-$V$ contributions over the corresponding SM curves is clearly seen. 
The origin of this growth at high energies will be identified below.

\begin{figure}
    \includegraphics[width=1\textwidth,clip]{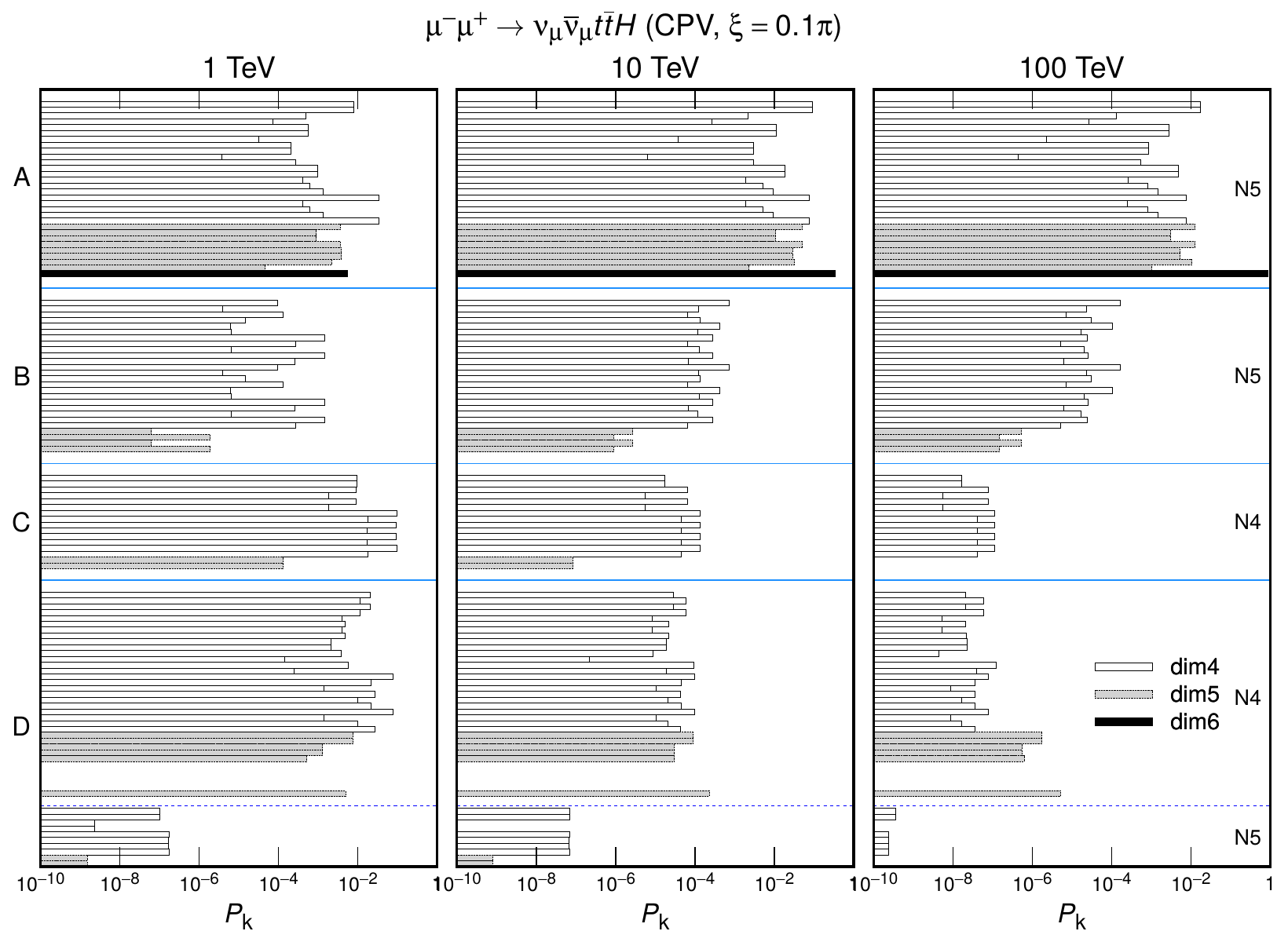}
\caption{
Fraction of the contribution of each FD amplitude channel, defined as $P_k = \sigma_k/\sigma_\mathrm{tot}$ with $\sigma_k$ in Eq.~\eqref{eq:sigma_mcps}, for the process $\mu^-\mu^+\rightarrow\nu_\mu\bar{\nu}_\mu t\bar{t}H$ in the CPV SMEFT with $(g,\xi)=(g_{\rm SM},0.1\pi)$ at three collision energies: 1, 10, and 100~TeV.
The diagrams are categorized as (A)--(D) in Fig.~\ref{fig:proc1}.
Open, grey, and filled histograms correspond to diagrams with dimension-4, 5, and 6 vertices, respectively, as listed in Table~\ref{tab:proc1}.
In each subprocess category, the contributions are also grouped according to the phase-space categories N4 and N5 in Eqs.~\eqref{eq:n4_proc1} and \eqref{eq:n5_proc1}, respectively.
The histogram bins are ordered according to the diagram number given by \MG\ in each category. 
}
  \label{fig:pk_proc1}
\end{figure}

Figure~\ref{fig:pk_proc1} shows the fraction of the contribution of each FD amplitude channel, defined as
\begin{align}
  P_k = \sigma_k/\sigma_\mathrm{tot},
\label{eq:pk}
\end{align}
at $\sqrt{s}=1$, 10, and 100~TeV.
The channels labeled by the diagram number $k$ are grouped according to the subprocess categories, as in Fig.~\ref{fig:proc1}.
Open, grey, and filled histograms correspond to diagrams with only dim4 vertices, those with a dim5, and with a dim6 vertices, respectively, as listed in Table~\ref{tab:proc1}.
For example, according to Table~\ref{tab:proc1}, the weak-boson-fusion amplitudes in (A) $W^-W^+$F have 21 open bins for dim4 vertices only, 8 grey bins for diagrams with one dim5 vertex, and 1 filled bin for a diagram with one dim6 vertex.
We do not show the SM case because the histograms are almost identical to those of the corresponding diagram with the dim4 vertices,
because of the small change in the coupling;
$|g_{\rm SM}-ge^{i\xi}|/g_{\rm SM}=\sqrt{2(1-\cos\xi)}\approx0.3$ for 
$(g,\xi)=(g_{\rm SM},0.1\xi)$.

As expected from Fig.~\ref{fig:xsec_proc1}, when the collision energy increases from 1 to 10 and 100~TeV, the $W^-W^+$F subamplitude contribution becomes relatively larger than those of  the other subamplitude categories.
Interestingly, at $\sqrt{s}=100$~TeV, only a single amplitude with a dim6 vertex in the $W^-W^+$F category (diagram ID=47) dominates the total cross section.
This amplitude has a dim6 contact $W^-W^+t\bar tH$ vertex, which 
originates from the dim6 $\pi^-\pi^+t\bar tH$ term in the SMEFT Lagrangian~\eqref{eq:SMEFTLag}.
The amplitude can be regarded as that of the Goldstone-boson collision process
\begin{align}
  \pi^-\pi^+\to t\bar t H,
  \label{pipi_tth}
\end{align}
as explained in Refs.~\cite{Barger:2023wbg,Hagiwara:2024xdh}.
We note that, unlike the amplitude with a dim6 vertex (ID=47) in (A) $W^-W^+$F,
the corresponding amplitude with a dim6 vertex (ID=7) in (D) ann-$V$ is zero at all energies.
Similarly, five (ID=17, 18, 19, 26, 32) of the 13 amplitudes with a dim5 vertex in (D) ann-$V$, listed in Table~\ref{tab:proc1}, are also zero.
This is because the (reduced) longitudinal polarization component of the virtual Z boson current is zero
for massless leptons in the $l\bar l$ collision rest frame.
See Ref.~\cite{Hagiwara:2024xdh} for further details.

The diagrams can also be reorganized by the number of the quasi on-shell resonances as 
\begin{align}
  &{\rm N4:}\  l\bar{l} \to  t\bar{t}H Z\ (Z\to \nu_l\bar{\nu}_l) ,\label{eq:n4_proc1}\\	
  &{\rm N5:}\  l\bar{l} \to  t\bar{t}H\nu_l\bar{\nu_l} .\label{eq:n5_proc1}
\end{align}
The phase space for the 
N4 diagram channels is given by the four-body phase space ($t\bar tHZ$ productions) times the two-body phase space (Z-boson decays into a neutrino pair),
whereas the phase space for the N5 channels does not have such factorization property.
All the diagrams in the categories (A) and (B) in Fig.~\ref{fig:proc1} are in the N5 category,
while all (C) diagrams are in N4. 
The diagrams in (D) fall into the two phase-space categories, i.e. N4 and N5.
As clearly seen in Fig.~\ref{fig:pk_proc1}, the N5 contributions in (D) ann-$V$ are negligible, comparing with all the other contributions.

\begin{figure}
    \includegraphics[width=1\textwidth,clip]{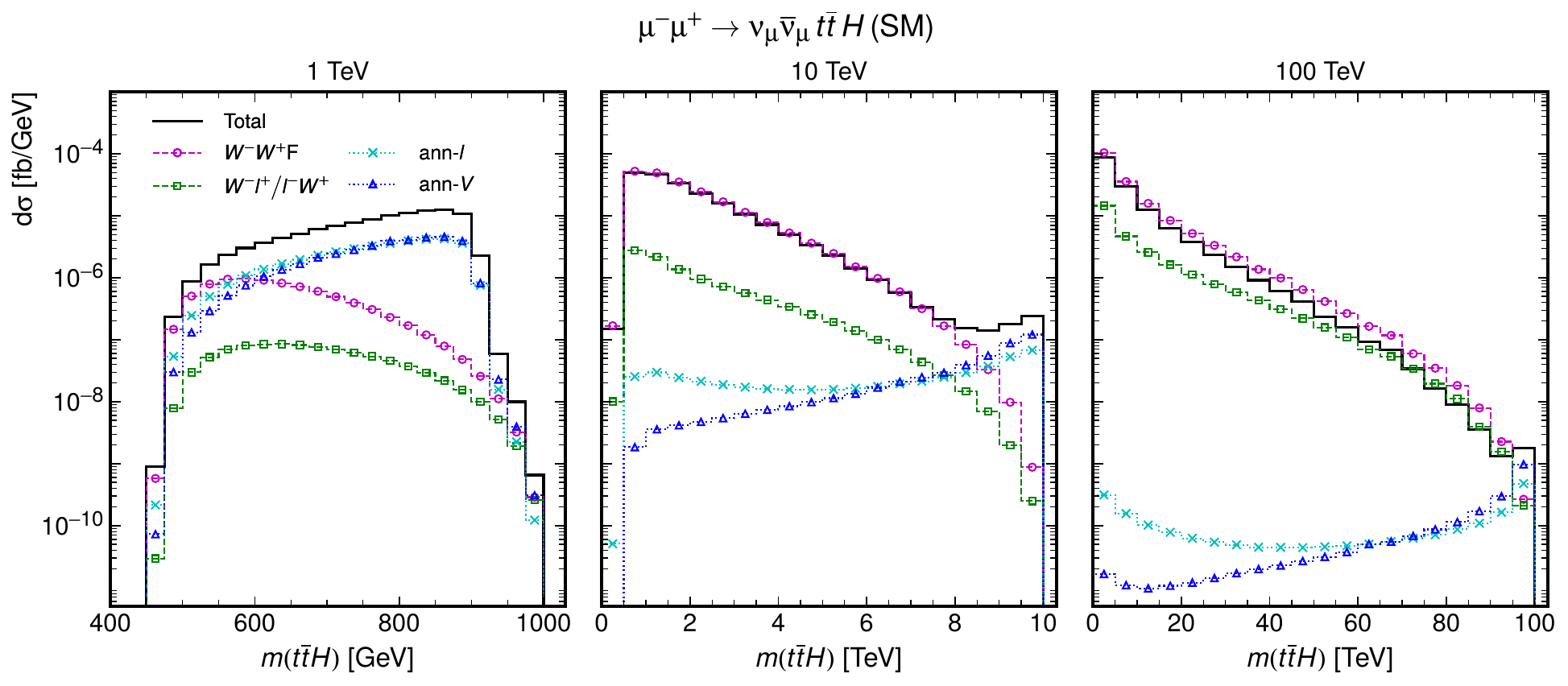}
    \includegraphics[width=1\textwidth,clip]{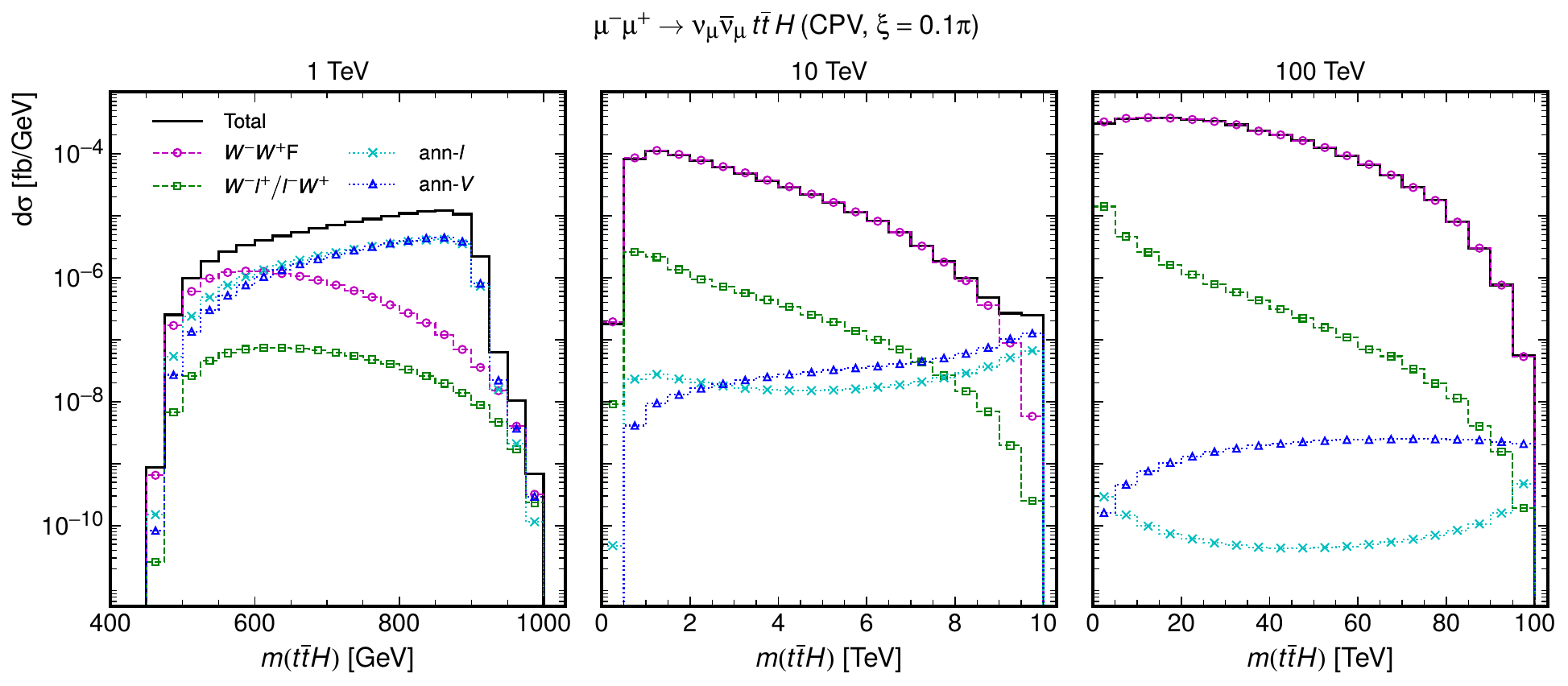}
\caption{Distributions of the invariant mass of the $t\bar{t}H$ system for $\mu^-\mu^+\rightarrow\nu_\mu\bar{\nu}_\mu t\bar{t}H$ in the SM (top) and in the CPV SMEFT with $(g,\xi)=(g_{\rm SM},0.1\pi)$ (bottom) at three collision energies: 1, 10, and 100~TeV. 
The line and marker styles are the same as those in Fig.~\ref{fig:xsec_proc1}.}
  \label{fig:xsec_proc1_mtth}
\end{figure}

We now turn to the differential cross sections.
Figure~\ref{fig:xsec_proc1_mtth} shows distributions of the invariant mass of the $t\bar{t}H$ system in the SM (top) and the CPV SMEFT with $(g,\xi)=(g_{\rm SM},0.1\pi)$ (bottom) at $\sqrt{s}=1$, 10, and 100~TeV.
At all energies, and particularly at higher energies, the $t$-channel $W$-exchange contributions, $W^-W^+$F (magenta with circles) and $W^-l^+/l^-W^+$ (green with squares), dominate the low invariant-mass region.
In contrast, the annihilation contributions, ann-$l$ (light blue with crosses) and ann-$V$ (blue with triangles), become significant in the high invariant-mass region, consistent with our naive expectation based on the vector-boson PDF.

Significant $\xi$ dependence is observed for the $W^-W^+$F subamplitudes. 
At $\sqrt{s}=1$~TeV, although the total cross section in Fig.~\ref{fig:xsec_proc1} shows little difference between the SM ($\xi=0$) and $\xi=0.1\pi$, 
we can observe slight enhancement of the $\xi=0.1\pi$ cross section at $m(t\bar tH)\sim550$~GeV, where the $W^-W^+$F contribution is largest. 
At higher energies $\sqrt{s}=10$ and 100~TeV, the $\xi=0.1\pi$ prediction for the $W^-W^+$F subamplitude contributions
is clearly larger than the SM one at all $m(t\bar tH)$.
At $\sqrt{s}=100$~TeV, the single amplitude (ID=7) with the dim6 contact $W^-W^+t\bar tH$ vertex dominates the entire spectrum
as observed in Ref.~\cite{Hagiwara:2024xdh}. 

We also observe significant enhancement of the ann-$V$ subamplitude contribution,
shown by blue-dashed curves with triangles,
for $\xi=0.1\pi$ at $\sqrt{s}=100$~TeV.
This is due to the 6 amplitudes with a dim5 vertex in the N4 category; see also Fig.~\ref{fig:pk_proc1}.
The dim5 vertex contributing to the non-vanishing ann-$V$ amplitudes has the form 
$\pi^0t\bar tH$, where $\pi^0$ is the fifth component of the virtual Z boson decaying into a pair of neutrinos.
However, this enhancement due to longitudinally polarized Z boson
is overshadowed by the dominant $W^-W^+$F contribution in the inclusive $m(t\bar tH)$ distribution,
except at the highest $m(t\bar tH)$ bins.

\begin{figure}
  \includegraphics[width=1\textwidth,clip]{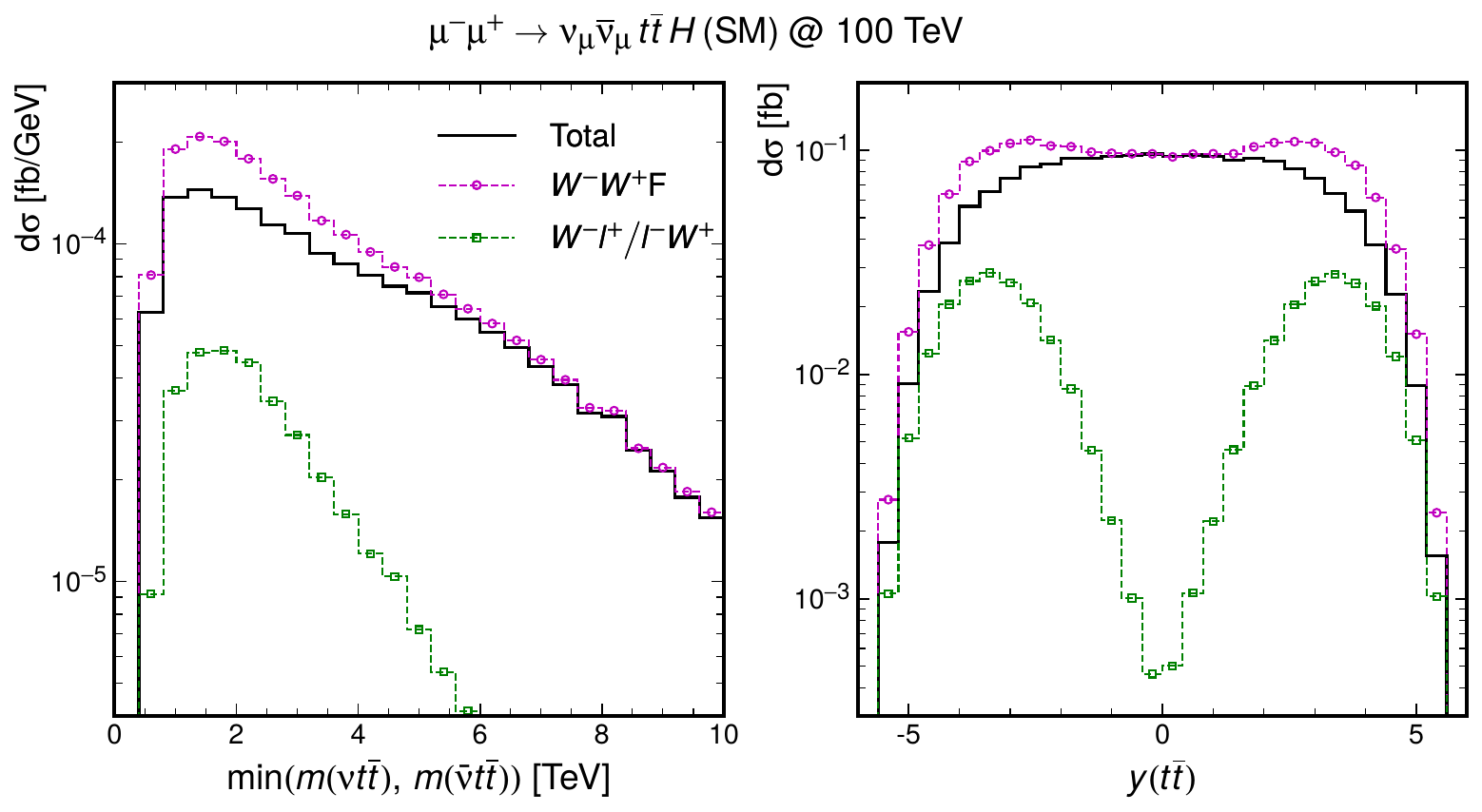}
  \caption{
    Distributions of the minimum of the invariant masses of the $\nu_l t\bar{t}$ and $\bar{\nu}_l t\bar{t}$ systems (left) and 
	rapidity distributions of the $t\bar{t}$ system (right)
	for the process $\mu^-\mu^+ \to \nu_\mu \bar{\nu}_\mu t\bar{t}H$ in the SM at a collision energy of 100~TeV.
    The line and marker styles are the same as those in Fig.~\ref{fig:xsec_proc1}.
	Note that  
	the ann-$l$ and ann-$V$ contributions are too small to appear in the range. 
    }
  \label{fig:fig:xsec_proc1_min_mttxv_yttx}
\end{figure}

At $\sqrt{s}=100$~TeV, for the SM, we observe that the $W^-W^+$F contribution slightly exceeds the physical distribution.
Actually, the excess can already be observed for the total cross-section level in Fig.~\ref{fig:xsec_proc1}(left).
This can be interpreted as destructive interference between the $W^-W^+$F and the $W^-l^+/l^-W^+$  subamplitudes:
The interference between the $W^-W^+$F and the $W^-l^+$ subamplitudes is significant when
$m(\bar\nu_l t\bar t)$ is small
since the $W^-l^+$ subamplitudes have an $s$-channel $\nu_l$ propagator with the invariant mass of $m(\bar\nu_l t\bar t)$, 
whereas the interference between the $W^-W^+$F and the $l^-W^+$ subamplitudes is significant when
$m(\nu_l t\bar t)$ is small.
Therefore, when one of the two invariant mass is small,
we expect significant interference,
which can indeed be observed in the left panel in Fig.~\ref{fig:fig:xsec_proc1_min_mttxv_yttx}.
At $\sqrt{s}=100$~TeV, 
the destructive interference is significant when min$(m(\nu_l t\bar t),m(\bar\nu_l t\bar t))\lesssim 5$~TeV,
while negligible when the both invariant masses are larger.

Although the invariant mass of a system with neutrinos is difficult to measure
even at lepton colliders,
we can observe localization of the significant interference in the rapidity distribution of the $t\bar t$ system, $y(t\bar t)$, 
which is shown in Fig.~\ref{fig:fig:xsec_proc1_min_mttxv_yttx}(right).
As we take the initial $\mu^-$ momentum in the positive $z$ direction, 
the $t\bar t$ system in the $W^-l^+$ ($l^-W^+$) collision tends to be produced in the negative (positive) $y(t\bar t)$ region.
At $\sqrt{s}=100$~TeV, the rapidity distribution of the $W^-l^+/l^-W^+$ contribution is peaked at $|y(t\bar t)|\sim3.5$ , where we observe significant destructive interference between $W^-W^+$F and $W^-l^+$ ($y(t\bar t)\sim-3.5$) or $l^-W^+$ ($y(t\bar t)\sim3.5$) subamplitudes. 

Figure~\ref{fig:fig:xsec_proc1_min_mttxv_yttx} suggests that 
interference effects among amplitudes can be localized in some specific kinematical region.
Use of a final-state cut, such as 
$|y(t\bar t)|<y_{\rm cut}$, 
may hence allow us to study detailed kinematical distributions based on the $W^-W^+$F subamplitudes.
It should be noted here that, in the unitary gauge, the distributions of the partial amplitudes are far from the physical distributions, and hence 
it is difficult to obtain insights from the interference patterns among contributing diagrams.

\section{Numerical results for $l\bar{l} \to l\bar{\nu}_l t\bar{b}H$}
\label{sec:lvtbh}

\begin{figure}
  \includegraphics[width=0.14\textwidth,clip]{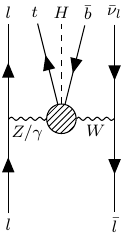} 
  \quad
  \includegraphics[width=0.145\textwidth,clip]{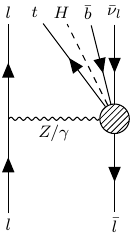} 
  \includegraphics[width=0.15\textwidth,clip]{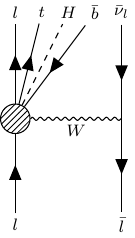} 
  \quad
  \includegraphics[width=0.145\textwidth,clip]{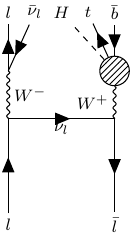} 
  \includegraphics[width=0.14\textwidth,clip]{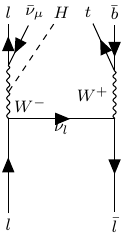} 
  \quad
  \includegraphics[width=0.145\textwidth,clip]{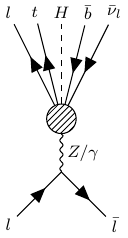} 
\hspace*{-0.cm}(A)\hspace*{3.6cm}(B)\hspace*{4.9cm}(C)\hspace*{3.5cm}(D)
\caption{ 
The Feynman diagrams for the process $l\bar{l}\to l\bar{\nu}_l t\bar{b} H$ are classified into four categories, as defined in Table~\ref{tab:ex_categories}: (A) $VW^+$F, (B) $Vl^+/l^-W^+$, (C) ann-$\nu$, and (D) ann-$V$, with $V=Z/\gamma$.}
\label{fig:proc2}
\end{figure}

This section demonstrates the SDE MCPS integration for the process
\begin{align}
    l\bar{l} \rightarrow l\bar{\nu}_l t\bar{b} H,
    \label{eq:ex2-1}
\end{align}
whose Feynman diagrams are shown in Fig.~\ref{fig:proc2}. 
They are classified into four categories: (A) $VW^+$F, (B) $Vl^+/l^-W^+$, (C) ann-$\nu$, and (D) ann-$V$, with $V=Z/\gamma$, as defined in Table~\ref{tab:ex_categories}. 
The number of Feynman diagrams in each category is summarized in Table~\ref{tab:proc2}. 

With currently available event-generation programs, it is difficult to calculate the cross section for the process accurately without any kinematical cuts, because of the lepton-mass singularities arising from $t$-channel photon-exchange diagrams, as seen in groups (A) and (B)-left in Fig.~\ref{fig:proc2}.

We have developed a phase-space parametrization that accurately reproduces the distributions of forward-emitted charged leptons.
As mentioned in Sec.~\ref{sec:PSnote}, by introducing the virtuality $q^2$ of the off-shell vector boson as the fifth element of the momentum vector and using its value directly, our library module \texttt{ph2t} can compute the kinematics of the phase space with high numerical precision, at an arbitrary high energies, as large as $E_l/m_l<\infty$.

In addition, we have modified the numerical code (\HELAS~\cite{Murayama:1992gi}) to evaluate helicity amplitudes in the singular region, so that vertices with very small (virtual) invariant momentum squared of order $m_l^2$ are treated accurately even at TeV energies.
Similar to the phase-space library modules, we added $q^2$ as the fifth element of the momentum vector in the \HELAS\ library and used its value directly when calculating amplitudes in \HELAS.
These modifications to \HELAS\ resulted in stable execution of the integrations and high numerical accuracy in the results.

\begin{table}
 \caption{Number of Feynman diagrams in each group classified in Fig.~\ref{fig:proc2}, where the SMEFT model in the FD gauge is considered. 
The numbers in parentheses indicate the diagram IDs generated by \MG.}
	\label{tab:proc2}
	\vspace*{2mm}
\begin{tabular}{c|lrllrlrllc}
\hline
 & \multicolumn{10}{c}{{\bf No. of Feynman diagrams} (diagram ID)} \\
 \multicolumn{1}{c|}{category} &\hspace*{0.1cm}& \multicolumn{2}{c}{total} &\hspace*{0.5cm} & dim4 &\hspace*{0.3cm}& \multicolumn{2}{c}{dim5} &\hspace*{0.3cm}& \multicolumn{1}{c}{dim6} \\ \hline
 (A) && {\bf 28} & (40-67) && {\bf 21} && {\bf 7} & (42, 44, 47, 49, 52, 66, 67) && {\bf 0}  \\
 (B) && {\bf 15} & (68--82) && {\bf 11} &&  {\bf 4} & (76--78, 82) && {\bf 0} \\
 (C) && {\bf 4}& (83--86) && {\bf 3} &&  {\bf 1} & (86) && {\bf 0}  \\ 
 (D) && {\bf 39}& (1--39) && {\bf 29} && {\bf 10} & (3, 5, 8, 10, 13, 27, 28, 37--39) && {\bf 0}  \\ \hline
\end{tabular}
\end{table}

\begin{figure}
    \includegraphics[width=0.495\textwidth,clip]{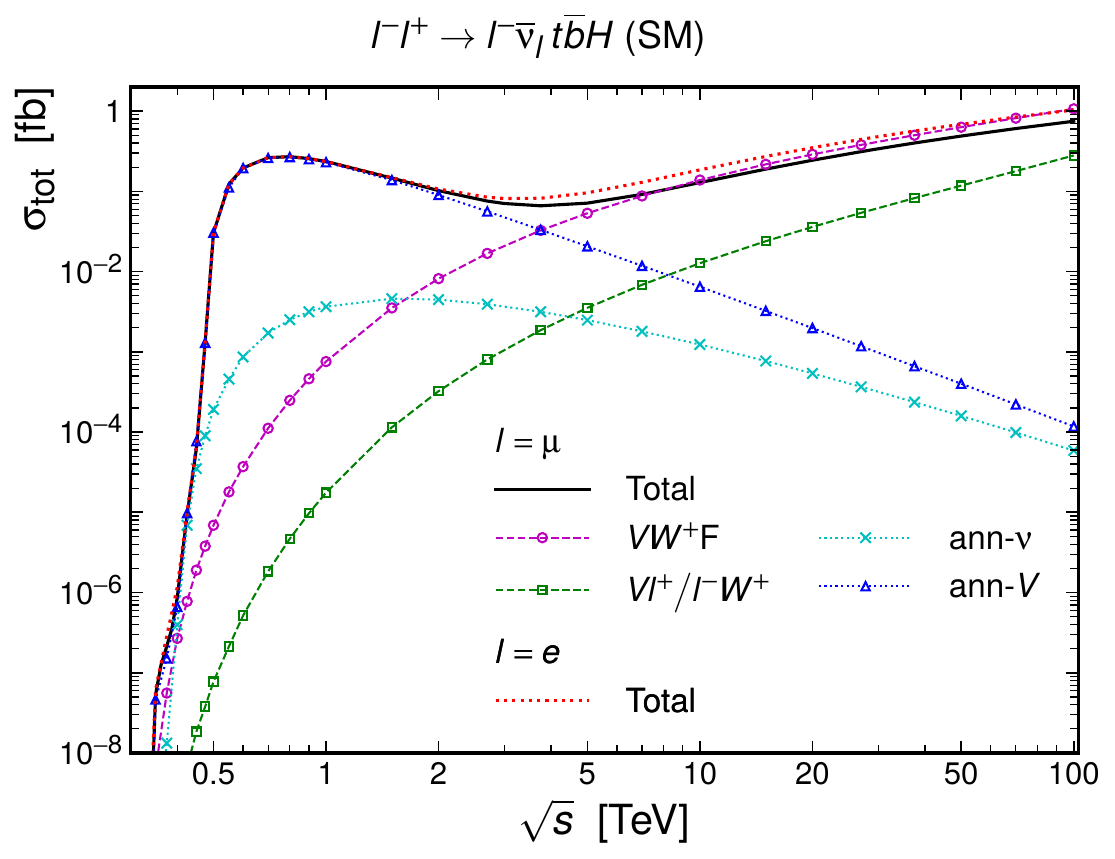}
    \includegraphics[width=0.495\textwidth,clip]{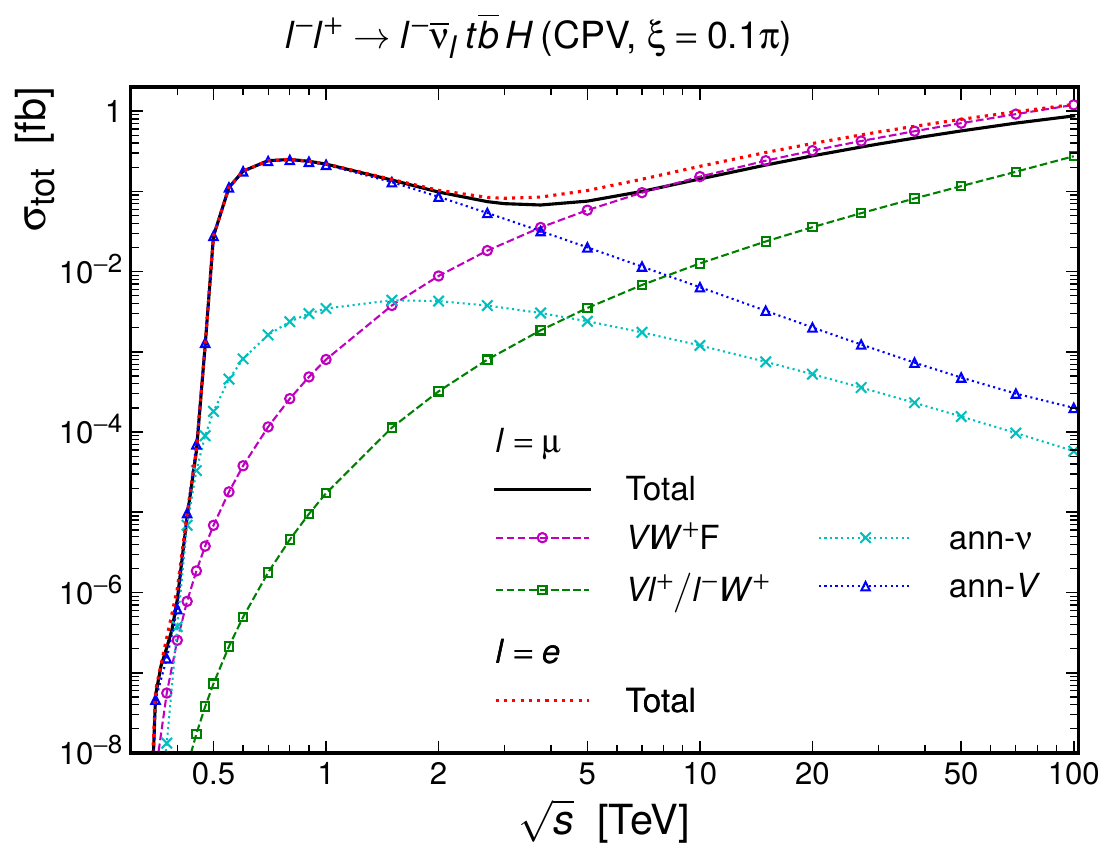}
    \caption{
      Collision-energy dependence of the total cross section for the process
$\mu^-\mu^+ \to \mu^-\bar{\nu}_{\mu} t\bar{b}H$ in the SM (left) and in the CPV SMEFT
with $(g,\xi)=(g_{\rm SM},0.1\pi)$ (right).
The solid line represents the observable total cross section, while the dashed and dotted lines represent the contributions from each diagram category, as defined in Table~\ref{tab:ex_categories}, calculated in the FD gauge.
The red dotted line represents the observable total cross section for
$e^-e^+ \to e^-\bar{\nu}_{e} t\bar{b}H$.
}
\label{fig:xsec_proc2}
\end{figure}

Figure~\ref{fig:xsec_proc2} shows the total cross section for $\mu^-\mu^+\to\mu^-\bar{\nu}_{\mu}t\bar{b}H$ as a function of the collision energy, in the SM (left) and in the CPV SMEFT with $(g,\xi)=(g_{\rm SM},0.1\pi)$ (right). 
The observable total cross section is shown by the black solid line, where all Feynman diagrams in the FD gauge, i.e., 64 (86) in the SM (SMEFT), are included.
Thanks to the developments described above, we can obtain accurate cross sections with the explicit muon or electron mass even at $\sqrt{s}=100$~TeV without applying any kinematical cuts.

\begin{figure}
  \includegraphics[width=1\textwidth,clip]{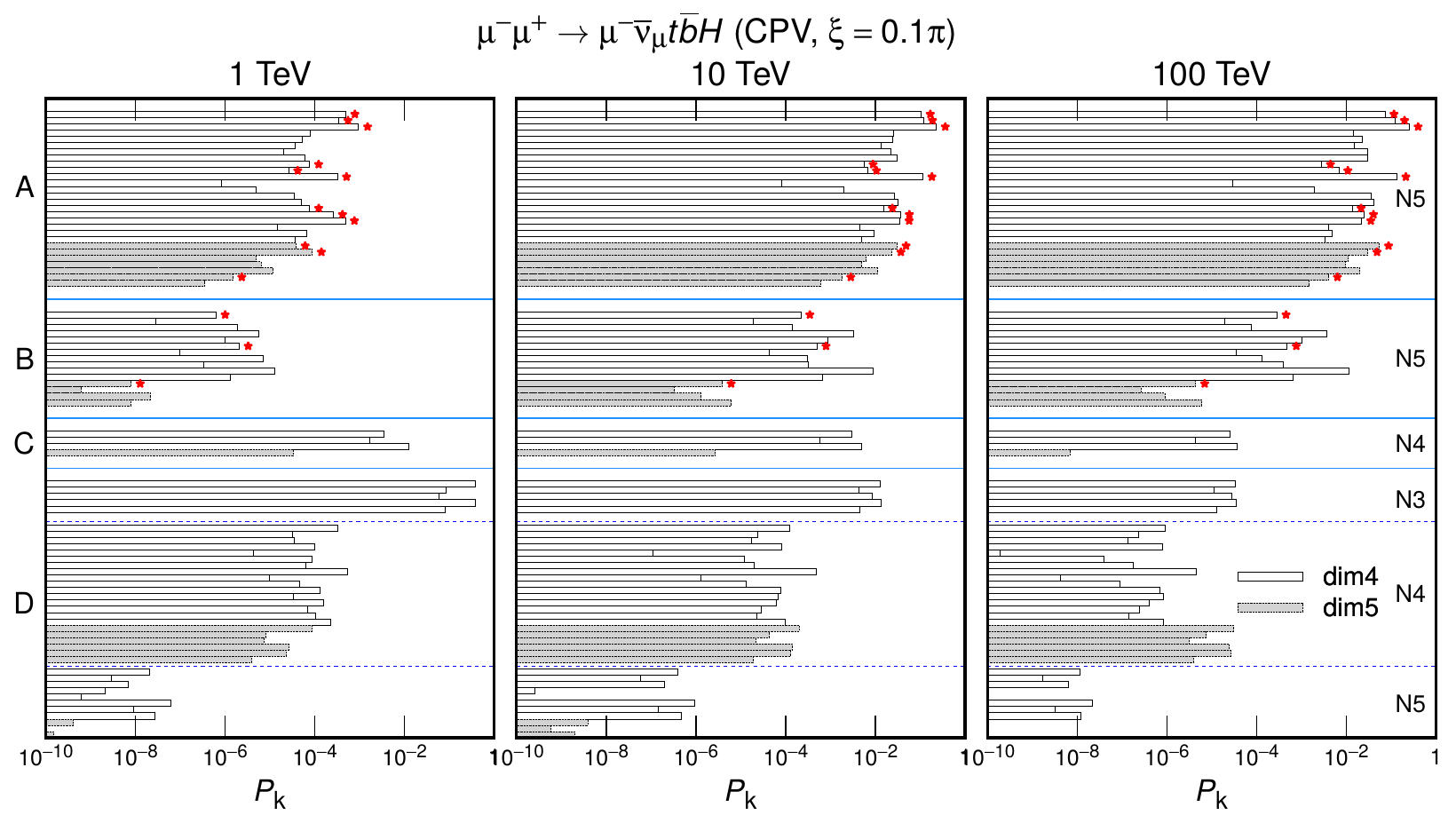}  
  \caption{
Fraction of the contribution of each FD amplitude, defined as $P_k = \sigma_k/\sigma_\mathrm{tot}$ with $\sigma_k$ in Eq.~\eqref{eq:sigma_mcps}, for the process $\mu^-\mu^+\rightarrow\mu^-\bar{\nu}_\mu t\bar{b}H$ in the CPV SMEFT with $(g,\xi)=(g_{\rm SM},0.1\pi)$ at three collision energies: 1, 10, and 100~TeV.
The diagrams are categorized as (A)--(D) in Fig.~\ref{fig:proc2}.
Open and grey histograms correspond to diagrams with dimension-4 and 5 vertices, respectively, as listed in Table~\ref{tab:proc2}.
In each subprocess category, the contributions are also grouped according to the phase-space categories N3, N4 and N5 in Eqs.~\eqref{eq:n3_proc2}, \eqref{eq:n4_proc2} and \eqref{eq:n5_proc2}, respectively.   
Red stars denote the amplitude including the $t$-channel photon propagator.
}
\label{fig:pk_proc2}
\end{figure}

A clear transition of the dominant contribution can be observed around $\sqrt{s}=3$–$4$~TeV. The ann-$V$ subamplitude contribution dominates in the low-energy region, whereas the $VW^+$F one becomes dominant at high energies.
Since, among the $VW^+$F contribution, the contribution from $t$-channel photon-exchange becomes significant, which is proportional to  
$\ln(\sqrt{s}/m_l)$, 
we find the enhancement of the cross section for $l=e$ (red dotted line) 
over the $l=\mu$ case (black solid line)
above $\sqrt{s}=3$–$4$~TeV.
Explicit values of the total cross sections and the $l=e$ over $l=\mu$ ratios are shown in Table~\ref{tab:proc2_xsec}.%
\footnote{
Those values are obtained for the default parameters in the \MG,
$1/\alpha_{\rm EM}=127.9$, $m_Z=91.1876\,$GeV, $G_F=1.166379\times10^{-5}$, $m_H=125\,$GeV, $m_t=173\,$GeV, $m_b=4.7\,$GeV 
with the finite charged lepton masses as $m_e=5.11\times10^{-4}\,$GeV and $m_\mu=1.0566\times10^{-1}\,$GeV.
}
The ratio $\sigma_e/\sigma_\mu$ agrees with the expectation
\begin{align}
  \frac{\sigma_e}{\sigma_\mu}\approx\frac{\ln(\sqrt{s}/m_e)}{\ln(\sqrt{s}/m_\mu)}
\end{align}
for $\sqrt{s}\gtrsim3$–$4$~TeV both for the SM and the CPV case.
We also confirm the dominance of the $t$-channel photon-exchange amplitudes 
in the $P_k$ distribution in Fig.~\ref{fig:pk_proc2}, 
where the corresponding channels are marked by red stars.

\begin{table}
 \caption{Values of the total cross sections in fb and the ratios, shown in Fig.~\ref{fig:xsec_proc2}.}
	\label{tab:proc2_xsec}
\begin{tabular}{c|ccc|ccc|cc}
\hline
 & \multicolumn{3}{c|}{SM} &  \multicolumn{3}{c|}{CPV ($\xi=0.1\pi$)} & \multicolumn{2}{c}{CPV/SM} \\
$\sqrt{s}$~[TeV] & $\sigma_e$ & $\sigma_\mu$ & $\sigma_e/\sigma_\mu$ & $\sigma_e$ & $\sigma_\mu$ & $\sigma_e/\sigma_\mu$ & $e$ & $\mu$\\ \hline
 1     & 0.24 & 0.24 & 1.00  & 0.22 & 0.22 & 1.01  & 0.93 & 0.93 \\
 10    & 0.19 & 0.13 & 1.44  & 0.21 & 0.14 & 1.44  & 1.10 & 1.11 \\
 100   & 1.04 & 0.75 & 1.38  & 1.21 & 0.87 & 1.39  & 1.16 & 1.16 \\
 \hline
\end{tabular}
\end{table}

We note here that,
similar to the $\nu_l\bar{\nu}_l t\bar{t}H$ final state shown in Fig.~\ref{fig:xsec_proc1}, 
the destructive interference effects are observable at high energies
for both the SM and the SMEFT with $(g,\xi)=(g_{\rm SM},0.1\pi)$ in Fig.~\ref{fig:xsec_proc2}.
This can be interpreted as destructive interference between the $VW^+$F and $Vl^+/l^-W^+$ subamplitudes.

Unlike the $l\bar l\rightarrow\nu_l\bar{\nu}_l t\bar{t}H$ process, no dim6 vertex contributes to this process, and the enhancement of the cross section from the CPV effect is not as large as for the $\nu_l\bar{\nu}_l t\bar{t}H$ case.
The total cross section of the SMEFT with $(g,\xi)=(g_{\rm SM},0.1\pi)$ is about
$-7$\%, 10\%, and 16\% different from the SM
at $\sqrt{s}=1$, 10, and 100~TeV; see Table~\ref{tab:proc2_xsec}.
Moderate $\xi$ dependence of the total cross section is 
a consequence of the absence of the dim6 vertex amplitudes in this process.
Amplitudes with a dim5 vertex give cross sections which grow logarithmically with energies,
as in the SM,
whereas amplitudes with a dim6 vertex give cross sections
which grow as a power of energy as shown by the filled histogram in Fig.~\ref{fig:pk_proc1}.

Similar to the previous section, 
the diagrams can be reorganized by the number of the quasi on-shell resonances as 
\begin{align}
  &{\rm N3:}\  l\bar{l} \to  H t\bar{t}\ (\bar t \to W^- \bar b, W^-\to l\bar{\nu}) ,\label{eq:n3_proc2}\\	 	
  &{\rm N4:}\  l\bar{l} \to  H t\bar{b} W^-\ (W^-\to l\bar{\nu}) ,\label{eq:n4_proc2}\\	
  &{\rm N5:}\  l\bar{l} \to  H t\bar{b}l\bar{\nu},
  \label{eq:n5_proc2}
\end{align}
where N3 diagrams are excluded from N4, and N3 and N4 diagrams are excluded from N5.
The phase space for 
N3 diagram channels is given by the three-body phase space times twice the two-body phase space,
whereas the phase space for the N4 channels is given by 
the four-body phase space times the two-body phase space.
All the diagrams in the categories (A) and (B) in Fig.~\ref{fig:proc2} are in the N5 category,
while all (C) diagrams are in N4. 
The diagrams in (D) fall into the three phase-space categories, i.e. N3, N4, and N5,
and the phase-space suppression effect can be clearly see in Fig.~\ref{fig:pk_proc2},
where the ann-$V$ subamplitude contribution (D) is dominated by the N3 contribution, i.e. the $t\bar tH$ production followed by the leptonic anti-top quark decay, at all energies.

\begin{figure}
 \includegraphics[width=0.85\textwidth,clip]{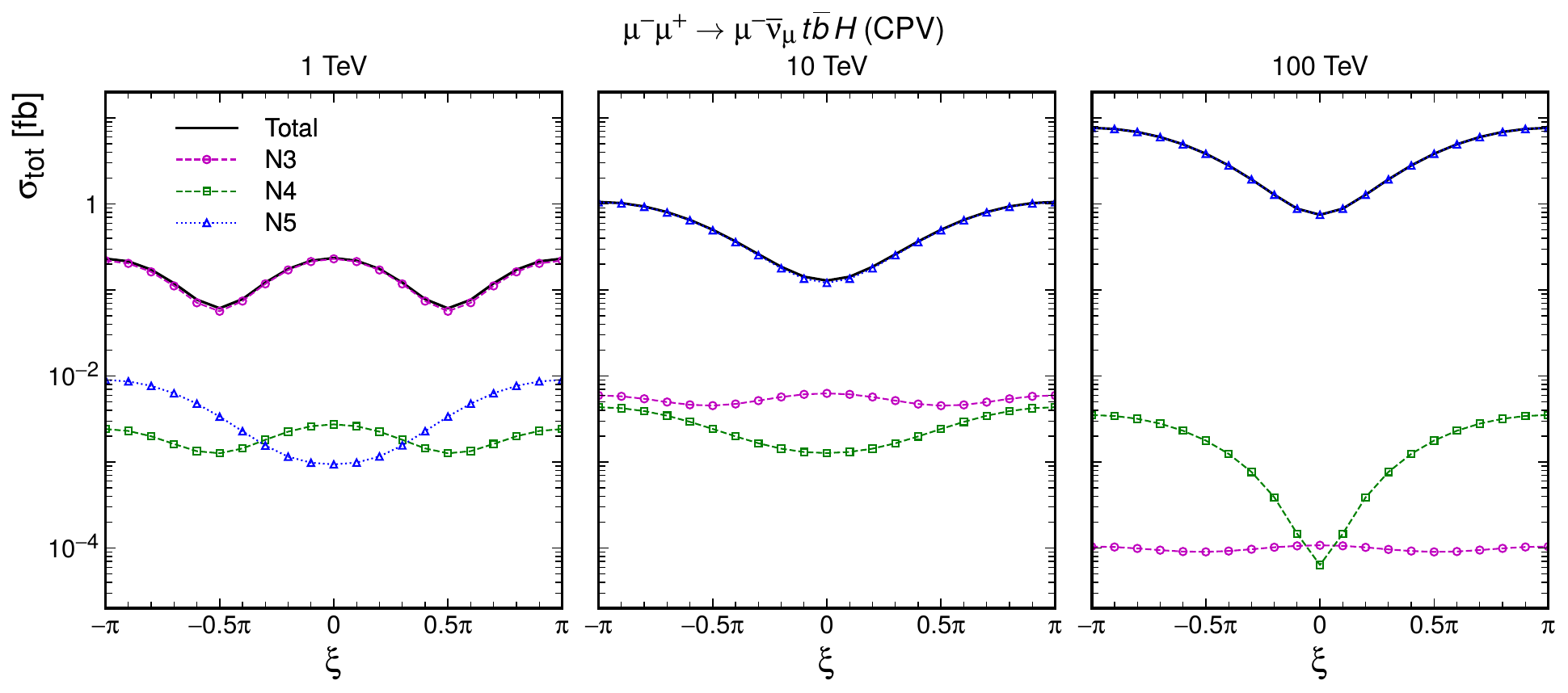}
  \caption{CP-phase dependence of the total cross section for the process $\mu^-\mu^+\rightarrow\mu^-\bar{\nu}_\mu t\bar{b}H$ in the CPV SMEFT at three collision energies: 1, 10, and 100~TeV.
  The black solid line represents the observable total cross section, while the dashed lines show the contributions from each phase-space category, calculated in the FD gauge. 
}
\label{fig:xidep_proc2}
\end{figure}

The $\xi$ dependence of the total cross section at $\sqrt{s}=1$, 10, and 100~TeV
is shown in Fig.~\ref{fig:xidep_proc2} for the entire range, $-\pi<\xi<\pi$, 
where the partial contributions from diagrams in the N3, N4, N5 categories are shown separately. 
At $\sqrt{s}=1$~TeV, the cross section is dominated by the N3 diagrams,
or by the subprocesses $l\bar{l} \to H t\bar{t}$~\eqref{eq:n3_proc2}, 
whose $\xi$ dependence is well known~\cite{Hagiwara:2017ban,Cassidy:2023lwd}.
At $\sqrt{s}=10$~TeV, the N5 $VW^+$F contribution dominates over the N3 and N4 contribution.
Because the $\xi$ dependence of the dominant N5 contribution and the subdominant N3 contribution is opposite
near $\xi=0$, 
the $\xi$ dependence of the total cross section is moderate at lower energies, 1~TeV$<\sqrt{s}<10$~TeV. 
At $\sqrt{s}=100$~TeV, the N5 $VW^+$F dominance is clearer and the ratio of
$\sigma(\xi=\pi)/\sigma(\xi=0)\approx10$ reflects the ten times larger cross section of the single top and Higgs production process ($bl^+\to\bar\nu tH$)
when the sign of the top-Yukawa coupling is reversed~\cite{Stirling:1992fx,Bordes:1992jy,Demartin:2015uha}.

\begin{figure}
  \includegraphics[width=1\textwidth,clip]{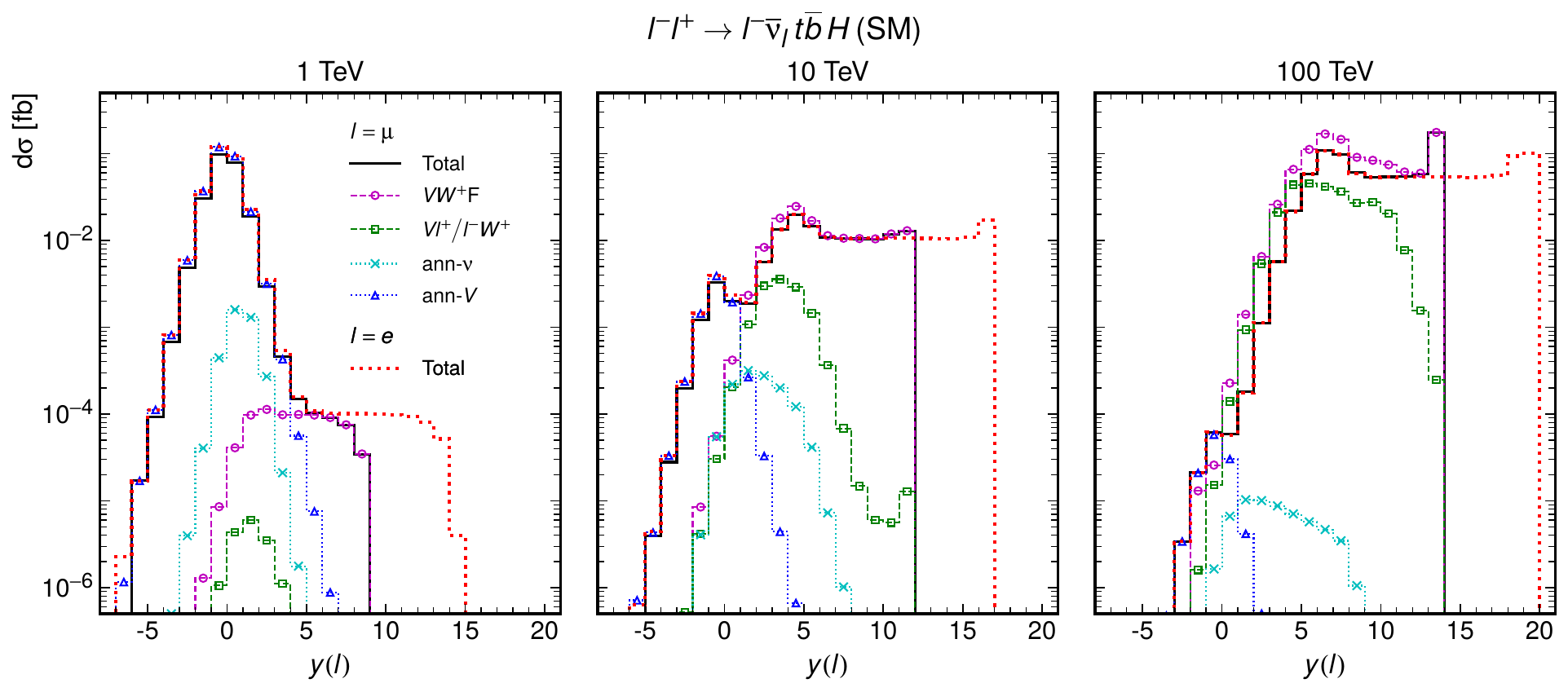}
  \caption{
Rapidity distributions of the outgoing $\mu^-$ for the process
$\mu^-\mu^+ \to \mu^-\bar{\nu}_{\mu} t\bar{b}H$ in the SM at three collision energies: 1, 10, and 100~TeV. 
The line and marker styles are the same as those in Fig.~\ref{fig:xsec_proc2}.
The red-dotted line indicates the rapidity distribution of the outgoing
$e^-$ for the process $e^-e^+ \to e^-\bar{\nu}_{e} t\bar{b}H$.
}
  \label{fig:mummup_mumvmxtbxh_yl}
\end{figure}

In order to see the enhancement from the $t$-channel photon exchange explicitly, 
in Fig.~\ref{fig:mummup_mumvmxtbxh_yl} we show rapidity distributions of the charged lepton ($l=\mu$ or $e$) for the process $l^-l^+\rightarrow l^-\bar{\nu}_{l}t\bar{b}H$ in the SM at three collision energies: 1, 10, and 100~TeV.
At all energies, especially at higher energies, the $VW^+$F contribution dominates in the forward region. 
Here, the $l=\mu$ case is shown by black solid curves, while the $l=e$ results are shown by red dotted curves. 
We can clearly see an effect of the lepton mass in the very forward region 
since the maximum of the rapidity is given by
\begin{align}
  y_{\rm max}\approx \ln\Big\{\frac{\sqrt{s}}{m_l}\Big(1-\frac{(m_t+m_b+m_H)^2}{s}\Big)\Big\},
\end{align}
and hence $y_{\rm max}\sim \ln(\sqrt{s}/m_l)$ at high energies.
The values of $y_{\rm max}$ for the muon and the electron are
\begin{align}
  y_{\rm max}(\mu) &=9.2,\ 11.5,\ 13.8, \\
  y_{\rm max}(e)   &=14.5,\ 16.8,\ 19.1, 
\end{align}
at $\sqrt{s}=1$, 10, 100~TeV, respectively, as one can observe clearly in Fig.~\ref{fig:mummup_mumvmxtbxh_yl}.  
This lepton-mass singularity explains the enhancement of the total cross section for $e^-e^+$ collisions at higher energies in Fig.~\ref{fig:xsec_proc2}.

\begin{figure}
  \includegraphics[width=0.975\textwidth,clip]{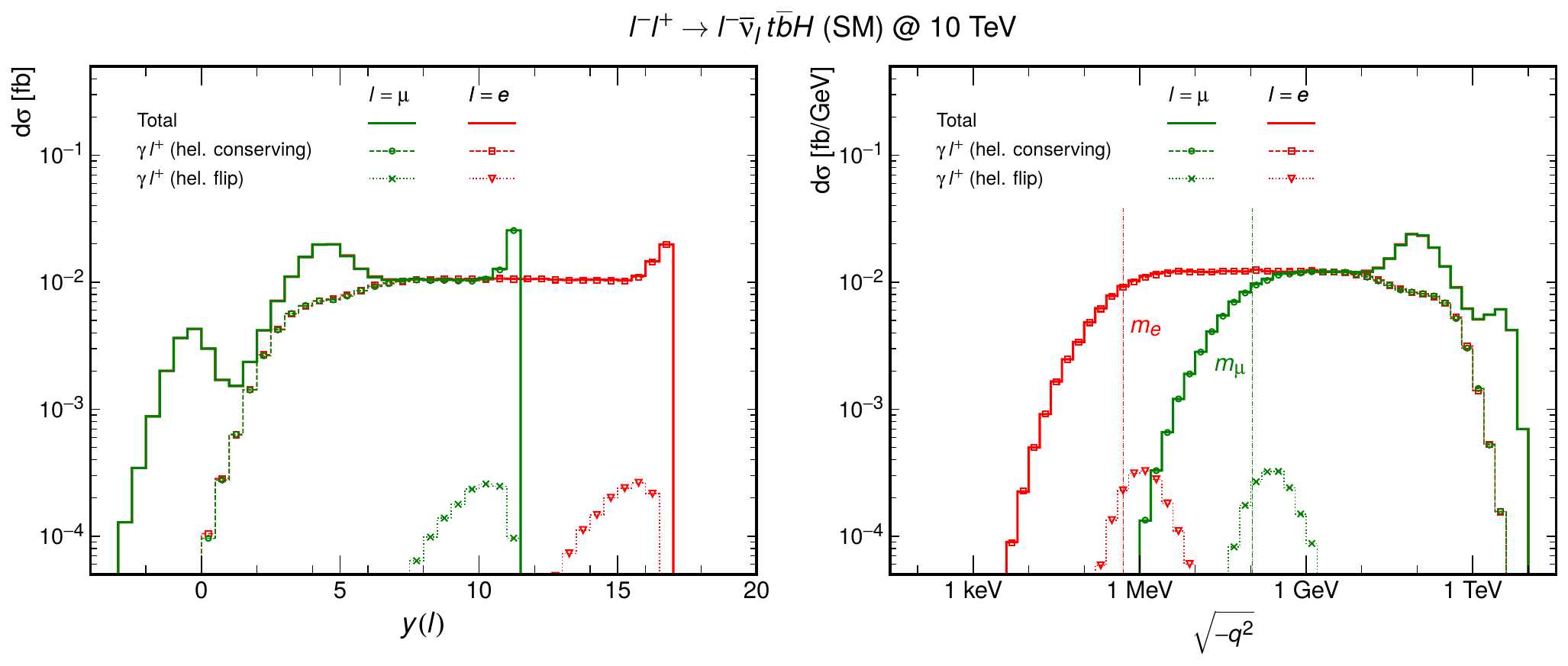}
  \includegraphics[width=0.975\textwidth,clip]{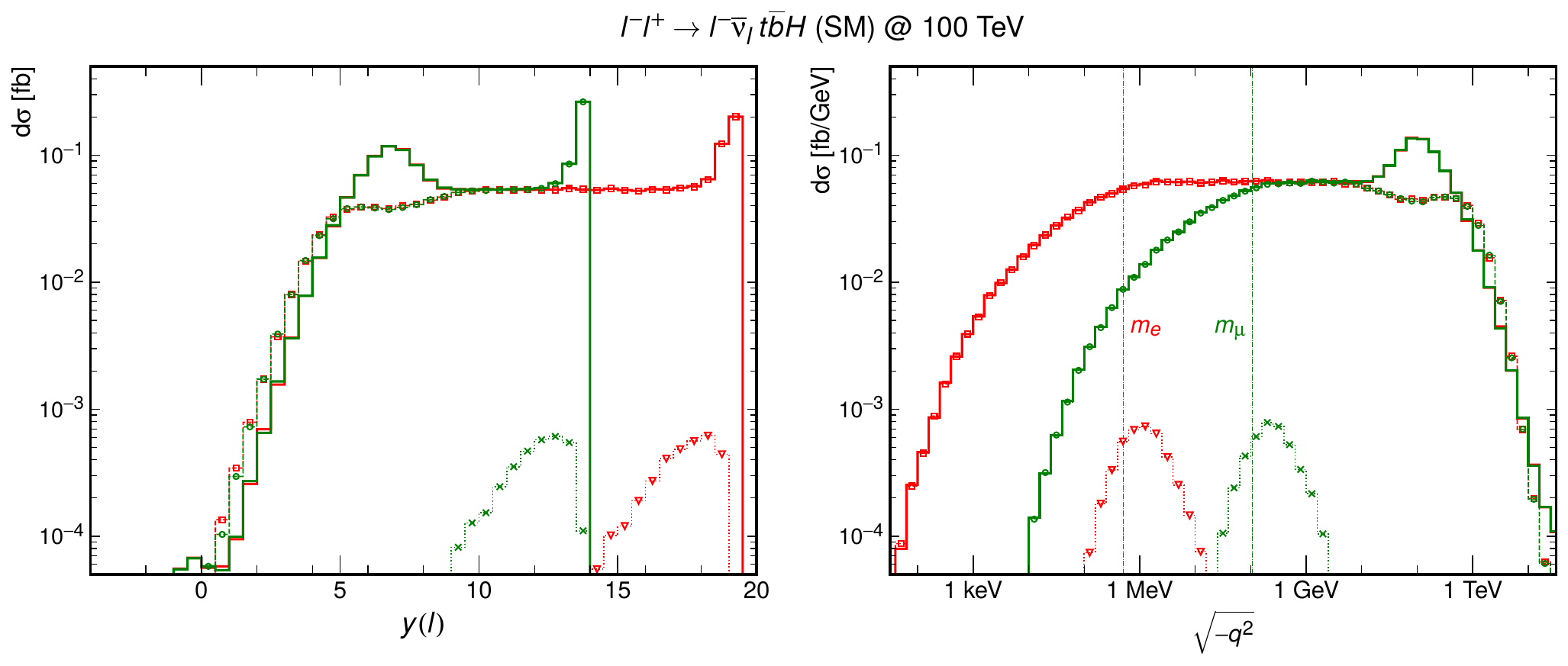}
  \caption{Distributions of the rapidity of the charged leptons (left) and $\sqrt{-q^2}$ (right) for the process $l^-l^+\to l^-\bar{\nu}_lt\bar{b}H$ in the SM at a collision energy of 10~TeV (top) and 100~TeV (bottom). 
  The squared momentum transfer is defined by $q^2=(p_{l^-}^{\rm initial}-p_{l^-}^{\rm final})^2$.
    Solid lines denote the total cross sections for electrons (red) and muons (green). 
    Dashed lines show contributions from the $\gamma l^+$-collision subprocesses with helicity-conserving interactions, while dotted lines correspond to the same subprocesses with helicity-flip interactions.
}
\label{fig:lmlp_lmvlxtbxh_helicity}
\end{figure}

We observe in Fig.~\ref{fig:mummup_mumvmxtbxh_yl} that there appears a peak in the lepton rapidity distribution near $y\approx y_{\rm max}$ at high energies, $\sqrt{s}=10$ and 100~TeV.
In order to understand the cause of the peak, we show in Fig.~\ref{fig:lmlp_lmvlxtbxh_helicity}  
distributions of the rapidity of the charged lepton (left) and those of $\sqrt{-q^2}$ (right) for the process $l\bar{l}\to l\bar{\nu}t\bar{b}H$ in the SM at a collision energy of 10~TeV (top) and 100~TeV (bottom).
The squared momentum transfer $q$ is defined by 
\begin{align}
  q^2=(p_l^{\rm initial}-p_l^{\rm final})^2=-Q^2,
\end{align}
which is the invariant mass of the virtual photon exchanged in the $t$-channel.
Solid lines denote the total cross sections for electrons (red) and muons (green), respectively.
The $\sqrt{-q^2}$ distributions extend down to $\sqrt{t_{\rm min}}\approx m_l x/\sqrt{1-x}$, Eq.~\eqref{eq:tmin3},
where $x$ is the energy fraction $E_\gamma/E_l$,
which can be significantly below $m_l$ for small $x$ at $\sqrt{s}>10$~TeV.
All the contributions from the $\sqrt{-q^2}\lesssim m_l$ give collinear splitting, and  
a peak at $y\approx y_{\rm max}$ appears at high energies where $x\approx m(t\bar b H\bar\nu)^2/s$ can be small.

In Fig.~\ref{fig:lmlp_lmvlxtbxh_helicity}, we show the contribution of lepton helicity-conserving ($h_l^{\rm final}=h_l^{\rm initial}$) and
lepton helicity-flip ($h_l^{\rm final}=-h_l^{\rm initial}$) amplitudes of the $\gamma l^+$ collision subprocesses,
the sum of $\gamma W^+$F (A) and $\gamma l^+$ (B-left) amplitudes in Fig.~\ref{fig:proc2},
by dashed and dotted lines, respectively. 
The helicity-conserving amplitudes dominate the total cross section in the region 
$\sqrt{t_{\rm min}}<\sqrt{-q^2}\lesssim10$~GeV at both $\sqrt{s}=10$ and 100~TeV,
whereas the helicity-flip amplitudes are significant only in the vicinity of $\sqrt{-q^2}\approx m_l$.
This tells that helicity-flip amplitudes contribute only to the collinear splitting,
where the transverse momentum of the final lepton is of the order of $m_l$.
Their contribution at large $y(l)$ is suppressed because the 
helicity-flip collinear emission is suppressed at small $x$,
where the final lepton energy $\sim (1-x)E$ is large.
These observations are consistent with the effective real photon distributions~\cite{Hagiwara:1990gk}:
\begin{align}\label{splittingfunc}
  D^{l_h\to l_{h'}\gamma_\lambda}(x,Q)=
  \left\{ 
  \begin{aligned}   
     & \frac{\alpha}{2\pi}\frac{1}{x}\Big(\ln\frac{Q^2}{t_{\rm min}(x)}-1\Big)  
	 && {\rm for}\ h=h'=\lambda/2, \\   
     & \frac{\alpha}{2\pi}\frac{(1-x)^2}{x}\Big(\ln\frac{Q^2}{t_{\rm min}(x)}-1\Big)  
	 && {\rm for}\ h=h'=-\lambda/2, \\  
     & \frac{\alpha}{2\pi}x
	 && {\rm for}\ h=-h'=\lambda/2, \\  
     & 0
	 && {\rm for}\ h=-h'=-\lambda/2, \\  	 	 
  \end{aligned} 
  \right.
\end{align}
where $h/h'$ and $\lambda$ are the helicities of initial/final leptons and the virtual photon, respectively,
and $t_{\rm min}(x)=m_l^2x^2/(1-x)$.
We can read off the hard scale of $Q\approx10$~GeV from the $\sqrt{-q^2}$ distribution in Fig.~\ref{fig:lmlp_lmvlxtbxh_helicity},
below which the real photon ($q^2=0$) approximation holds.
The scale is essentially the virtuality of the $b$-quark propagator
in the subamplitudes $\gamma\bar l\to\bar b tH \bar\nu$,
which gives rise to collinear emission of $\bar b$. 

The above effective real photon distributions are obtained from integrating the splitting function in the range 
 $t_{\rm min}<|t|<Q^2$~\cite{Hagiwara:1990gk};
\begin{align}
  D^{l_h\to l_{h'}\gamma_\lambda}(x,Q)=
  \int_{t_{\rm min}}^{Q^2} d|t|\, P_{\sigma\sigma'\lambda}(x,t)
  \label{int_t}
\end{align}
with $t_{\rm min}=m_l^2x^2/(1-x)$ and
\begin{subequations}
\begin{align}
	P_{\pm\pm\pm}(x,t)&=\frac{\alpha}{4\pi}\frac{1}{x}\frac{(|t|-t_{\rm min})}{t^2}, \\
	P_{\pm\pm\mp}(x,t)&=\frac{\alpha}{4\pi}\frac{(1-x)^2}{x}\frac{(|t|-t_{\rm min})}{t^2}, \\ 
	P_{\pm\mp\pm}(x,t)&=\frac{\alpha}{4\pi}x\frac{t_{\rm min}}{t^2}, \label{Pc} \\ 
	P_{\pm\mp\mp}(x,t)&=\frac{\alpha}{4\pi}\frac{1}{x(1-x)}\frac{m_l^2}{s^2}\frac{(|t|-t_{\rm min})^2}{t^2}, \label{Pd}
\end{align}
\end{subequations}
where $(\sigma,\sigma')=2(h,h')$ gives the sign of lepton helicities.
In the lepton helicity-flip cases, \eqref{Pc} and \eqref{Pd},
we keep only the leading term in $m_l^2/s$.
The $|t|$ dependence in the numerator shows the square of the overlapping of the spherical harmonic wavefunctions 
(Wigner's $d$-functions), via
\begin{align}
	\frac{1-\cos\theta}{2}=\frac{|t|-t_{\rm min}}{sx},
\end{align}
where $\theta$ is the scattering angle of the final lepton in the laboratory frame, see Eq.~\eqref{eq:4mom}.
When $h=h'+\lambda$ ($\sigma\sigma'\lambda=\pm\mp\pm$) in~\eqref{Pc},
angular momentum along the colliding lepton-beam momentum direction is conserved at $\theta=0$,
and the angular integral~\eqref{int_t} gives finite distribution in the $m_l\to0$ limit. 

\begin{figure}
\includegraphics[width=1\textwidth,clip]{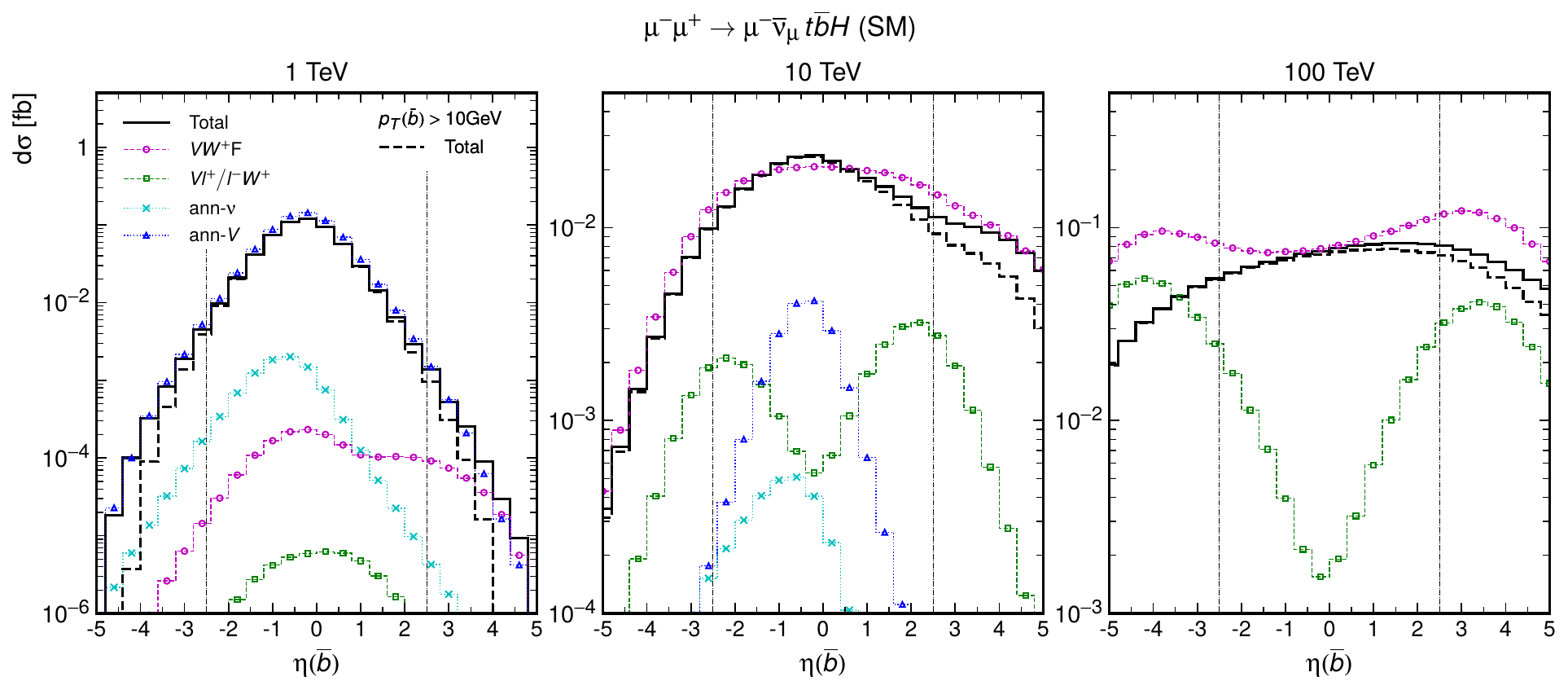}
  \caption{
Pseudo-rapidity distributions of the $\bar b$ quark for the process
$\mu^-\mu^+ \to \mu^-\bar{\nu}_{\mu} t\bar{b}H$ in the SM at three collision energies: 1, 10, and 100~TeV. 
The line and marker styles are the same as those in Fig.~\ref{fig:xsec_proc2}.
The black-dashed lines show the rapidity distributions with a $p_T(\bar b) > 10~\mathrm{GeV}$ cut.
}
   \label{fig:mummup_mumvmxtbxh_yb_with_and_without_ptb_cut_sm}
\end{figure}

In Fig.~\ref{fig:mummup_mumvmxtbxh_yb_with_and_without_ptb_cut_sm},
we show the pseudo-rapidity distribution of the final $\bar b$ quark.
We observe clearly the forward emission of $\bar b$ at $\sqrt{s}=10$ and 100~GeV,
where the $\gamma W^+$F subamplitudes (shown by the magenta dashed lines) dominate the total cross section.
At $\sqrt{s}=1$~TeV, the $\gamma W^+$F contribution is over-shadowed by the dominant ann-$V$ contribution,
$l\bar l\to t \bar t H$ with $\bar t\to \bar bl\bar\nu$, Eq.~\eqref{eq:n3_proc2}.
In these plots, we show by thick black-dashed curves the contribution of events that satisfy the cut 
\begin{align}
  p_T(\bar b)>10~{\rm GeV}.
\label{cut_ptb}  
\end{align}
By comparing the black-solid curve for the total cross section and the black-dashed curve,
we observe that most of the final $\bar b$-jet satisfy the $p_T(\bar b)$ cut.
At $\sqrt{s}=100$~TeV, 
however, significant fraction of the $\bar b$-jets are emitted outside of the pseudo-rapidity cut
\begin{align}
  |\eta(\bar b)|<2.5,
\label{cut_etab}  
\end{align}
which are shown by two thin vertical lines at $\eta(\bar b)=\pm2.5$.
We find that only 47\% of the total events satisfy the $b$-jet tagging conditions
\eqref{cut_ptb} and \eqref{cut_etab}.%
\footnote{
The above results are obtained in the tree-level approximation with $m_b=4.7$~GeV.
Not only the cross section with a tagged $\bar b$-jet but also 
non-tagged inclusive cross section with the effective real $b$-quark PDF inside the lepton
should also be improved by QCD higher order corrections 
which can be significant at small $p_T(\bar b)$ and small $\sqrt{-q^2}$.
}

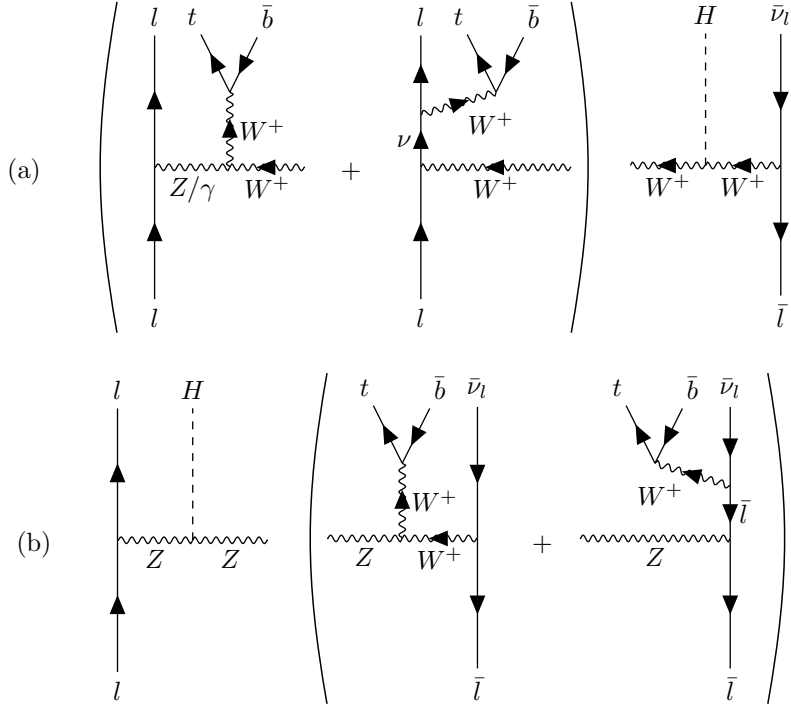
\begin{figure}
	\center
	\setlength{\feynhandblobsize}{5mm} 
\begin{footnotesize}	
\raisebox{21mm}{(a)}\qquad
\begin{tikzpicture}[scale=1]
\begin{feynhand}
    \vertex (i1) at (-1,-2) {$l$};
    \vertex (f1) at (-1, 2) {$l$};
	\vertex (b) at (0,0);
	\vertex (wtb) at (0,1);
    \vertex (w1) at (-1, 0);
    \vertex (w2) at ( 1, 0);
    \vertex (w3) at (-1,-0.75);
    \propag [fermion](i1) to (w1);
	\propag [fermion] (w1) to (f1);
    \propag [bos] (w1) to [edge label'=$Z/\gamma$] (b);
	\propag [chabos] (w2) to [edge label=$W^+$] (b);
	\vertex (f4) at (-0.5,2) {$t$};
    \vertex (f5) at ( 0.5,2) {$\bar b$};	
	\propag [chabos] (b) to [edge label'=$W^+$] (wtb);
	\propag [fermion] (wtb) to (f4);
	\propag [fermion] (f5) to (wtb);
	\vertex (l1) at (-2,-2.2);
	\vertex (l2) at (-2,2.2);
	\vertex (l3) at (-1.5,-2.2);
	\vertex (l4) at (-1.5,2.2);
	\draw [line width=0.2mm] (l3) to [out=100, in=260] (l4);
\end{feynhand}
\end{tikzpicture}
\quad\raisebox{21mm}{+}\quad
\begin{tikzpicture}[scale=1]
\begin{feynhand}
	\vertex (b) at (-1,0);
	\vertex (nl) at (-1,0.7);
	\vertex (wtb) at (0,1);
    \vertex (i1) at (-1,-2) {$l$};
    \vertex (f1) at (-1, 2) {$l$};
    \vertex (w1) at (-1, 0);
    \vertex (w2) at ( 1, 0);
    \vertex (w3) at (-1,-0.75);
    \propag [fermion](i1) to (b);
    \propag [fermion] (b) to [edge label=$\nu$] (nl) ;
	\propag [fermion] (nl) to (f1);
    \propag [chabos] (w2) to [edge label=$W^+$] (b);
	\vertex (f4) at (-0.5,2) {$t$};
    \vertex (f5) at ( 0.5,2) {$\bar b$};	
	\propag [chabos] (nl) to [edge label'=$W^+$] (wtb);
	\propag [fermion] (wtb) to (f4);
	\propag [fermion] (f5) to (wtb);
	\vertex (l1) at (1.5,-2.2);
	\vertex (l2) at (1.5,2.2);
	\vertex (l3) at (1.,-2.2);
	\vertex (l4) at (1.,2.2);
	\draw [line width=0.2mm] (l3) to [out=80, in=280] (l4);	
\end{feynhand}
\end{tikzpicture}
\quad
\begin{tikzpicture}[scale=1]
\begin{feynhand}
    \vertex (i2) at ( 1,-2) {$\bar{l}$};
    \vertex (f2) at ( 1, 2) {$\bar\nu_l$};
	\vertex (b) at (0,0) ;
    \vertex (w1) at (-1, 0);
    \vertex (w2) at ( 1, 0);
    \vertex (w3) at (-1,-0.75);
    \propag [fermion] (f2) to (w2);
	\propag [fermion] (w2) to (i2);
    \propag [chabos] (b) to [edge label=$W^+$] (w1);
	\propag [chabos] (w2) to [edge label=$W^+$] (b);
	\vertex (f3) at (0,2) {$H$};
	\propag [scalar] (b) to (f3);
\end{feynhand}
\end{tikzpicture}\\[4mm]

\raisebox{21mm}{(b)}\qquad
\begin{tikzpicture}[scale=1]
\begin{feynhand}
    \vertex (i1) at (-1,-2) {$l$};
    \vertex (f1) at (-1, 2) {$l$};
	\vertex (b) at (0,0) ;
    \vertex (w1) at (-1, 0);
    \vertex (w2) at ( 1, 0);
    \vertex (w3) at (-1,-0.75);
    \propag [fermion](i1) to (w1);
	\propag [fermion] (w1) to (f1);
    \propag [bos] (w1) to [edge label'=$Z$] (b);
	\propag [bos] (w2) to [edge label=$Z$] (b);
	\vertex (f3) at (0,2) {$H$};
	\propag [scalar] (b) to (f3);
\end{feynhand}
\end{tikzpicture}
\quad
\begin{tikzpicture}[scale=1]
\begin{feynhand}
    \vertex (i2) at ( 1,-2) {$\bar{l}$};
    \vertex (f2) at ( 1, 2) {$\bar\nu_l$};
	\vertex (b) at (0,0);
	\vertex (wtb) at (0,1);
    \vertex (w1) at (-1, 0);
    \vertex (w2) at ( 1, 0);
    \vertex (w3) at (-1,-0.75);
    \propag [fermion] (f2) to (w2);
	\propag [fermion] (w2) to (i2);
    \propag [bos] (w1) to [edge label'=$Z$] (b);
	\propag [chabos] (w2) to [edge label=$W^+$] (b);
	\vertex (f4) at (-0.5,2) {$t$};
    \vertex (f5) at ( 0.5,2) {$\bar b$};	
	\propag [chabos] (b) to [edge label'=$W^+$] (wtb);
	\propag [fermion] (wtb) to (f4);
	\propag [fermion] (f5) to (wtb);
	\vertex (l1) at (-1.5,-2.2);
	\vertex (l2) at (-1.5,2.2);
	\vertex (l3) at (-1.,-2.2);
	\vertex (l4) at (-1.,2.2);
	\draw [line width=0.2mm] (l3) to [out=100, in=260] (l4);
\end{feynhand}
\end{tikzpicture}
\quad\raisebox{21mm}{+}\quad
\begin{tikzpicture}[scale=1]
\begin{feynhand}
	\vertex (b) at (1,0);
	\vertex (nl) at (1,0.7);
	\vertex (wtb) at (0,1);
    \vertex (i2) at ( 1,-2) {$\bar{l}$};
    \vertex (f2) at ( 1, 2) {$\bar\nu_l$};
    \vertex (w1) at (-1, 0);
    \vertex (w2) at ( 1, 0);
    \vertex (w3) at (-1,-0.75);
    \propag [fermion] (f2) to (nl);
	\propag [fermion] (b) to (i2);
    \propag [fermion] (nl) to [edge label=$\bar l$] (b);
    \propag [bos] (w1) to [edge label'=$Z$] (b);
	\vertex (f4) at (-0.5,2) {$t$};
    \vertex (f5) at ( 0.5,2) {$\bar b$};	
	\propag [chabos] (nl) to [edge label=$W^+$] (wtb);
	\propag [fermion] (wtb) to (f4);
	\propag [fermion] (f5) to (wtb);
	\vertex (l1) at (2,-2.2);
	\vertex (l2) at (2,2.2);
	\vertex (l3) at (1.5,-2.2);
	\vertex (l4) at (1.5,2.2);
	\draw [line width=0.2mm] (l3) to [out=80, in=280] (l4);	
\end{feynhand}
\end{tikzpicture}
	
\end{footnotesize}
\caption{Feynman diagrams contributing to the destructive interference for 
$l\bar{l}\to l\bar{\nu}_l t \bar{b} H$.
 }
\label{fig:diagrams_int}
\end{figure}

In the $\eta(\bar b)$ distribution of Fig.~\ref{fig:mummup_mumvmxtbxh_yb_with_and_without_ptb_cut_sm}, 
especially at $\sqrt{s}=100$~TeV,
we observe significant destructive interference between the 
$VW^+$F and $Vl^+/l^-W^+$ amplitudes, 
which are shown by magenta and green dashed curves, respectively.
We identify the origin of the interference as patterns represented by 
Figs.~\ref{fig:diagrams_int} (a) and (b).
In case (a), the initial $l$ collides with the virtual $W^+$ emitted from $\bar l$,
to produce $l$ and $W^{+*}$ via $t$-channel $Z/\gamma$ exchange or $s$-channel $\nu$ exchange,
which interfere destructively. 
In case (b), the virtual $Z$ boson emitted from $l$ collides with $\bar l$,
to produce $\bar\nu$ and $W^{+*}$ via $t$-channel $W$ exchange or $s$-channel $\bar l$ exchange,
which also interfere destructively.  

\begin{figure}
  \includegraphics[width=0.975\textwidth,clip]{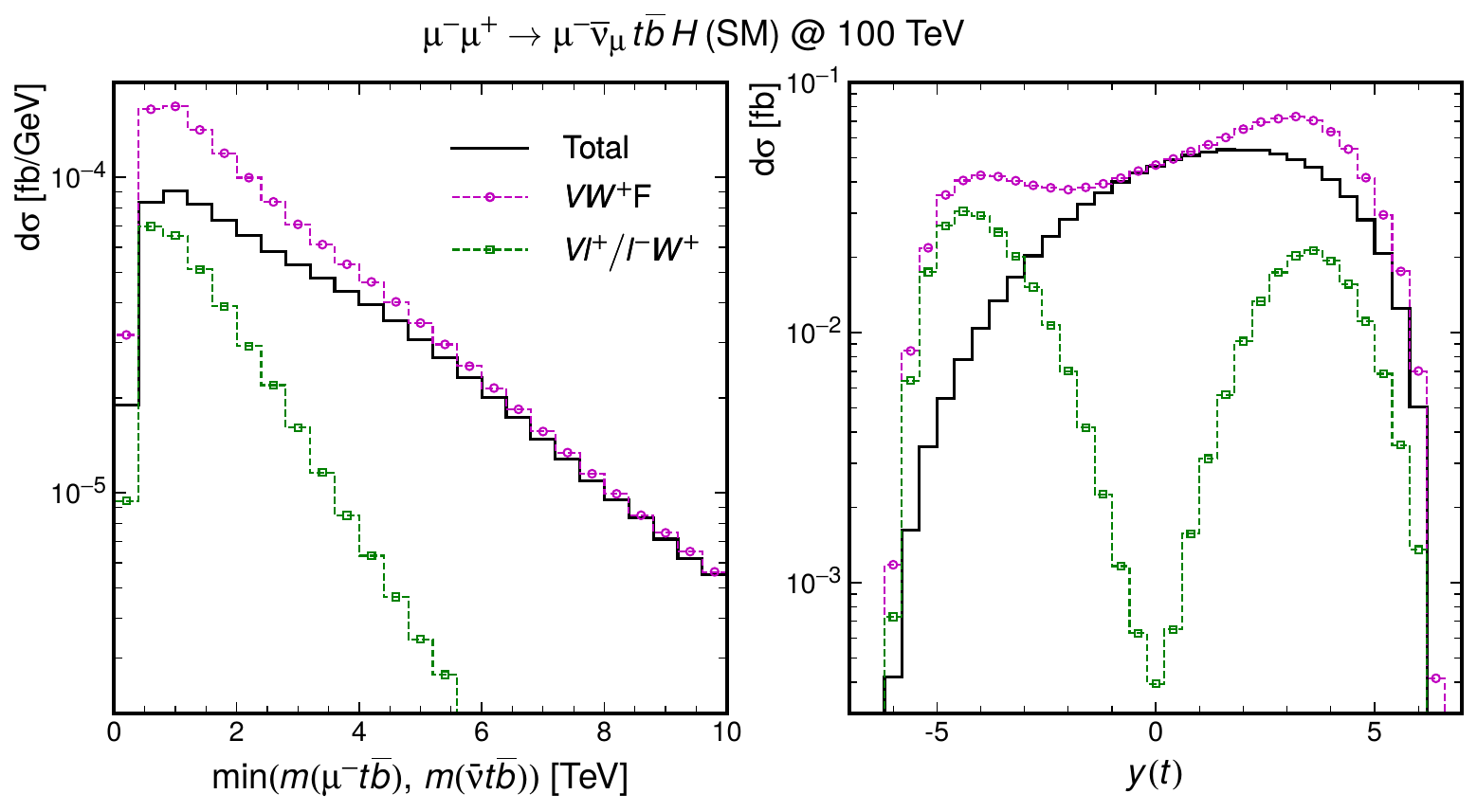}
  \caption{
    Distributions of the minimum of the invariant masses of the $\mu^- t\bar{b}$ and $\bar{\nu}_\mu t\bar{b}$ systems (left) and of the rapidity of the top quark (right) for the process $\mu^-\mu^+ \to \mu^- \bar{\nu}_\mu t\bar{b}H$ in the SM at a collision energy of 100~TeV.
    The line and marker styles are the same as those in Fig.~\ref{fig:xsec_proc2}.
	Note that  
	the ann-$l$ and ann-$V$ contributions are too small to appear in the range. 
	}
    \label{fig:mcps_mummup_mumvmxtbxh_min_mtbx_yt}
\end{figure}

\begin{figure}
  \includegraphics[width=0.975\textwidth,clip]{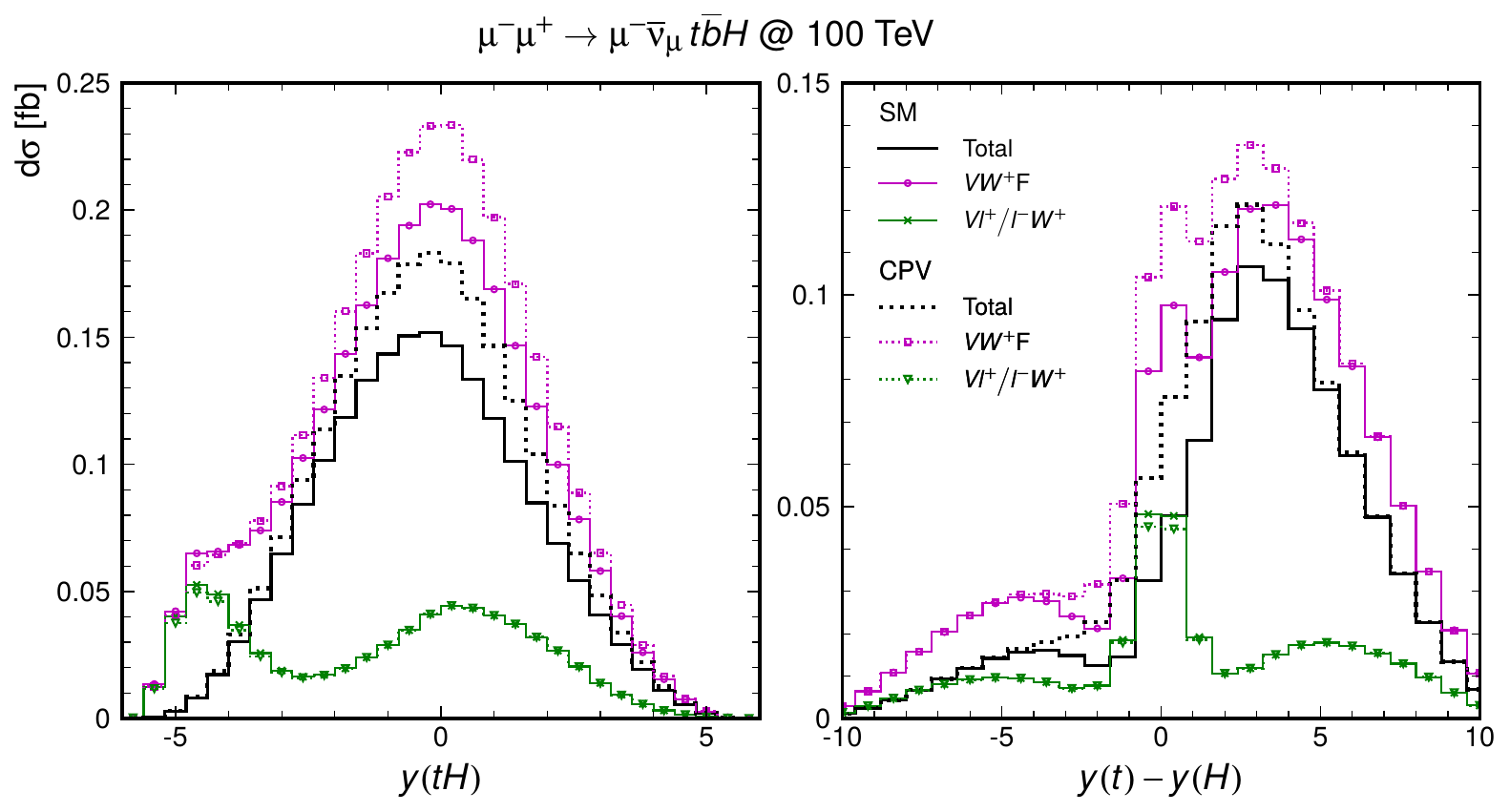}
  \caption{
    Distributions of the rapidity of the $tH$ system (left) and the rapidity difference between the top quark and the Higgs boson (right) for the process $\mu^-\mu^+ \to \mu^- \bar{\nu}_\mu t\bar{b}H$ in the SM and in the CPV SMEFT with $(g,\xi)=(g_{\rm SM},0.1\pi)$ at a collision energy of 100~TeV.
    The solid lines denote the contributions from the full amplitudes, and its subcategories, $VW^+$F and $Vl^+/l^-W^+$, in the SM.
    The dotted lines show the corresponding contributions in the CPV SMEFT.
  }
    \label{fig:mcps_mummup_mumvmxtbxh_yth_dyth_sm_cpv_lin}
\end{figure}

\begin{figure}
	\center
\includegraphics[width=1\textwidth,clip]{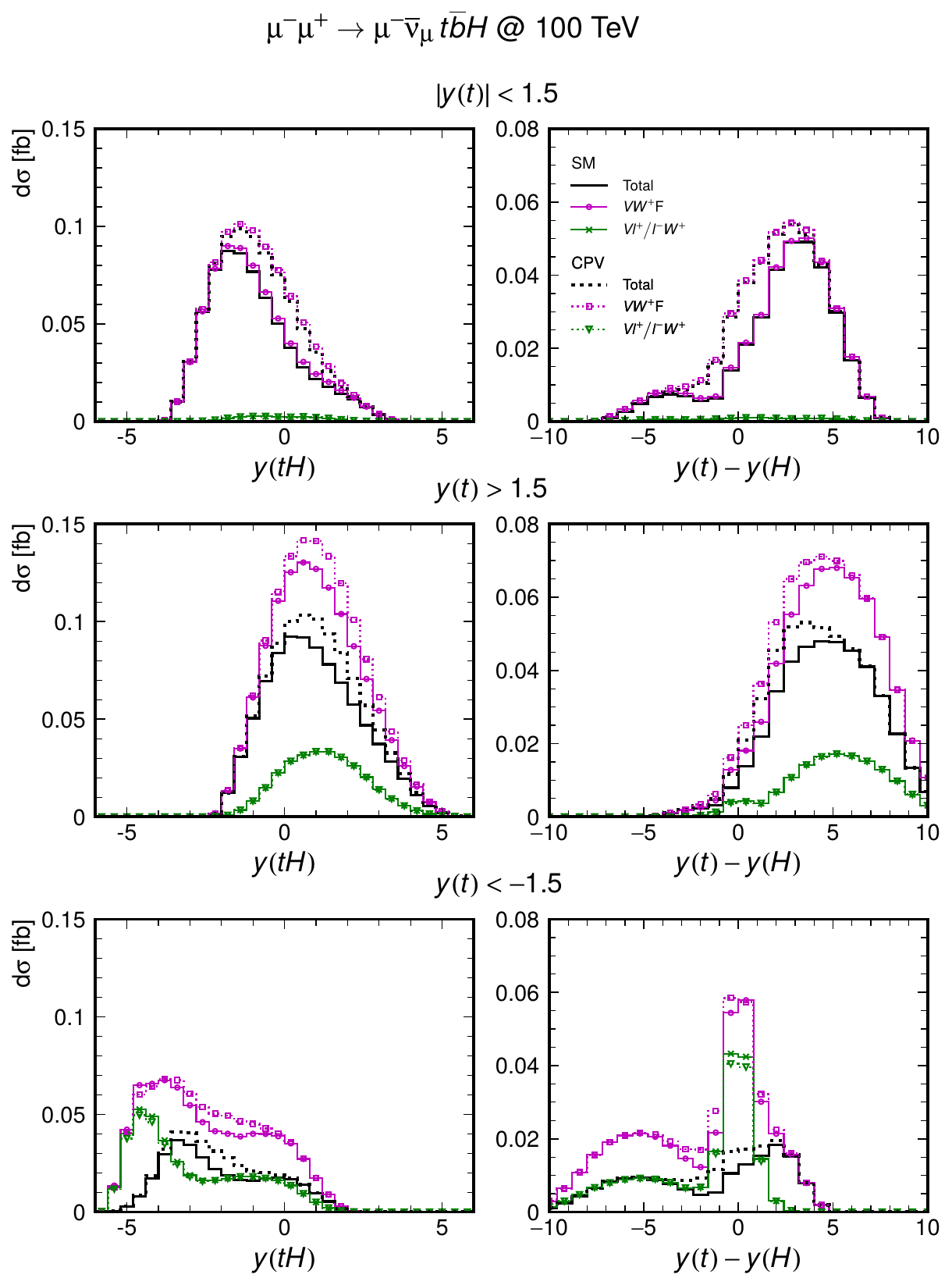}	
  \caption{
Same as Fig.~\ref{fig:mcps_mummup_mumvmxtbxh_yth_dyth_sm_cpv_lin}, but with the
 $|y(t)|< 1.5$ (top), $y(t)>1.5$  (middle), and $y(t)<-1.5$ (bottom) cut.
  }
    \label{fig:mcps_mummup_mumvmxtbxh_yth_dyth_sm_cpv_lin_ytcut}
\end{figure}

The destructive interference is significant only when both amplitudes are large.
Therefore, either the $s$-channel $\nu$ or $l$ virtuality,
i.e. $m(lt\bar b)$ or $m(t\bar b\bar\nu_l)$, respectively,
should be small in Fig.~\ref{fig:diagrams_int}. 
This is confirmed in Fig.~\ref{fig:mcps_mummup_mumvmxtbxh_min_mtbx_yt}(left),
where we show the differential cross section as a function of
\begin{align}
  \min(m(\mu^-t\bar b),m(\bar\nu t\bar b)),
\end{align}
where destructive interference is significant only when one of the two invariant masses 
is below 5~TeV at $\sqrt{s}=100$~TeV.
Since the invariant masses including $\bar\nu$ and $b$-jets are difficult to measure,
we show in the right-hand-side of Fig.~\ref{fig:mcps_mummup_mumvmxtbxh_min_mtbx_yt}
the interference pattern in the $y(t)$ distribution.
We can clearly observe the destructive interference pattern Fig.~\ref{fig:diagrams_int}(a)
in the positive rapidity region $y(t)\gtrsim2$ and 
the pattern Fig.~\ref{fig:diagrams_int}(b)
in the negative rapidity region $y(t)\lesssim-2$. 

In Fig.~\ref{fig:mcps_mummup_mumvmxtbxh_yth_dyth_sm_cpv_lin}, we show the CPV-phase dependence of the rapidity distributions of $y(tH)$ and $y(t)-y(H)$.
The black-solid curves show the SM predictions, while the black-dotted ones show the predictions of the CPV model with $(g,\xi)=(g_{\rm SM},0.1\pi)$.
For both distributions, enhancement of the cross section is observed in the central region for the CPV case.
However, the destructive interference between the $VW^+$F (magenta curves) and the $Vl^+/l^-W^+$ (green curves) contributions
makes it difficult to interpret the results.

We show in Fig.~\ref{fig:mcps_mummup_mumvmxtbxh_yth_dyth_sm_cpv_lin_ytcut}
the rapidity distributions separately for the three kinematical region
\begin{align} 
  |y(t)|< 1.5,\quad y(t)>1.5,\quad y(t)<-1.5.
\end{align}
From the interference pattern shown in Fig.~\ref{fig:mcps_mummup_mumvmxtbxh_min_mtbx_yt}(right),
the region $|y(t)|< 1.5$ is dominated by the $VW^+$F subamplitudes, 
and the corresponding results shown in the top of Fig.~\ref{fig:mcps_mummup_mumvmxtbxh_yth_dyth_sm_cpv_lin_ytcut}
clearly shows that the $\xi$ dependence of the $VW^+$F amplitudes gives the total $\xi$ dependence.
In case of $y(t)>1.5$, the destructive interference  between the $VW^+$F and the $l^-W^+$ amplitudes is significant.
The middle pair of Fig.~\ref{fig:mcps_mummup_mumvmxtbxh_yth_dyth_sm_cpv_lin_ytcut} tells us that
the $\xi$ dependence is largely coming from the $VW^+$F amplitudes.
Finally, for $y(t)<-1.5$, $\xi$ dependence appears both in the $VW^+$F and $Vl^+$ amplitudes.
By examining all the channels contributing to the $VW^+$F and $Vl^+$ categories,
we find that the $\gamma W^+$F and the $\gamma l^+/l^-W^+$ amplitudes
with a dim5 $tbHW$ vertex contribute significantly to the $\xi$ dependence of the 
magenta and green curves, respectively.

\section{Numerical results for $l\bar{l} \to l\bar{l} t\bar{t} H$}
\label{sec:lltth}

\begin{figure}
  \includegraphics[width=0.1425\textwidth,clip]{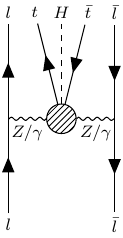} 
  \quad
  \includegraphics[width=0.15\textwidth,clip]{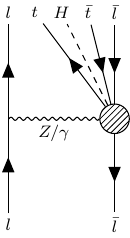} 
  \includegraphics[width=0.15\textwidth,clip]{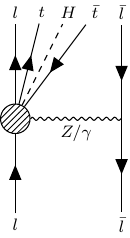} 
  \quad
  \includegraphics[width=0.15\textwidth,clip]{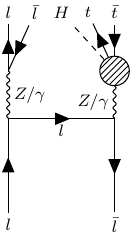} 
  \includegraphics[width=0.15\textwidth,clip]{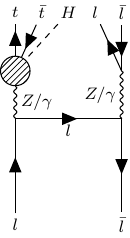} 
  \quad
  \includegraphics[width=0.1425\textwidth,clip]{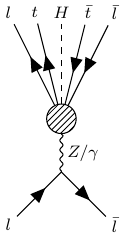} 
  
\hspace*{0.0cm}(A)\hspace*{3.6cm}(B)\hspace*{5cm}(C)\hspace*{3.7cm}(D)
\caption{
  The Feynman diagrams for the process $l\bar{l}\to l\bar{l} t \bar{t} H$ are classified into four categories, as define in Table~\ref{tab:ex_categories}: (A) $VV$F, (B) $Vl^+/l^-V$, (C) ann-$l$, and (D) ann-$V$, with $V=Z/\gamma$.
In (C), there are additional diagrams where the Higgs boson is emitted from the $Z$ boson that gives $l\bar{l}$.
 }
\label{fig:proc3}
\end{figure}

This section demonstrates the SDE MCPS integration of the process
\begin{align}
    l\bar{l} \rightarrow l\bar{l} t\bar{t} H,
    \label{eq:ex3-1}
\end{align}
whose Feynman diagrams are illustrated in Fig.~\ref{fig:proc3}.
They are classified into four categories: 
(A) $VV$F, (B) $Vl^+/l^-V$, (C) ann-$l$, and (D) ann-$V$, with $V=Z/\gamma$, as defined in Table~\ref{tab:ex_categories}. 
The number of the Feynman diagrams for each category is summarized in Table~\ref{tab:proc3}. 

The phase-space integration for this process \eqref{eq:ex3-1} are more challenging than for the process studied in the previous section, $l\bar{l}\to l\bar{\nu}t\bar{b}H$, especially at very high collision energies without any kinematical cuts. 
This difficulty arises because the lepton-mass singularities from the $t$-channel photon-exchange diagrams, as seen in groups (A) and (B) of Fig.~\ref{fig:proc3}, originate from both charged leptons in the final state. 
Thanks to the phase-space library we developed and the corresponding improvements to the original \HELAS\ code~\cite{Hagiwara:1990dw,Murayama:1992gi}, we are now able to evaluate the cross section for this process with high precision even up to collision energies of 100~TeV
without applying any kinematical cuts.

\begin{table}
\caption{Number of Feynman diagrams in each group classified in Fig.~\ref{fig:proc3}, where the SMEFT model in the FD gauge is considered. 
The numbers in parentheses indicate the diagram IDs generated by \MG.}
	\label{tab:proc3}
\begin{tabular}{c|lrllrlrllrl}
\hline
 & \multicolumn{11}{c}{{\bf No. of Feynman diagrams} (diagram ID)} \\
 \multicolumn{1}{c|}{category} &\hspace*{0.1cm}& \multicolumn{2}{c}{total} &\hspace*{0.cm} & dim4 &\hspace*{0.1cm}& \multicolumn{2}{c}{dim5} &\hspace*{0.1cm}& \multicolumn{2}{c}{dim6} \\ \hline
 (A) && {\bf 52} & (81--132)  && {\bf 38} && {\bf 13} & (85, 88, 93, 94, 98, 101, 104--107, 118, 128, 132) && {\bf 1} & (95)  \\
 (B) && {\bf 56} & (133--188) && {\bf 48} &&  {\bf 8} & (157--160, 185--188) && {\bf 0} \\
 (C) && {\bf 28}& (189--216) && {\bf 24} &&  {\bf 4} & (213--216) && {\bf 0} & \\ 
 (D) && {\bf 80}& (1--80) && {\bf 62} && {\bf 17} & (5, 8, 13, 14, 18, 21, 24--27, 38, 48, 52, 77--80) && {\bf 1} & (15) \\ \hline
\end{tabular}
\end{table}

There is an additional singularity in this process, requiring modifications to \HELAS\ subroutines. 
That is a collinear lepton-pair splitting from a nearly on-shell virtual photon, $\gamma^*\to l\bar l$, which appears in diagram groups (C) and (D) in Fig.~\ref{fig:proc3}.
The modifications of the \HELAS~\ code are described explicitly in Appendix~\ref{app:helas}.

\begin{figure}
  \includegraphics[width=0.495\textwidth,clip]{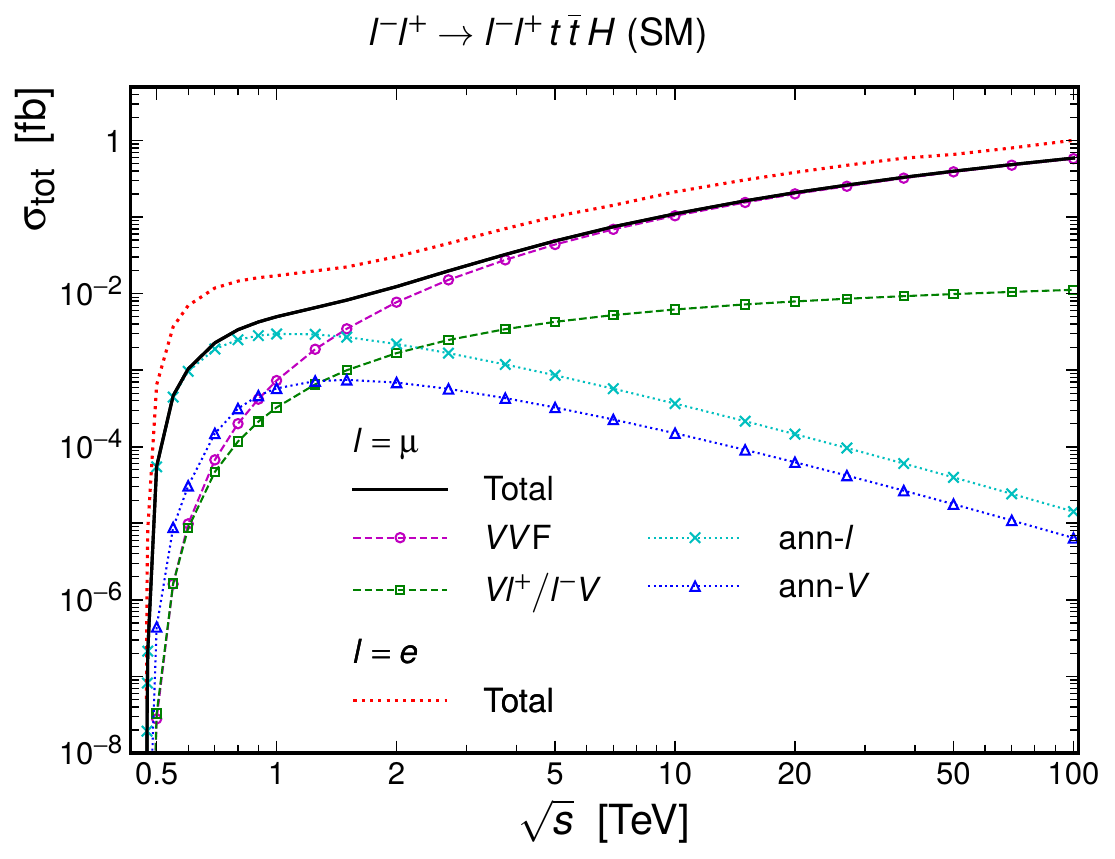}
  \includegraphics[width=0.495\textwidth,clip]{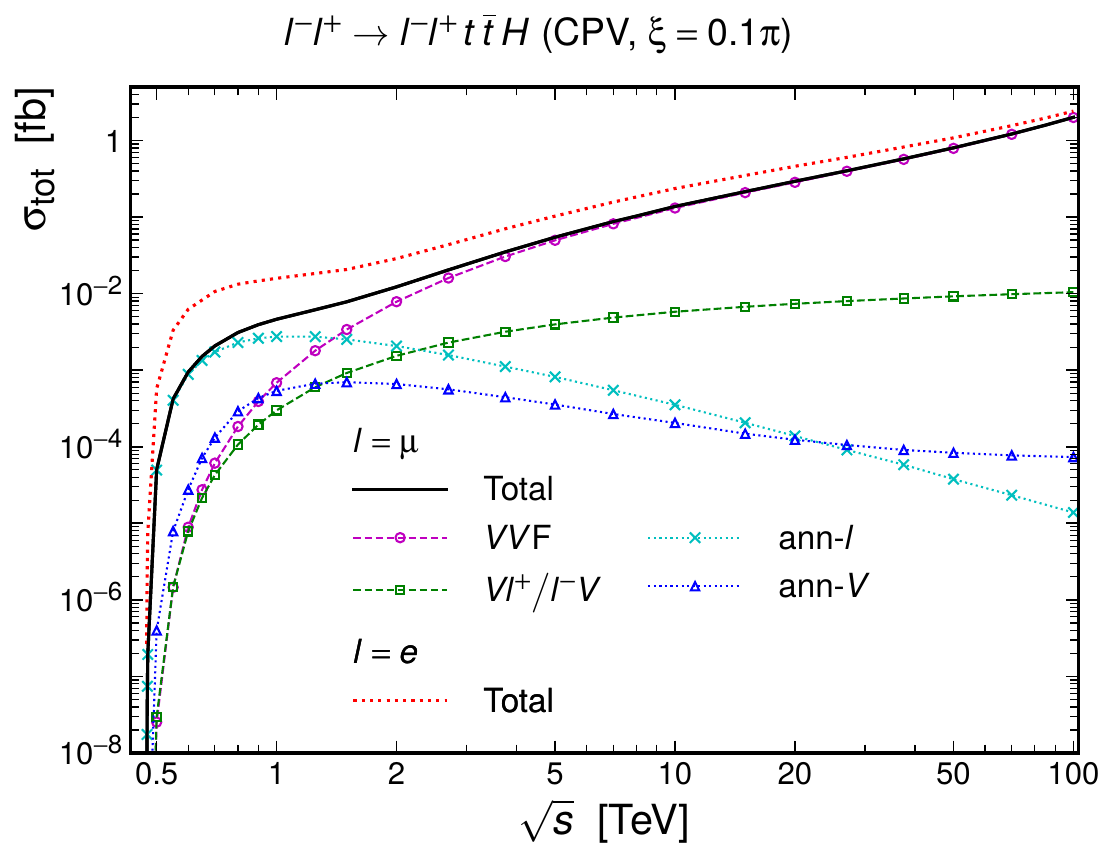}
\caption{Collision-energy dependence of the total cross section for the process $\mu^-\mu^+\rightarrow \mu^-\mu^+t\bar{t}H$ in the SM (left) and in the CPV SMEFT with $(g,\xi)=(g_{\rm SM},0.1\pi)$ (right).
The black solid line represents the observable total cross section, while the dashed and dotted lines show the contributions from each diagram category, as defined in Table~\ref{tab:ex_categories}, calculated in the FD gauge.
The red dotted line represents the observable total cross section for $e^-e^+\rightarrow e^-e^+t\bar{t}H$.}
\label{fig:xsec_proc3}
\end{figure}

Figure~\ref{fig:xsec_proc3} shows the total cross section of $\mu^-\mu^+\to\mu^-\mu^+t\bar{t}H$ as a function of the collision energy in the SM (left) and in the CPV SMEFT with $(g,\xi)=(g_{\rm SM},0.1\pi)$ (right).
The observable total cross section is shown by the black solid line, where all the Feynman amplitudes, 172 (216) in the SM (SMEFT) as listed in Table~\ref{tab:ngraphs}, are summed before squaring. 
Also shown are the partial contributions from subsets of diagrams classified in Fig.~\ref{fig:proc3}, in the same manner as in the previous processes: Fig.~\ref{fig:proc1} for $l\bar{l}\to\nu_l\bar{\nu}_lt\bar{t}H$ and Fig.~\ref{fig:proc2} for $l\bar{l}\to l\bar{\nu}_lt\bar{b}H$.  

\begin{figure}
    \includegraphics[width=0.95\textwidth,clip]{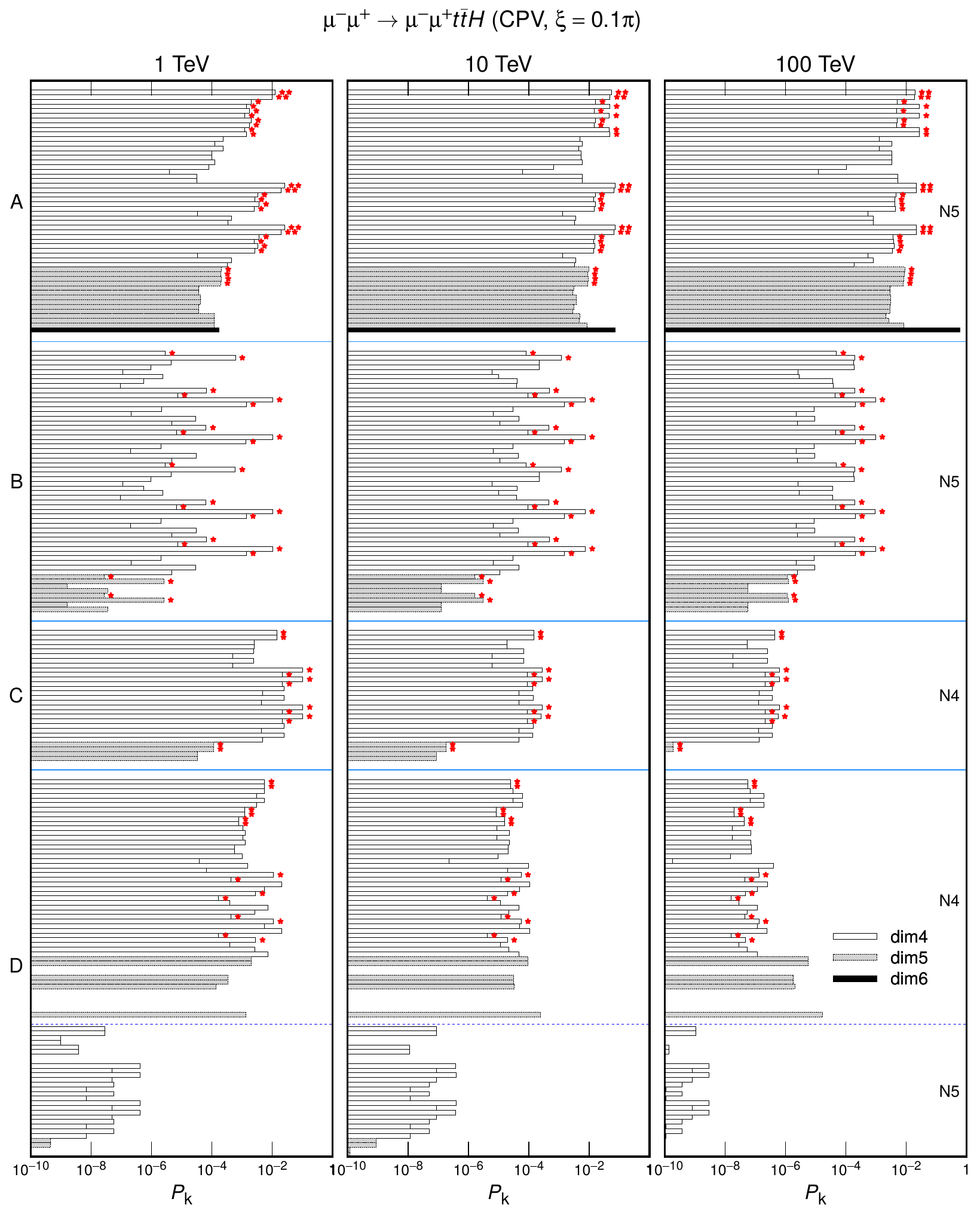}
\caption{
Fraction of the contribution of each FD amplitude channel, defined as $P_k = \sigma_k/\sigma_\mathrm{tot}$ with $\sigma_k$ in Eq.~\eqref{eq:sigma_mcps}, for the process $\mu^-\mu^+\rightarrow \mu^-\mu^+ t\bar{t}H$ in the CPV SMEFT with $(g,\xi)=(g_{\rm SM},0.1\pi)$ at three collision energies: 1, 10, and 100~TeV.
The diagrams are categorized as (A)--(D) in Fig.~\ref{fig:proc3}.
Open, grey, and filled histograms correspond to diagrams with dimension-4, 5, and 6 vertices, respectively, as listed in Table~\ref{tab:proc3}.
In each subprocess category, the contributions are also grouped according to the phase-space categories N4 and N5 in Eqs.~\eqref{eq:n4_proc3} and \eqref{eq:n5_proc3}, respectively.
Red stars denote the amplitude including the $t$-channel photon propagator in (A) and (B)
or the $s$-channel $\gamma^*\to l^-l^+$ splitting in (C) and (D).
}
\label{fig:mummup_mummupttxh_pk}
\end{figure}

A clear transition of the dominant contribution can be observed at around $\sqrt{s}\sim1.5$~TeV. The ann-$l$ subamplitude contribution dominates in the low-energy region, whereas the $VV$F one becomes dominant at high energies.
Different from the total cross section for $l\bar{l}\to l\bar{\nu}t\bar{b}H$,
shown in Fig.~\ref{fig:xsec_proc2}, 
we find the enhancement of the cross section for $l=e$ (red dotted line) 
over the $l=\mu$ case (black solid line)
not only in the high-energy region but also in the low-energy region.
This is because 
the lepton-mass singularities appear not only in the category (A) $VV$F 
from the $t$-channel photon exchange but also in (C) ann-$l$
from the $s$-channel photon splitting to the lepton pair,
$\gamma^*\to l\bar l$.
We also confirm the dominance of the double/single $t$-channel photon-exchange amplitudes
in (A) and (B) as well as of the amplitudes with $\gamma^*\to l\bar l$ in (C) and (D)
in the $P_k$ distribution in Fig.~\ref{fig:mummup_mummupttxh_pk}, where the corresponding channels are marked by red stars.

For the CPV case for $(g,\xi)=(g_{\rm SM},0.1\pi)$, shown in Fig.~\ref{fig:xsec_proc3}(right), the overall behavior is similar to that in the SM, 
and one can hardly observe difference at $\sqrt{s}\lesssim 1$~TeV. 
On the other hand, for $\sqrt{s}\gtrsim 3$~TeV, the enhancement of the $VV$F and ann-$V$ contributions over the corresponding SM curves is clearly seen. 
The origin of this growth at high energies will be identified below.

Figure~\ref{fig:mummup_mummupttxh_pk} shows the fractional contribution of each FD diagram channel, defined in Eq.~\eqref{eq:pk}.
This process contains two diagrams involving a dim6 vertex, as shown in Table~\ref{tab:proc3}; 
one (diagram ID=95) in the $VV$F category (A) and the other (ID=15) in the ann-$V$ category (D).
Both of them are due to the $ttHZZ$ vertex, 
where the 5th (Goldstone boson) component of the Z boson in the FD gauge appears
in the dim6 SMEFT operator in Eq.~\eqref{eq:SMEFTLag}.
At very high collision energies, the $VV$F diagram with the dim6 vertex dominates the total cross section in the CPV SMEFT.
This can be observed in the total cross section shown in Fig.~\ref{fig:xsec_proc3},
where the $VV$F subamplitude contribution (magenta-dashed) grows significantly
at high energies as compared to the SM case.
Also, the difference between the $l=e$ and $\mu$ curves,
the red-dotted and the black-solid curves, respectively, becomes small
at higher energies.
This is because the dim6 vertex appears only in the $ZZ$ fusion amplitudes,
which has no sensitivity to the charged lepton mass.
We note that, similar to the $l\bar l\to\nu_l\bar\nu_lt\bar tH$ process as 
discussed in Sec.~\ref{sec:vvtth},
the amplitude with a dim6 vertex (ID=15) and 
seven (ID=13, 14, 25--27, 38, 48) of the 17 amplitudes with a dim5 vertex in (D) ann-$V$, listed in Table~\ref{tab:proc3}, are zero at all energies.

The diagrams can also be reorganized by the number of the quasi on-shell resonances as 
\begin{align}
  &{\rm N4:}\  l\bar{l} \to  t\bar{t}H Z\ (Z\to l\bar l) ,\quad
               l\bar{l} \to  t\bar{t}H \gamma^*\ (\gamma^*\to l\bar l) ,\label{eq:n4_proc3}\\	
  &{\rm N5:}\  l\bar{l} \to  t\bar{t}H l\bar{l} .\label{eq:n5_proc3}
\end{align}
The phase space for the 
N4 diagram channels is given by the four-body phase space ($t\bar tHZ$ productions) times the two-body phase space (Z-boson decays into a charged-lepton pair).
In addition, we also identify the production of the nearly on-shell virtual photon ($l\bar{l} \to  t\bar{t}H \gamma^*$) 
and its splitting into a lepton pair ($\gamma^*\to l\bar l$), marked as red stars in Fig.~\ref{fig:proc3}, as the N4 category
since this contribution is similar or larger to the on-shell $Z$ boson production when any kinematical cuts for the leptons are not applied.
The phase space for the N5 channels does not have such factorization property.
All the diagrams in the categories (A) and (B) in Fig.~\ref{fig:proc3} are in the N5 category,
while all (C) diagrams are in N4. 
The diagrams in (D) fall into the two phase-space categories, i.e. N4 and N5.
As clearly seen in Fig.~\ref{fig:mummup_mummupttxh_pk}, the N5 contributions in ann-$V$ channels (D) are negligible, comparing with all the other contributions.

Also notable point in Fig.~\ref{fig:xsec_proc3} is the increase of the ann-$V$ subamplitude contribution (blue-dotted curve) at high energies for the CPV case.
This is due to the dim5 $ttHH$ and $ttHZ$ vertices, which contributes to the N4 process~\eqref{eq:n4_proc3}.
Since the dim5 $ttHZ$ vertex has only the 5th (Goldstone boson) component of the Z boson
in the FD gauge, 
final lepton pair should have the longitudinally polarized $Z\to l\bar l$ decay distributions,
when $|ge^{i\xi}-g_{\rm SM}|$ is significant.

\begin{figure}
    \includegraphics[width=1\textwidth,clip]{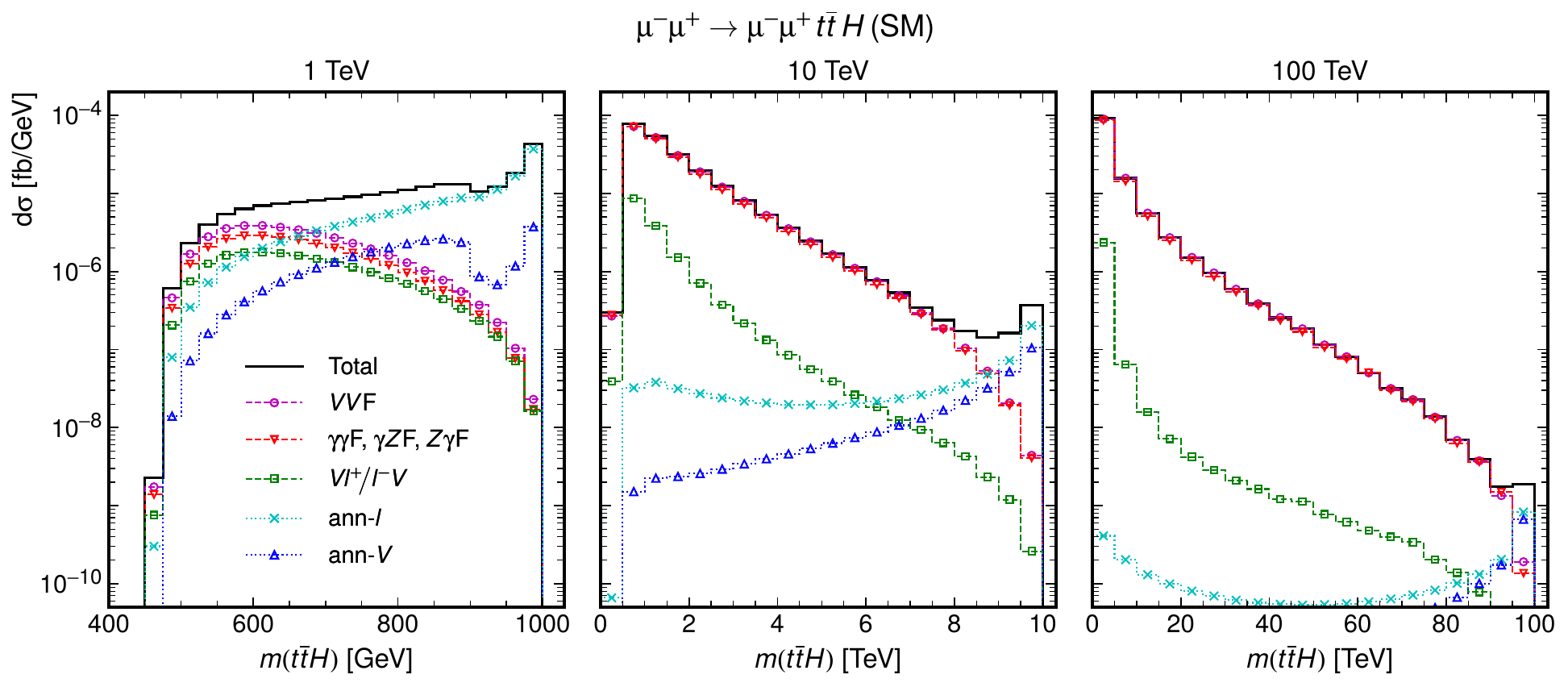}
    \includegraphics[width=1\textwidth,clip]{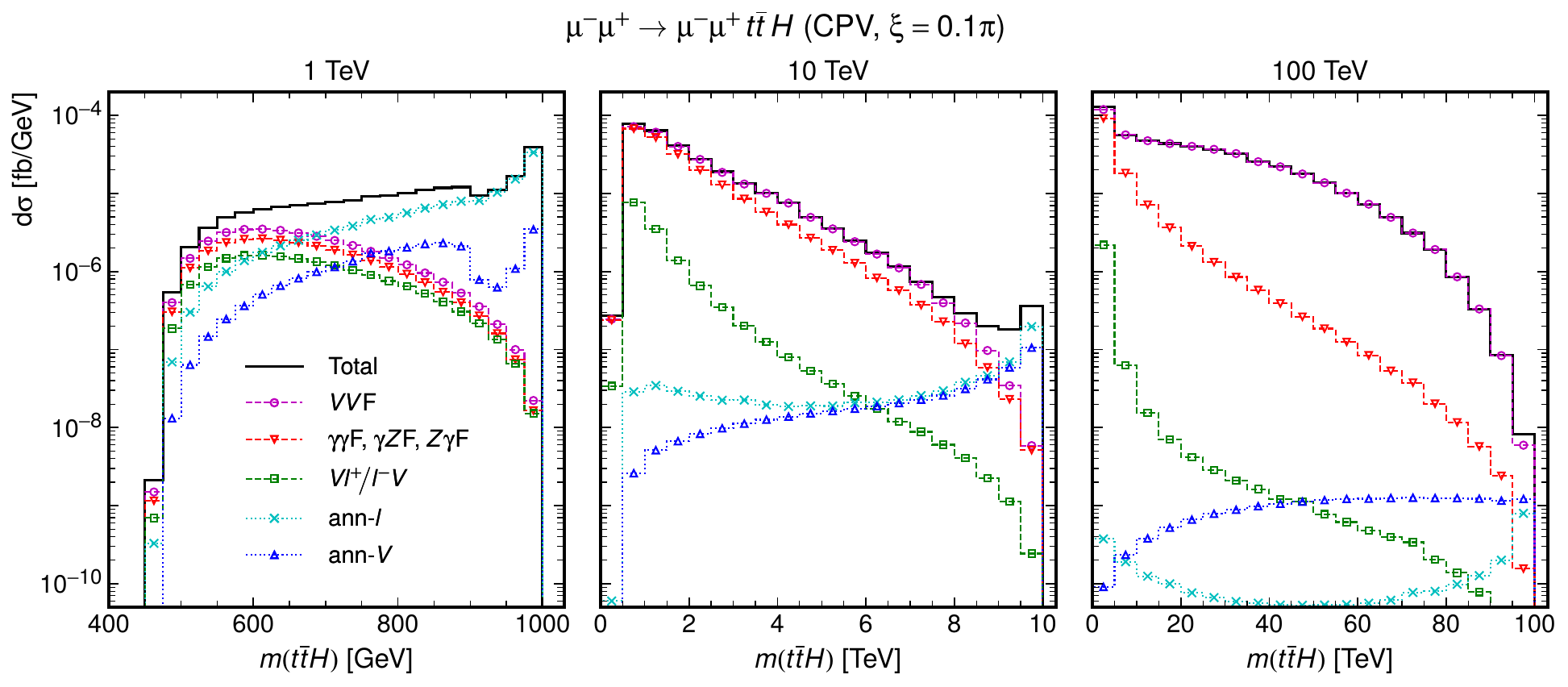}
    \caption{Distributions of the invariant mass of the $t\bar{t}H$ system for the process $\mu^-\mu^+\rightarrow \mu^-\mu^+ t\bar{t}H$ in the SM (top) and in the CPV SMEFT with $(g,\xi)=(g_{\rm SM},0.1\pi)$ (bottom), shown for three collision energies: 1, 10, and 100~TeV.
      The line and marker styles are the same as in Fig.~\ref{fig:xsec_proc3}.
The contributions from the $\gamma\gamma$F, $\gamma Z$F and $Z\gamma$F subamplitudes are shown additionally by red-dashed lines.  
}
  \label{fig:xsec_proc3_mtth}
\end{figure}

Shown in Fig.~\ref{fig:xsec_proc3_mtth} are the invariant mass distribution $d\sigma/dm(t\bar tH)$ for the $\mu^-\mu^+\rightarrow \mu^-\mu^+ t\bar{t}H$ 
at three energies, $\sqrt{s}=1$, 10, 100~TeV, for the SM (top) and 
for the CPV top-Yukawa coupling (bottom), $(g,\xi)=(g_{\rm SM},0.1\pi)$.
The dominance of the $VV$F subamplitude contributions, shown by magenta-dashed curves,
is clear at $m(t\bar tH)\lesssim700$~GeV at $\sqrt{s}=1$~TeV,
whereas at almost the whole range except the highest $m(t\bar tH)$ bins at $\sqrt{s}=10$ and 100~TeV.
Within the $VV$F amplitude channel, we show by red-dashed curves
the contribution of the $\gamma\gamma$F, $\gamma Z$F and $Z\gamma$F subamplitudes. 
The difference between the magenta and red dashed curves hence shows
the contribution of the $ZZ$F amplitudes.
In the SM case, shown in the top of Fig.~\ref{fig:xsec_proc3_mtth}, 
the $VV$F contribution (magenta-dashed curves)
is dominated by the sum of the $\gamma\gamma$F, $\gamma Z$F and $Z\gamma$F channels 
in the entire $m(t\bar tH)$ range at high energies, $\sqrt{s}=10$ and 100~TeV.
On the other hand, for the CPV case with $(g,\xi)=(g_{\rm SM},0.1\pi)$, 
the difference, i.e. the $ZZ$F contribution is significant at high energies.
This is due to the amplitude with the dim6 $ttHZZ$ vertex, 
which is given by the filled histogram in Fig.~\ref{fig:mummup_mummupttxh_pk}.
We also observe the increase in the ann-$V$ amplitudes (blue-dotted curves),
whose contribution is significant only at the highest $m(t\bar tH)$ bins
with small cross section.
This is due to amplitudes with the dim5 $ttHH$ or $ttHZ$ vertex. 

\begin{figure}
	\includegraphics[width=0.6\textwidth,clip]{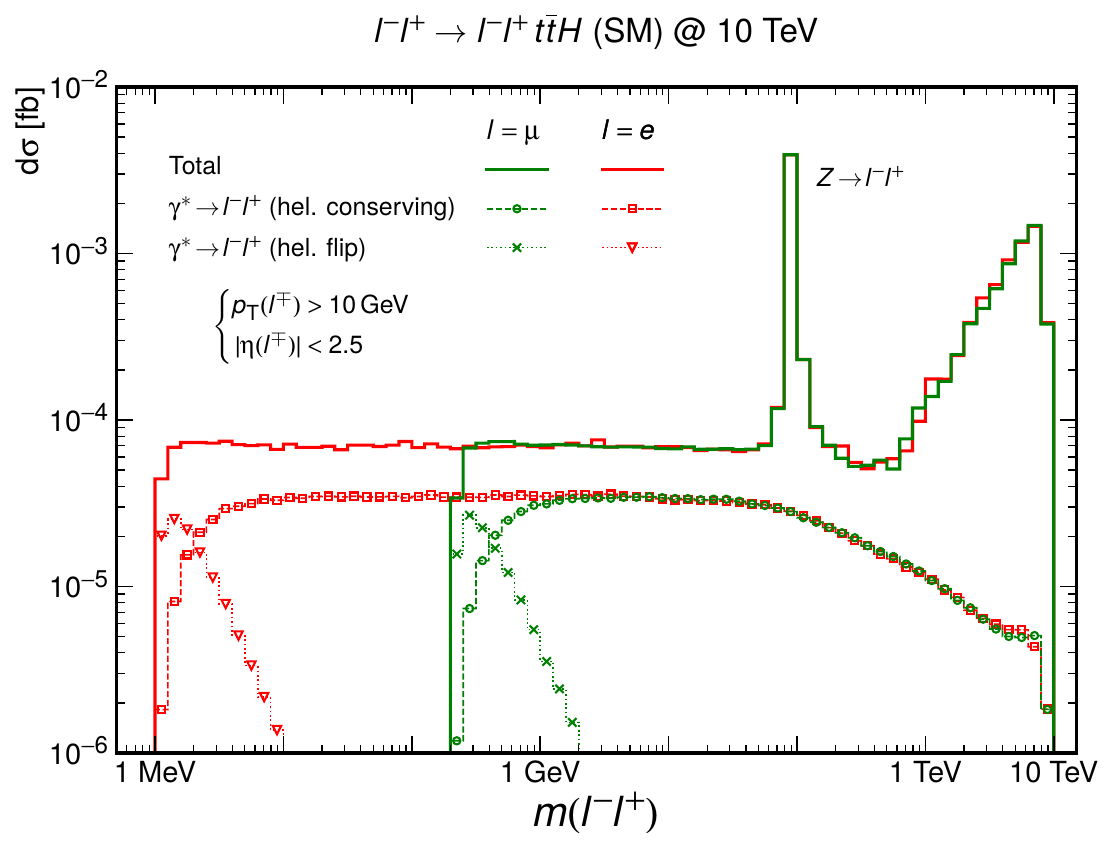}
\caption{Distributions of the $l^-l^+$ invariant mass for $l^-l^+\to l^-l^+t\bar{t}H$ in the SM at a collision energy of 10~TeV, where the kinematical cuts for the leptons in Eq.~\eqref{cut} are imposed.
    Solid lines denote the total cross sections for electrons (red) and muons (green). 
	Dashed lines show contributions from the $\gamma^*\to l^-l^+$ subprocesses with helicity-conserving
	interactions between the final charged leptons, while dotted lines correspond to the same subprocesses with
	helicity-flip interactions.
}
  \label{fig:lmlp_lmlpttxh_helicity_a2ll_log_mll_cut}
\end{figure}

Finally, 
in Fig.~\ref{fig:lmlp_lmlpttxh_helicity_a2ll_log_mll_cut}, 
we show the $l\bar l$ invariant mass distribution $d\sigma/dm(l^-l^+)$
for $l\bar{l}\to l\bar{l}t\bar{t}H$ 
at $\sqrt{s}=10$~TeV in the SM.
Here,
we show the distribution for events with observable final state $l^-$ and $l^+$,
which satisfy the cuts 
\begin{align}
  p_T(l^\mp)>10~{\rm GeV\quad and\quad} |\eta(l^\mp)|<2.5
  \label{cut}
\end{align}
by solid curves.
The contributions of all the diagrams with an $s$-channel $\gamma^*\to l^-l^+$ splitting amplitude 
are shown separately 
for the helicity-conserving, $h^{\rm final}_{l^-}=-h^{\rm final}_{l^+}$,
and helicity-flip, $h^{\rm final}_{l^-}=h^{\rm final}_{l^+}$, cases 
by dashed and dotted curves, respectively.
The channels with the $\gamma^*\to l^-l^+$ splitting vertex are marked by red stars in Fig.~\ref{fig:mummup_mummupttxh_pk}.
The red curves are for $l=e$, and the green curves are for $l=\mu$.
We observe that the amplitudes with a final state $\gamma^*\to l^-l^+$ splitting vertex dominates at low $m(l^-l^+)$ region,
and the helicity-flip amplitudes contribution is significant only when $m(l^-l^+)\sim2m_l$.

\section{Summary}
\label{sec:summary}

In this work, we have revisited the single-diagram-enhanced (SDE) multi-channel phase-space (MCPS) integration method of Ref.~\cite{Maltoni:2002qb} in the context of the recently proposed FD-gauge amplitudes~\cite{Hagiwara:2020tbx,Chen:2022gxv,Chen:2022xlg}. 
FD-gauge amplitudes are particularly well suited for SDE MCPS integration, since they avoid subtle cancellations among interfering contributions associated with the propagation of nearly on-shell vector bosons, including gluons, photons, and electroweak gauge bosons. 
Moreover, partial summation of FD-gauge amplitudes that share a common vector boson propagator saturates the physical scattering amplitude on and near the corresponding mass shell. 
As a result, the kinematical dependence of the squared individual amplitudes, $|{\cal M}_k|^2$, is largely dictated by invariant propagator factors and vertex functions, which reduce to Wigner $d$-functions in the rest (Breit) frame of the time-like (space-like) virtual vector boson and exhibit only mild residual dependence on phase-space variables.

To achieve numerically stable cross-section evaluations in the presence of collinear splittings between nearly on-shell photons and light charged leptons ($e,\mu$), we have improved the modular phase-space parametrization. 
All (virtual) $2\to2$ subprocesses are parametrized in the rest frame of the colliding particles, with the $z$-axis aligned along the three-momentum of one of them. 
This choice allows a stable integration over the full polar-angle range, $-1 \le \cos\theta \le 1$, by employing logarithmic variables of the inverse propagator factor, $\ln t'$, as in Eq.~\eqref{eq:lnt'}. 
In addition, the minimum value of the inverse propagator factor, $t'_{\rm min}$ at $\cos\theta=1$, is evaluated using expressions that avoid cancellations among terms of order $E^2/m_l^2$.

Further refinements were implemented in the \HELAS\ subroutines~\cite{Hagiwara:1990dw,Murayama:1992gi} to prevent loss of numerical precision in collinear configurations. 
The invariant four-momentum squared, $p^2$, of both on-shell and off-shell particles is stored explicitly as a fifth component of the momentum vector, thereby avoiding large cancellations in the evaluation of $p^2=(p^0)^2-(p^1)^2-(p^2)^2-(p^3)^2$. 
For time-like splittings $\gamma^* \to l\bar l$, transverse momenta of the leptons with respect to the virtual-photon direction are computed first, and the longitudinal components are subsequently reconstructed using the virtual-photon invariant mass and three-momentum.

With these improvements in place, we assessed the performance of the framework by studying Higgs- and top-quark production in $l\bar l$ collisions ($l=e,\mu$),
\begin{align}
  l\bar l \to \nu_l\bar\nu_l t\bar tH,\qquad
  l\bar\nu_l t\bar bH,\qquad
  l\bar l t\bar tH,
\end{align}
both in the SM and in the SMEFT including a dimension-six operator that modifies the top-Yukawa coupling.

We find that SDE MCPS integration achieves efficient Monte Carlo convergence for all processes considered, despite the presence of hundreds of contributing channels. 
By decomposing the full amplitudes into subamplitudes characterized by distinct propagator structures, such as $VV$ fusion ($VV$F), $Vl^+/l^-V$ scattering, and $l^-l^+$ annihilation, we identify sizable destructive interference effects between the $VV$F and $Vl^+/l^-V$ subamplitudes when the invariant mass of the latter subprocess is small. 
Contributions from diagrams involving dimension-six or dimension-five operators exhibit the expected power-like growth with increasing center-of-mass energy. 
The phase-space integration remains smooth even in regions where charged leptons are emitted collinearly with the beam, enabling reliable evaluation of total cross sections at the exact tree level. 
Helicity amplitudes involving lepton chirality flips are found to contribute at the $\mathcal{O}(0.1\%)$ level when the virtual-photon invariant mass is comparable to the charged-lepton mass.

In conclusion, the SDE MCPS integration framework, when combined with FD-gauge amplitudes and improved numerical stability in collinear regions, provides a reliable and efficient basis for event generation based on exact tree-level matrix elements.
It is therefore highly suitable for precision phenomenological studies at present and future colliders.

\section*{Acknowledgement}

This work is supported in part by the Japan Society for the Promotion of Science (JSPS) Grants-in-Aid for Scientific Research (KAKENHI), Grant Nos. 21H01077, 21K03583, 23K20840, and 24K07032.
F. M. acknowledges support from the Fonds de la Recherche Scientifique–FNRS (F.R.S.–FNRS) through the Interuniversity Institute of Nuclear Sciences (IISN) Project No. 4.4503.16 (MaxLHC), from the European Union (EU) through the European Cooperation in Science and Technology (COST) Action CA22130, “Comprehensive Multiboson Experiment-Theory Action” (COMETA), and from the Italian Ministry of University and Research (MUR) through the 2022 Projects of Relevant National Interest (PRIN 2022), Grant No. 2022RXEZCJ. F. M. also acknowledges support from the project “QIHEP—Exploring the foundations of quantum information in particle physics,” financed through the National Recovery and Resilience Plan (PNRR) with NextGenerationEU funds, in the context of the extended partnership PE00000023, National Quantum Science and Technology Institute (NQSTI) (CUP J33C24001210007).
Y.-J. Z. is supported by JSPS KAKENHI Grant No. 23K03403.
Feynman diagrams were drawn using TikZ-FeynHand [54,55].

\bibliography{MCPS}
\bibliographystyle{utphys}

\newpage
\begin{appendices}
\renewcommand{\thesubsection}{\arabic{subsection}}
\section{Modular phase-space subroutines}
\label{app:modular_units}
     
\subsection{Elements of the modular library units}

\paragraph{\texttt{dshat}}

\begin{Verbatim}[commandchars=\\\{\}]
  \textbf{call dshat(z_shat,what_min,what_max,mv,wv, q_out,dshat_out,ierr)}
\end{Verbatim}

\begin{figure}[htbp]
  \centering
    \includegraphics[width=0.45\textwidth]{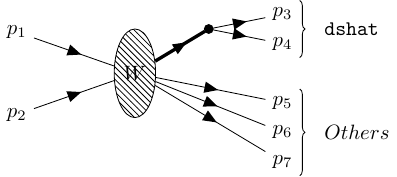}\qquad
    \includegraphics[width=0.45\textwidth]{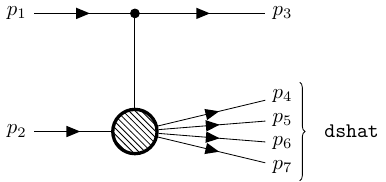}
  \caption{\texttt{dshat} usages.}
  \label{fig:dshat_usage}
\end{figure}

This module generates the $\hat s=M^2_{(m,\cdots,n)}$ value of a subsystem of the Feynman diagram using a given random number.  
It corresponds to the integration of $d\hat s/2\pi$ in Eq.~\eqref{eq:general-nbody-ps4}.  
It is used both for $s$-channel and $t$-channel elements (Fig.~\ref{fig:dshat_usage}).
\begin{enumerate}
	\setlength{\itemsep}{0cm}
\item To determine the \shat\ of the $s$-channel elements (Fig.~\ref{fig:dshat_usage}(left)).
  It includes the case of the Breit-Wigner resonance in the propagator. 
\item To determine the \shat\ of the subsystem in the $t$-channel element (Fig.~\ref{fig:dshat_usage}(right)).
\end{enumerate}

\noindent
\textsc{Inputs}
\begin{description}
	\setlength{\itemsep}{0cm}
\item[\Rnum{1}. \texttt{z\_shat}] Random number to generate \shat.
\item[\Rnum{2}. \texttt{what\_min}] Square root of the minimum \shat.  
  Typically the sum of masses of the subsystem particles.
\item[\Rnum{3}. \texttt{what\_max}] Square root of the maximum \shat.  
  Typically the collision energy minus the masses of other particles.
\item[\Rnum{4}. \texttt{mv}] Mass of the $s$-channel propagator (zero if no obvious propagator in the subsystem as $t$-channel case).
\item[\Rnum{5}. \texttt{wv}] Width of the $s$-channel propagator  (zero if no obvious propagator in the subsystem as $t$-channel case).
\end{description}

\noindent
\textsc{Outputs}
\begin{description}
	\setlength{\itemsep}{0cm}
\item[\Rnum{1}. \texttt{q\_out(0:4)}] Five-vector of the subsystem.  
  The calculated \shat\ is stored in the fifth element \texttt{q\_out(4)}.
\item[\Rnum{2}. \texttt{dshat\_out}] Jacobian of the \shat\ integration.
\item[\Rnum{3}. \texttt{ierr}] Error flag.
\end{description}

\paragraph{\texttt{ph2s}}

\begin{Verbatim}[commandchars=\\\{\}]
  \textbf{call ph2s(z_cos,z_phi,p, p1,p2, ph2s_out)}
\end{Verbatim}

\begin{figure}
  \centering
    \includegraphics[width=0.35\textwidth]{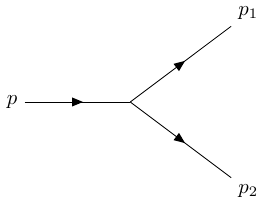}\qquad
    \includegraphics[width=0.35\textwidth]{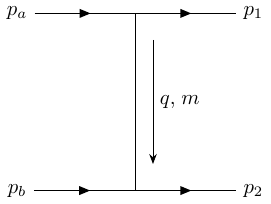}
  \caption{Left: Two-body splitting of a subsystem handled with \texttt{ph2s}.
  Right: Two-body collision in $t$-channel element handled with \texttt{ph2t}.}
  \label{fig:ph2}
\end{figure}

This module generates two five-vectors of a two-body $s$-channel splitting using two random numbers.  
It corresponds to Eq.~\eqref{eq:two-body}.
The parent system's five-vector which contains \shat\ must be given in advance (Fig.~\ref{fig:ph2}(left)).

\noindent
\textsc{Inputs}
\begin{description}
	\setlength{\itemsep}{0cm}
\item[\Rnum{1}. \texttt{z\_cos}, \texttt{z\_phi}] Random numbers for the two angular variables.
\item[\Rnum{2}. \texttt{p(0:4)}] Five-vector of the parent system.  
  \texttt{p(4)} contains \shat\ calculated by \texttt{dshat}.
\end{description}

\noindent
\textsc{Outputs}
\begin{description}
	\setlength{\itemsep}{0cm}
\item[\Rnum{1}. \texttt{p1(0:4)}, \texttt{p2(0:4)}] Five-vectors of the two final-state particles.
\item[\Rnum{2}. \texttt{ph2s\_out}] Jacobian of the two-body $s$-channel phase-space generation.
\end{description}

\paragraph{\texttt{ph2t}}

\begin{Verbatim}[commandchars=\\\{\}]
  \textbf{call ph2t(z1,z2,pa,pb, p1,p2,q_out, mv, ph2t_out,ierr)}
\end{Verbatim}

This module generates two five-vectors of a two-body $t$-channel process using two random numbers.  
It also corresponds to Eq.~\eqref{eq:two-body}.
Two input five-vectors of the subsystem must be given (Fig.~\ref{fig:ph2}(right)).

\noindent
\textsc{Inputs}
\begin{description}
	\setlength{\itemsep}{0cm}
\item[\Rnum{1}. \texttt{z1}, \texttt{z2}] Random numbers for phase-space generation.
\item[\Rnum{2}. \texttt{pa(0:4)}, \texttt{pb(0:4)}] Input five-vectors of the subsystem.
\item[\Rnum{3}. \texttt{mv}] Mass of the $t$-channel propagator.
\end{description}

\noindent
\textsc{Outputs}
\begin{description}\setlength{\itemsep}{0cm}
\item[\Rnum{1}. \texttt{p1(0:4)}, \texttt{p2(0:4)}] Five-vectors of the two final-state particles.
\item[\Rnum{2}. \texttt{q\_out(0:4)}] Five-vector of the $t$-channel propagator.
\item[\Rnum{3}. \texttt{ph2t\_out}] Jacobian of the two-body $t$-channel phase-space generation.
\item[\Rnum{4}. \texttt{ierr}] Error flag.
\end{description}

\subsection{Auxiliary modules for multi-point vertices: \texttt{ph2c}, \texttt{ph3c}, \texttt{ph4c}}

\begin{figure}
    \centering
\begin{tabular}{ccc}	     
    \includegraphics[width=0.32\textwidth]{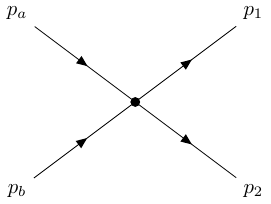} &
    \includegraphics[width=0.32\textwidth]{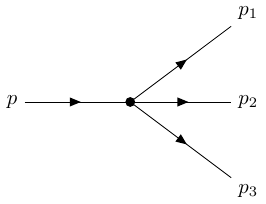} &
    \includegraphics[width=0.32\textwidth]{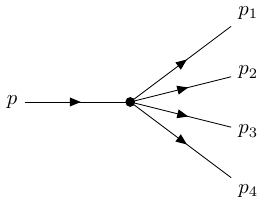} \\
    \begin{tabular}{l}
    Two-body splitting of \\ a subsystem at a \\ four-point vertex \\ handled with \texttt{ph2c}.
    \end{tabular} &
    \begin{tabular}{l}
    Three-body splitting of \\ a subsystem at a \\ four-point vertex \\ handled with \texttt{ph3c}.
    \end{tabular} &
    \begin{tabular}{l}
     Four-body splitting of \\ a subsystem at a \\ five-point vertex \\ handled with \texttt{ph4c}.
     \end{tabular}
\end{tabular}
  \caption{Multi-body splittings at four- and five-point vertices.}
  \label{fig:ph2c-ph4c}
\end{figure}

These are convenience modules for generating phase space at contact vertices (Fig.~\ref{fig:ph2c-ph4c}).  
They are internally composed of \texttt{dshat} and \texttt{ph2s}.

\paragraph{\texttt{ph2c}}

\begin{Verbatim}[commandchars=\\\{\}]
  \textbf{call ph2c(z_cos,z_phi, pa,pb, p1,p2, ph2c_out)}
\end{Verbatim}

\noindent
\textsc{Inputs}
\begin{description}\setlength{\itemsep}{0cm}
\item[\Rnum{1}. \texttt{z\_cos}, \texttt{z\_phi}] Random numbers for two-body phase space.
\item[\Rnum{2}. \texttt{pa(0:4)}, \texttt{pb(0:4)}] Initial five-vectors of the subsystem.  
  Their fifth components contain \shat.
\end{description}

\noindent
\textsc{Outputs}
\begin{description}\setlength{\itemsep}{0cm}
\item[\Rnum{1}. \texttt{p1(0:4)}, \texttt{p2(0:4)}] Generated five-vectors of the two particles.
\item[\Rnum{2}. \texttt{ph2c\_out}] Jacobian of the two-body contact phase-space generation.
\end{description}

\paragraph{\texttt{ph3c}}

\begin{Verbatim}[commandchars=\\\{\}]
  \textbf{call ph3c(z_shat,z_c123,z_p123,z_c23,z_p23, p, p1,p2,p3, ph3c_out)}
\end{Verbatim}

\noindent
\textsc{Inputs}
\begin{description}\setlength{\itemsep}{0cm}
\item[\Rnum{1}. \texttt{z\_shat}, \texttt{z\_c123}, \texttt{z\_p123}, \texttt{z\_c23}, \texttt{z\_p23}]  
  Five random numbers for three-body phase space.
\item[\Rnum{2}. \texttt{p(0:4)}] Five-vector of the splitting system.  
  \texttt{p(4)} contains \shat\ from \texttt{dshat}.
\end{description}

\noindent
\textsc{Outputs}
\begin{description}\setlength{\itemsep}{0cm}
\item[\Rnum{1}. \texttt{p1(0:4)}, \texttt{p2(0:4)}, \texttt{p3(0:4)}] Five-vectors of the three particles.
\item[\Rnum{2}. \texttt{ph3c\_out}] Jacobian of the three-body contact phase-space generation.
\end{description}

\paragraph{\texttt{ph4c}}

\begin{Verbatim}[commandchars=\\\{\}]
  \textbf{call ph4c(z_shat234,z_shat34,z_c1234,z_p1234,z_c234,z_p234,} &
                 \textbf{z_c34,z_p34, p, p1,p2,p3,p4, ph4c_out)}
\end{Verbatim}

\noindent
\textsc{Inputs}
\begin{description}\setlength{\itemsep}{0cm}
\item[\Rnum{1}. \texttt{z\_shat234}, \texttt{z\_shat34}, \texttt{z\_c1234}, \texttt{z\_p1234},  
              \texttt{z\_c234}, \texttt{z\_p234}, \texttt{z\_c34}, \texttt{z\_p34}]  
  Eight random numbers for four-body phase space.
\item[\Rnum{2}. \texttt{p(0:4)}] Five-vector of the splitting system.  
  \texttt{p(4)} contains \shat\ from \texttt{dshat}.
\end{description}

\noindent
\textsc{Outputs}
\begin{description}\setlength{\itemsep}{0cm}
\item[\Rnum{1}. \texttt{p1(0:4)}, \texttt{p2(0:4)}, \texttt{p3(0:4)}, \texttt{p4(0:4)}]  
  Five-vectors of the generated four particles.
\item[\Rnum{2}. \texttt{ph4c\_out}] Jacobian of the four-body contact phase-space generation.
\end{description}

\section{Modifications to \HELAS\ subroutines for singular vertices}
\label{app:helas}

The general-purpose phase-space generation library presented in this paper enables precise cross-section calculations for lepton collisions up to 100 TeV.  
However, achieving accurate results requires not only reliable phase-space generation but also precise evaluation of the scattering amplitudes.  
In this work, the amplitudes are computed using \HELAS, a helicity-amplitude calculation library.  
If the precision of \HELAS's internal kinematic calculations in singular regions is insufficient, the resulting cross sections in these regions will inevitably be inaccurate.  
To resolve this issue, we have modified the existing \HELAS\ library to improve the precision of kinematic calculations in singular regions, thereby enhancing the accuracy of both the amplitudes and the corresponding cross sections.  

\begin{figure}
  \centering
    \includegraphics[width=0.4\textwidth,clip]{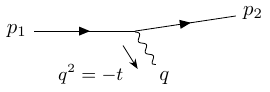}\qquad
    \includegraphics[width=0.4\textwidth,clip]{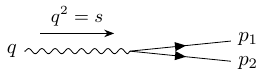}
\caption{Singular kinematics requiring modifications to \HELAS.}
\label{fig:singular_kinematics}
\end{figure}

The regions requiring modification include cases with very small invariant momentum-squared emissions (Fig.~\ref{fig:singular_kinematics}(left)), as well as cases where lepton pairs are emitted with extremely small angular separations (Fig.~\ref{fig:singular_kinematics}(right)).  
In both situations, it is essential to determine with high accuracy the value of $q^2$ of the photon associated with the two leptons.  
This value must match as closely as possible the one obtained during phase-space generation.  
Therefore, it is crucial first to determine the minimum value of $q^2$ accurately and then to establish a reliable procedure for computing $q^2$ from the two lepton momenta.  
The specific modifications implemented for each case are described below.  

\subsection{Emission of a nearly on-shell photon}

For the case of a nearly on-shell photon emitted from a charged lepton,
\begin{align}
  l(p_1) \to l(p_2) + \gamma(q),
  \label{eq:singular_t}
\end{align}
the invariant momentum squared, $q^2$, of the emitted photon is
\begin{align}
  q^2 = (p_1-p_2)^2.
  \label{eq:singular_t_qsq}
\end{align}
We introduce a positive variable, $t$, defined as
\begin{align}
  t & \equiv -q^2 = -(p_1-p_2)^2 \nonumber \\
    & = -p_1^2-p_2^2+2(E_1 E_2-|{\vec{p}}_1||{\vec p}_2|\cos\theta_{12}) \nonumber \\
    & = -p_1^2-p_2^2+2(E_1 E_2-|{\vec{p}}_1||{\vec p}_2|) + 2|{\vec{p}}_1||{\vec p}_2|(1-\cos\theta_{12}) \nonumber\\
    & = t_\mathrm{min} + 2|{\vec{p}}_1||{\vec p}_2|(1-\cos\theta_{12}).
  \label{eq:singular_t_definition}
\end{align}
Here, $\theta_{12}$ is the relative angle between $p_1$ and $p_2$, and $t_\mathrm{min}$ is the minimum value of $t$:
\begin{align}
 t_\mathrm{min}
  = -p_1^2-p_2^2+2(E_1 E_2-|{\vec{p}}_1||{\vec p}_2|).
\end{align}
The expression for $t_\mathrm{min}$ can also be written, using Eq.~\eqref{eq:tmin2} in Sec.~\ref{sec:PSgen}, as
\begin{align}
  t_\mathrm{min} & =  -p_1^2-p_2^2+2(E_1 E_2-|\vec{p}_1||\vec{p}_2|) \nonumber\\
  & = -p_1^2-p_2^2+2 \frac{m_1^2 E_2^2+m_2^2 E_1^2}{E_1 E_2+|\vec{p}_1||\vec{p}_2|}.
\end{align}

\subsection{Emission of a lepton pair with a small opening angle}

For the case of a nearly on-shell photon splitting into a lepton pair with a very small opening angle,
\begin{align}
  \gamma(q) \to l(p_1) + \overline{l}(p_2),
  \label{eq:singular_s}
\end{align}
the invariant momentum squared of the photon is
\begin{align}
  q^2 = (p_1+p_2)^2.
  \label{eq:singular_s_qsq}
\end{align}
We define $s$ as
\begin{align}
  s & \equiv q^2 = (p_1+p_2)^2 \nonumber\\
    & = p_1^2+p_2^2+2(E_1 E_2-|{\vec{p}}_1||{\vec p}_2|\cos\theta_{12}) \nonumber\\
    & = p_1^2+p_2^2+2(E_1 E_2-|{\vec{p}}_1||{\vec p}_2|) + 2|{\vec{p}}_1||{\vec p}_2|(1-\cos\theta_{12}) \nonumber\\
    & = s_\mathrm{min} + 2|{\vec{p}}_1||{\vec p}_2|(1-\cos\theta_{12}).
  \label{eq:singular_s_definition}
\end{align}
Here, $\theta_{12}$ is the opening angle between $p_1$ and $p_2$, and $s_\mathrm{min}$ is the minimum value of $s$:
\begin{align}
 s_\mathrm{min}
  = 4m_l^2.
\end{align}

The relative angle $\theta_{12}$ between two leptons must also be calculated with high precision.  
In particular, when two charged leptons are emitted with a very small opening angle, $\theta_{12}$ becomes extremely small, and special care is required in its evaluation.  
To stably calculate $\theta_{12}$ over a wide range of values, we use the following formula:
\begin{align}
\theta_{12} = \mathtt{atan2}(|\vec{p}_1\times\vec{p}_2|,\: \vec{p}_1\cdot\vec{p}_2),
\end{align}
where $|\vec{p}_1\times\vec{p}_2|$ is the magnitude of the cross product of $\vec{p}_1$ and $\vec{p}_2$, and $\vec{p}_1\cdot\vec{p}_2$ is their dot product.  
The function \texttt{atan2} computes the angle as
\begin{align}
\theta = \mathtt{atan2}(y,x),
\end{align}
and is numerically stable, returning angles in the range $[0,\pi]$.
\end{appendices}

\end{document}